\documentclass[12pt,aps,preprint,nofootinbib]{revtex4}

\usepackage{graphicx}
\usepackage{cancel}
\usepackage{amssymb}
\usepackage{textcomp}
\usepackage{amsmath}
\usepackage{bm}
\usepackage{times}
\usepackage{epsfig}
\usepackage{epstopdf}
\usepackage{color}
\usepackage{multirow}

\def\lsim{\mathrel{\raise.3ex\hbox{$<$\kern-.75em\lower1ex\hbox{$\sim$}}}}
\def\gsim{\mathrel{\raise.3ex\hbox{$>$\kern-.75em\lower1ex\hbox{$\sim$}}}}

\def\beq{\begin{equation}}
\def\eeq{\end{equation}}
\def\be{\begin{equation}}
\def\ee{\end{equation}}
\def\bea{\begin{eqnarray}}
\def\eea{\end{eqnarray}}

\def\mz{m_Z}

\def\mh{m_{h^0}}
\def\mH{m_{H^0}}
\def\mHpm{m_{H^\pm}}
\def\ma{m_{A}}
\def\BR{\rm Br}

\def\gev{\,{\rm GeV}}

\def\to{\rightarrow}

\def\ww{W^{+}W^{-}}
\def\tautau{\tau^{+}\tau^{-}}

\def\gaga{\gamma\gamma}

\begin{document}

\preprint{~~PITT-PACC 1201}

\title{MSSM Higgs Bosons at The LHC }

%\vskip -1cm

\author{Neil Christensen$^{\bf a}$}
%\email{neilc@pitt.edu}

\author{Tao Han$^{\bf a}$}
%\email{than@pitt.edu}

\author{Shufang Su$^{\bf b}$}
%\author{Shufang Su}
%\email{shufang@physics.arizona.edu}

\affiliation{
$^{\bf a}$  Pittsburgh Particle physics, Astrophysics, and Cosmology Center, Department of Physics $\&$ Astronomy, University of Pittsburgh, 3941 O'Hara St., Pittsburgh, PA 15260, USA\\
$^{\bf b}$  Department of Physics, University of Arizona, P.O.Box 210081, Tucson, AZ 85721, USA
}

\begin{abstract}
The recent results on Higgs boson searches from LHC experiments provide significant guidance in exploring the Minimal Supersymmetric (SUSY) Standard Model (MSSM) Higgs sector.
If we accept the existence of a SM-like Higgs boson in the mass window of 123 GeV$-$127 GeV as indicated by the observed $\gamma\gamma$ events, there are two distinct mass regions (in $\ma$) left in the MSSM Higgs sector:  (a) the lighter CP-even Higgs boson being SM-like and the non-SM-like Higgs bosons all heavy and nearly degenerate above 300 GeV (an extended decoupling region); (b) the heavier CP-even Higgs boson being SM-like and the neutral non-SM-like Higgs bosons all nearly degenerate around 100 GeV (a small non-decoupling region). 
On the other hand, due to the strong correlation between the Higgs decays to $\ww$ and to $\gaga$ predicted in the MSSM, the apparent absence of a $\ww$ final state signal is in direct conflict with the $\gaga$ peak.  
If we consider the $\ww$ channel on its own, the absence of the $\ww$ signal would imply that the SM-like Higgs boson has reduced coupling to $W^\pm$, and that the other non-SM-like Higgs bosons should not be too heavy and do not decouple.
%If we consider the $\ww$ channel on its own, if the absence of  the $\ww$ signal persists, it would imply that the SM-like Higgs boson has reduced couplings to $W^{\pm}$, and that the other non-SM-like Higgs bosons should not be too heavy. 
%
If both the $\gaga$ excess and the absence of a $\ww$ signal  continue, new physics beyond the MSSM will be required.
A similar correlation exists between the $W^+W^-$ and $\tau^+\tau^-$ channels: a reduced  $W^+W^-$ channel would force the $\tau^+\tau^-$ channel to be larger.
Future searches for the SM-like Higgs boson at the LHC will provide critical tests for the MSSM prediction. We also study the signals predicted for the non-SM-like Higgs bosons and emphasize the potential importance of the electroweak processes $pp\to H^{+}H^{-},\ H^{\pm} A^{0}$, 
which are independent of the SUSY parameters except for their masses. In addition, there may be sizable contributions from $pp\to H^{\pm} h^{0},\ A^{0} h^{0}$ and $W^{\pm} H^{0},\ Z H^{0}$ in the low-mass non-decoupling region, which may serve to discriminate the model parameters. 
We allow variations of the relevant SUSY parameters in a broad range and demonstrate the correlations and constraints on these parameters and associated SUSY particles. 
\end{abstract}

\maketitle

\section{Introduction}

The outstanding performance of the LHC experiments has led the field of high energy physics into unprecedented territory in the energy and luminosity frontier. Major discoveries at the Tera-scale are highly anticipated. 
One of the primary motivations for LHC experiments is the exploration for the mechanism of electroweak symmetry breaking. 
Among the many possibilities for new physics beyond the Standard Model (SM), Supersymmetry (SUSY) provides a natural framework for electroweak symmetry breaking. Although the signals for SUSY are still elusive at the LHC, significant progress has been made in the search for the Higgs boson. Recently, the ATLAS and CMS Collaborations have reported their updated searches for the SM Higgs boson \cite{ATLASH,CMSH,ATLASHnew,CMSHnew}.  Continuously extending the previous LEP2 mass bound for a SM Higgs (114.4 GeV) \cite{LEP2H}, the LHC search has reached an impressively wide coverage for the mass parameters. The main conclusions are
\begin{itemize}
\item A SM-like Higgs boson was excluded at $95\%$ C.L.~in the mass range of $<117.5$ GeV, in  $118.5$ GeV $-122.5$ GeV \cite{ATLASHnew}
and in $127.5$ GeV$-600$ GeV \cite{CMSHnew}, thus leaving a  $95\%$ C.L.~mass window
\begin{equation}
117.5 \gev - 118.5\gev,\quad\quad 122.5 \gev - 127.5\gev.
\label{eq:mhwide}
\end{equation}
\item An excess of events above the background expectation was observed in the final state of $\gamma\gamma$,  at 126 GeV with 2.5$\sigma$ by the ATLAS Collaboration \cite{ATLASHnew} and at 125 GeV with 2.8$\sigma$ by the CMS Collaboration \cite{CMSHnew}, thus giving a tantalizing hint for a Higgs boson in the mass range
\begin{equation}
\sim 125 \gev \pm 2 \gev .
\label{eq:mh1}
\end{equation}
\item No significant excess of events above the SM backgrounds was observed in the final states of $W^{+}W^{-},\ \tau^{+}\tau^{-},\ b\bar b$, however, a small excess has been seen in $ZZ\to 4\ell$ \cite{ATLASHnew,CMSHnew}.
\end{itemize}
Although inconclusive with the current data, each one of the statements above 
has significant impact on our understanding of electroweak symmetry breaking and thus guides us for the next step of the Higgs search. 

In this paper, we study the consequences of the above findings on the Higgs sector within the framework of the Minimal Supersymmetric Standard Model (MSSM) \cite{Gunion:1989we,Djouadi:2005gj}.  
We first recollect the existing constraints from all the current bounds of the direct searches from LEP2 \cite{LEP2H}, the Tevatron \cite{CDFD0} and the LHC \cite{ATLASH,CMSH,ATLASHnew,CMSHnew,CMS-tautau,CMSA0,HpmCMS,HpmATLAS}.
If we accept the existence of a CP-even Higgs boson in the mass range of Eq.~(\ref{eq:mh1}) 
as observed in the $\gamma\gamma$ mode, 
we then find very interesting features for the MSSM Higgs sector and some other relevant SUSY parameters. There are two distinctive scenarios, both of which incorporate a SM-like Higgs boson.
\begin{itemize}
\item[(a)] ``Decoupling" regime with $m_A \gtrsim 300$ GeV \cite{HHaber}:  
The light CP-even Higgs $h^0$ is in the mass range of Eq.~(\ref{eq:mh1}) and SM-like.  The non-SM-like Higgs bosons (heavy CP-even state $H^0$, CP-odd state $A^0$ and the charged state $H^\pm$) are all heavy and nearly degenerate, with masses around $m_A $. 
\item[(b)] ``Non-decoupling" regime with $m_A$ around  $95 - 130$ GeV: The heavy CP-even Higgs $H^0$ is in the mass range of
 Eq.~(\ref{eq:mh1}) and SM-like, while the light CP-even Higgs $h^0$ is non-SM-like.  The masses of the light CP-even Higgs and the CP-odd Higgs are nearly degenerate while the charged Higgs is nearly degenerate with $\mH$ \cite{Gerard:2007kn}. 
\end{itemize}
Each of these two cases predicts unique signatures to establish the nature of the MSSM at the LHC. 
While the current searches continue to improve in the future runs and the standard electroweak production processes 
\begin{equation}
pp\to W^{\pm} h^{0} (H^{0}),\ Z h^{0} (H^{0}),\ {\rm and}\ q\bar q h ^{0} (H^{0}), 
\end{equation}
are still available, we would like to point out the potential importance of the electroweak processes 
\begin{equation}
pp\to H^{+}H^{-},\ H^{\pm} A^{0}, 
\end{equation}
which are via pure gauge interaction and independent of the SUSY parameters except for their masses.
In addition, there may be sizable contributions from 
\begin{equation}
pp\to H^{\pm} h^{0},\ A^{0} h^{0}
\end{equation}
in the low-mass non-decoupling region, which may be used to distinguish the model-parameters. 

On the other hand, due to the strong positive correlation between the Higgs decays to $\ww$ and to $\gaga$ predicted in the MSSM, the observed $\gaga$ signal and the apparent absence of the $\ww$ final state signal near the peak would be mutually exclusive to each other. Namely, the suppression of the $\ww$ channel would automatically reduce the $\gaga$ channel, in direct conflict with the observed $\gaga$ excess. 
We also found another interesting inverse correlation between the Higgs decays to $\ww$ and to $\tautau$. 
In this case, the suppression to the $\ww$ channel would automatically force the $\tautau$ channel to be bigger.
 If the deficit in the $\ww$ channel persists and the result is strengthened for an extended mass range in the future run at the LHC, it  would imply that the SM-like Higgs boson has reduced couplings to $W^{\pm},\ Z$, rendering it less SM-like. Consequently, the other non-SM-like Higgs bosons cannot be deeply into the decoupling regime, and thus cannot be too heavy, typically below 350 GeV, making them more accessible at the LHC.
Moreover, if the excess in the $\gaga$ channel and the absence of an excess in the $\ww$ channel continue to be strengthened at the LHC, new physics beyond the MSSM will be required\footnote{We note that there may be an exception \cite{Carena:2011aa} when a light stau with large mixing and large $\tan\beta$
could help to enhance the branching fraction of the $h^0 \rightarrow \gamma\gamma$ channel.}.

In the current study, we wish to focus on the essentials of the Higgs sector in the MSSM and to minimize the effects from other SUSY sectors \cite{Baer:2011ab, Feng:2011ew}. 
Nevertheless, a few other SUSY parameters, the Higgs mixing $\mu$, the stop mixing $A_{t}$ and the stop soft SUSY masses $M_{3SQ}$ and $M_{3SU}$, play crucial roles in the Higgs sector. 
We explore the effects of the Higgs searches on those SUSY parameters by scanning them in a wide range and we find clear correlations and thus predictions on them. 

The rest of the paper is organized as follows. 
In Sec.~\ref{sec:MSSMHiggs}, we give a brief introduction to the MSSM Higgs sector, focusing on the mass corrections as well as the coupling structures that are relevant for our discussion below. 
In Sec.~\ref{sec:ParameterRegion}, we discuss our broad scanning of the relevant MSSM parameters by imposing the existing constraints of the direct searches from LEP2, the Tevatron and the LHC. We obtain the surviving regions for the Higgs mass and the other parameters.   %
With the further improvement expected at the LHC with 8 TeV and 14 TeV, 
we discuss the consequence of the SM-like Higgs boson searches on the 
MSSM Higgs sector in Sec.~\ref{sec:Future}.
In light of the current direct search,  we present the dominant production and decay channels as well as the characteristic channels for the non-SM Higgs bosons in Sec.~\ref{sec:LHCSearch} to test the MSSM in the future runs. We conclude in Sec.~\ref{sec:Conclude}. 

%%%%%%%%%%%%%%%%%%%%%%%%%%%%%%%%%%%%%%%%%%%%%%%%

\section{MSSM Higgs Sector}
\label{sec:MSSMHiggs}

\subsection{Masses}

Unlike in the Standard Model where the Higgs mass is a free parameter in the theory, in the MSSM with two Higgs doublets, the masses of the five physical Higgs bosons (two CP-even Higgs bosons $h^{0}$ and $H^{0}$, one CP-odd state $A^{0}$ and a pair of charged Higgs  $H^{\pm}$) at tree level and the mixing angle of the CP-even Higgs bosons $\alpha$, 
can be expressed in terms of two parameters \cite{Gunion:1989we,Djouadi:2005gj},  
conventionally chosen as
the mass of $A^0\ (m_{A})$ and the ratio of the two vacuum expectation values ($\tan\beta=v_u/v_d$): 
% \cite{Gunion:1989we}:
\begin{eqnarray}
&&
m_{h^0, H^0}^2 = \frac{1}{2} \left( (m_A^2 + m_Z^2) \mp \sqrt{(m_A^2 - m_Z^2)^2 + 4 m_A^2 m_Z^2 \sin^22 \beta} \right),\\ 
&&
m_{H^\pm}^2 = m_A^2 + m_W^2, \quad \cos^{2}(\beta-\alpha) = {
{\mh^{2} (m_{Z}^{2} - \mh^{2} )} \over {\ma^{2} (\mH^{2} - \mh^{2})}
 } .
\end{eqnarray}
We will call the CP-even Higgs boson that couples to $\ww/ZZ$ more strongly the ``Standard Model-like'' Higgs as we discuss it's properties further in the next section. For a low-mass $m_A \lesssim m_Z/2$, or a high mass $m_A \gtrsim 2m_Z$, the Higgs boson masses can be approximated by
\begin{eqnarray}
&&m_{h^0} \approx  {\rm min}\ \{m_A, m_Z\}  |\cos 2 \beta|, 
\ \ \ m_{H^0} \approx {\rm max}\ \{m_A, m_Z\}, \ \ \ m_{H^\pm} \approx {\rm max}\ \{m_A, m_W\}.
\end{eqnarray}

Because of the large Yukawa coupling of the top quark and the possible large mixing of the left-right top squark, the CP-even Higgs boson masses receive significant radiative corrections.  For nearly degenerate soft SUSY breaking parameters in the stop sector: $M_{3SQ}^2 \sim M_{3SU}^2 \sim M_S^2$, the correction to the mass of the SM-like Higgs can be approximately expressed as \footnote{For the non-decoupling case when $H^0$ is SM-like, this expression also applies to the correction of $m_{H^0}$.} \cite{Carena:1995wu, Carena:1995bx}
\begin{eqnarray}
&& \Delta m_{h^0}^{2} \approx  \frac{3}{4 \pi^2}\frac{m_t^4}{v^2}\left[\ln \left(\frac{M_{S}^2}{m_t^2} \right)
+ \frac{\tilde{A}_t^2}{M_S^2} \left( 1- \frac{\tilde{A}_t^2}{12 M_S^2}\right) \right] + \ldots,
\label{eq:deltamh}
\end{eqnarray}
where the mixing in the stop sector is given by 
\begin{eqnarray}
 \quad \tilde{A}_{t} = A_{t} - \mu \cot\beta.
 \label{eq:At}
\end{eqnarray}
For $\tilde{A}_t=0$, the corrections to the Higgs mass from the stop sector is minimized, this is the so-called ``$m_{h}^{\rm min}$" scenario \cite{Carena:2002qg}, where the radiative contributions could give rise to a Higgs mass as high as 117 GeV including a dominant two-loop corrections for a stop mass up to about 2 TeV. %
For $\tilde{A}_t=\sqrt{6}M_S$, the second term in Eq.~(\ref{eq:deltamh}) is maximized, leading to the so-called ``$m_{h}^{\rm max}$" scenario \cite{Carena:2002qg}, where a maximum Higgs mass of about 127 GeV can be reached in such a scenario. To obtain a relatively large correction to the 
light CP-even Higgs mass, relatively heavy stop masses (at least for one of the stops) as well as large LR mixing in the stop sector is needed.  When two-loop corrections of the oder of ${\cal O}(\alpha \alpha_s)$ are included, there is an asymmetric contribution to the Higgs mass from the $A_t$ term, where postitive $A_t$ gives a few GeV larger correction compared to the negative $A_t$ case. 
Note that there are  uncertainties of a few GeV coming from higher loop orders, as well as from the uncertainties in $m_t$, $\alpha_s$, etc..  For detailed calculations and results on the Higgs mass corrections in the MSSM, see Refs.~\cite{Carena:1995bx, Heinemeyer:1998np,Degrassi:2002fi}.  
 
 \subsection{Couplings to SM particles}

 Another important aspect is the couplings of the Higgs bosons to the SM particles \cite{Gunion:1989we,Djouadi:2005gj}.  
The couplings to gauge bosons behave like 
\begin{eqnarray}
\nonumber
&& W^{+}W^{-}h^{0},\  ZZh^{0},\  ZH^{0}A^{0},\ WH^{\pm}H^{0} \propto g \sin(\beta-\alpha),  \\
\nonumber
&& W^{+}W^{-}H^{0},\  ZZH^{0},\  Zh^{0}A^{0},\ WH^{\pm}h^{0} \propto g \cos(\beta-\alpha), \\
%\nonumber
&&  \gamma H^{+}H^{-},\  Z H^{+}H^{-},\ WH^{\pm} A^{0} \propto g.
\end{eqnarray}
where $g$ is the weak coupling.
Either $h^{0}$ or $H^{0}$ can be SM-like when it has a stronger coupling to $\ww$ and $ZZ$. 
In the ``decoupling limit'' $m_A \gg m_Z$, 
$\sin(\beta-\alpha) \sim 1,\  \cos(\beta-\alpha) \sim 0$. Then $h^{0}$ is light and SM-like, while all the other Higgs bosons are heavy, nearly degenerate, and the $H^{0}$ coupling to $\ww,\ ZZ$ is highly suppressed. 
In the non-decoupling region $m_A \sim m_Z$, 
$\sin(\beta-\alpha) \sim 0,\  \cos(\beta-\alpha) \sim 1$. Then $H^{0}$ is SM-like, while all the other neutral Higgs bosons are lighter, nearly degenerate, and the $h^{0}$ coupling to $\ww$ and $ZZ$ are highly suppressed. 
Note that the couplings of the pair of Higgs bosons $H^{+}H^{-},\ H^{\pm} A^0$ to a gauge boson are of pure gauge coupling strength and are independent of the model parameters.

The tree-level couplings of the Higgs bosons to the SM fermions scale as
\begin{eqnarray}
\nonumber
&& h^{0}d\bar d: m_{d} [\sin(\beta-\alpha) - \tan\beta \cos(\beta-\alpha)],\ \quad 
h^{0}u\bar u: m_{u} [\sin(\beta-\alpha) + \cot\beta \cos(\beta-\alpha)], \\
\nonumber
&& H^{0} d\bar d: m_{d} [\cos(\beta-\alpha) + \tan\beta \sin(\beta-\alpha)],\ \quad 
H^{0} u \bar u: m_{u} [\cos(\beta-\alpha) - \cot\beta \sin(\beta-\alpha)], \\
%\nonumber
&& A^{0}d \bar d: m_{d}\tan\beta\  \gamma_{5},\quad A^{0}u\bar u: m_{u}\cot\beta\ \gamma_{5}, 
\quad H^{\pm}d \bar{u} :  m_{d} \tan\beta\  P_{R} + \ m_{u} \cot\beta\ P_{L}, 
\end{eqnarray}
where $P_{L,R}$ are the left-  and right-projection operators. In the decoupling limit, these result in the branching fractions for the leading channels, 
 \begin{eqnarray}
 \nonumber
&& {\BR}(b\bar b): {\BR}(\tau\bar \tau): {\BR}(t\bar t) \approx
3m_{b}^{2}\tan^{2}\beta : m_{\tau}^{2}\tan^{2}\beta : 3m_{t}^{2}/\tan^{2}\beta\quad {\rm for}\ H^0,\ A^0,\\
&& {\BR}(t\bar b): {\BR}(\tau\bar \nu) \approx
3(m_{b}^{2}\tan^{2}\beta + m_{t}^{2}/\tan^{2}\beta) : m_{\tau}^{2}\tan^{2}\beta \quad {\rm for}\ H^\pm.
\label{eq:brs}
 \end{eqnarray}
In the non-decoupling limit, 
the couplings of $H^0$ to the SM fermions become SM-like, while the above branching fraction relations still approximately hold for $h^0$, $A^0$ and $H^\pm$, except that the top quark channel would not be kinematically open.

Radiative corrections can change the above relations \cite{Carena:1995bx,Carena:1999bh,Carena:1998gk, deltahb}, in particular for the channels involving $b$ and $t$.
Both the mixing in the Higgs sector, as well as the top and bottom Yukawa couplings could receive relatively large loop corrections in certain regions of the MSSM parameter space.  
In particular, a large positive MSSM correction to $\Delta m_b$, defined as \cite{deltahb}
\begin{equation}
m_b = h_b v_d ( 1+ \Delta m_b), 
\label{eq:deltamb}
\end{equation}
where $h_b$ is the bottom Yukawa coupling,  
leads to a suppression in $h^0/H^0 \rightarrow  b \bar b$ decay, resulting in an enhancement in $h^0/H^0 \rightarrow \gamma\gamma, \ww$ and $ZZ$.  For more discussion on this, see Sec.~\ref{sec:Discussion}.   

%%%%%%%%%%%%%%%%%%%%
\subsection{Parameter scan}

We wish to examine the theoretical parameter space of the MSSM Higgs sector as generally as possible. To do so, we  
study the 6-dimensional parameter space in the ranges
\begin{eqnarray}
\nonumber
& 3 < \tan\beta < 55, \quad
 50\ {\rm GeV}< m_A < 500\ {\rm GeV}, \quad 100\gev <  \mu < 1000\ {\rm GeV}, &\\
&   100\gev < M_{3SU}, M_{3SQ} < 2000\ {\rm GeV}, \quad  
    {-4000\ \gev}< A_t < 4000\  {\rm GeV}. &
\label{eq:para}
\end{eqnarray}
The lower limit of $\tan\beta$ is chosen based on the LEP2 Higgs search exclusion \cite{LEP2H}, while the upper limit takes into account the perturbativity of the bottom Yukawa coupling.  We limit $m_A$ within 500 GeV since it already reaches  the decoupling region. A higher value for $m_A$ simply pushes up the nearly degenerate masses for $H^0$, $A^0$ and $H^\pm$ while it does not affect the phenomenology of the light CP-even Higgs $h^0$. The ranges of $M_{3SU}, M_{3SQ}$ and $\mu$ are motivated by the naturalness consideration, as well as the current collider search limits for SUSY particles. The range of $A_t$ is chosen to cover both the limiting scenarios of $m_{h}^{\rm min}$ and $m_{h}^{\rm max}$ as mentioned below Eq.~(\ref{eq:At}).
It turns out that $A_{t}$ is of critical importance. It dictates the mixing of the stop sector. In turn, it has significant effects  on the radiative corrections to the Higgs mass, mixing in the CP-even Higgs sector, $gg\rightarrow h^0/H^0 \rightarrow \gamma\gamma$ via stop loops, as well as a contribution to $b\to s \gamma$ through a chargino-stop loop. Indeed, we find that changing the sign of $A_{t}$ could lead to potentially distinctive results.

The effects of the other SUSY parameters on the Higgs sector phenomenology is small. Therefore we take the simplified approach in our analyses to decouple their effects by setting the other SUSY soft mass scales to be 3 TeV. Some notable effects in special cases will be discussed in Sec.~\ref{sec:Discussion}.

%%%%%%%%%%%%%%%%%%%%%%%%%%%%%%%%%%%%%%%%%%%%%%%%%

\section{The Higgs Sector in light of direct searches}
\label{sec:ParameterRegion} 

We used FeynHiggs 2.8.6
\cite{Frank:2006yh,Degrassi:2002fi,Heinemeyer:1998np,Heinemeyer:1998yj}
to calculate the mass spectrum and other SUSY parameters, as well as the Higgs decay widths and branching fractions and dominant Higgs production cross sections. We used HiggsBound 3.6.1beta \cite{Bechtle:2008jh,Bechtle:2011sb} to check the exclusion constraints from LEP2 \cite{LEP2H}, the Tevatron \cite{CDFD0} and the LHC \cite{ATLASH,CMSH,CMS-tautau,CMSA0,HpmCMS,HpmATLAS}.
In practice, we generated a large Monte Carlo sample to scan over the multiple dimensional parameter region and test against the experimental constraints. 
For the following presentation, the allowed points (or regions) in the plots are indicative of consistent theoretical solutions satisfying experimental constraints, but are not meant to span the complete space of possible solutions.

\begin{figure}[tb]
\includegraphics[scale=1,width=8cm]{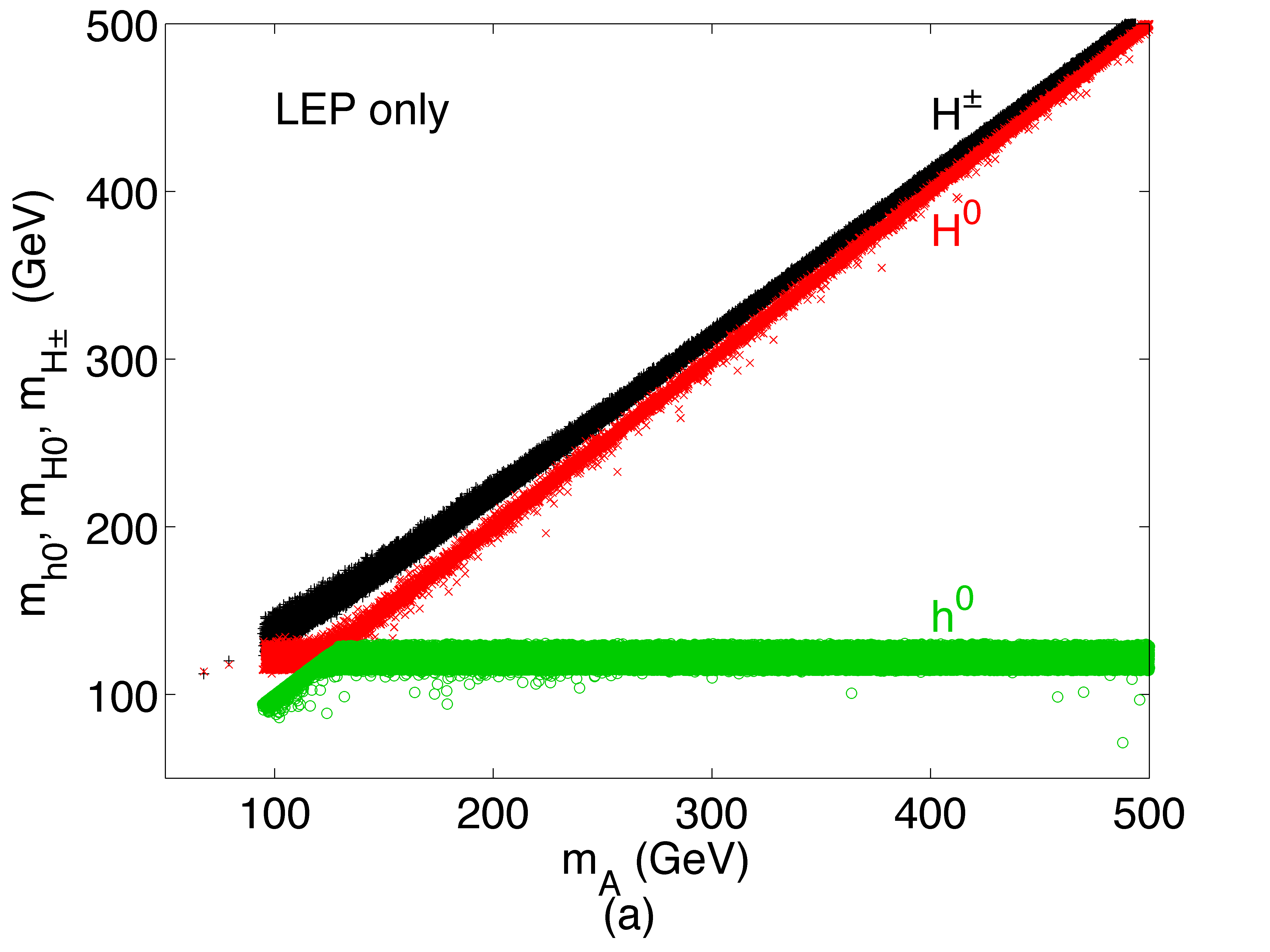}
\includegraphics[scale=1,width=8cm]{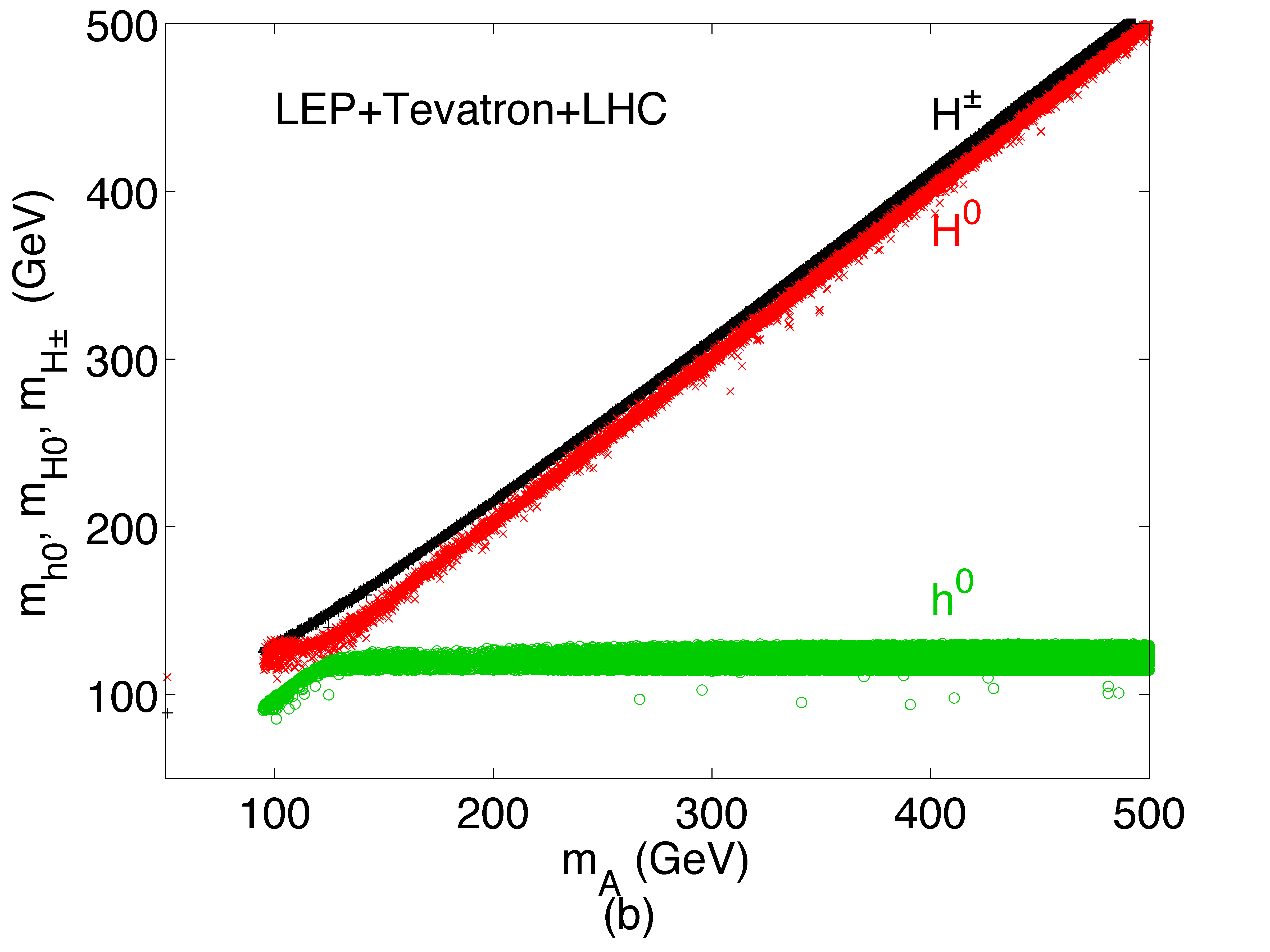}
\caption{Allowed mass regions versus $\ma$ for the light CP even $h^0$ (green circles), and 
the heavy CP even $H^0$ (red crosses), and the charged $H^{\pm}$ (black pluses),
scanned over the parameter ranges given in Eq.~(\ref{eq:para}), 
for (a) satisfying the LEP2 bounds, and (b) further including the bounds from the Tevatron and the LHC.
 }
\label{fig:mass}
\end{figure}

\begin{figure}[tb]
\includegraphics[scale=1,width=8cm]{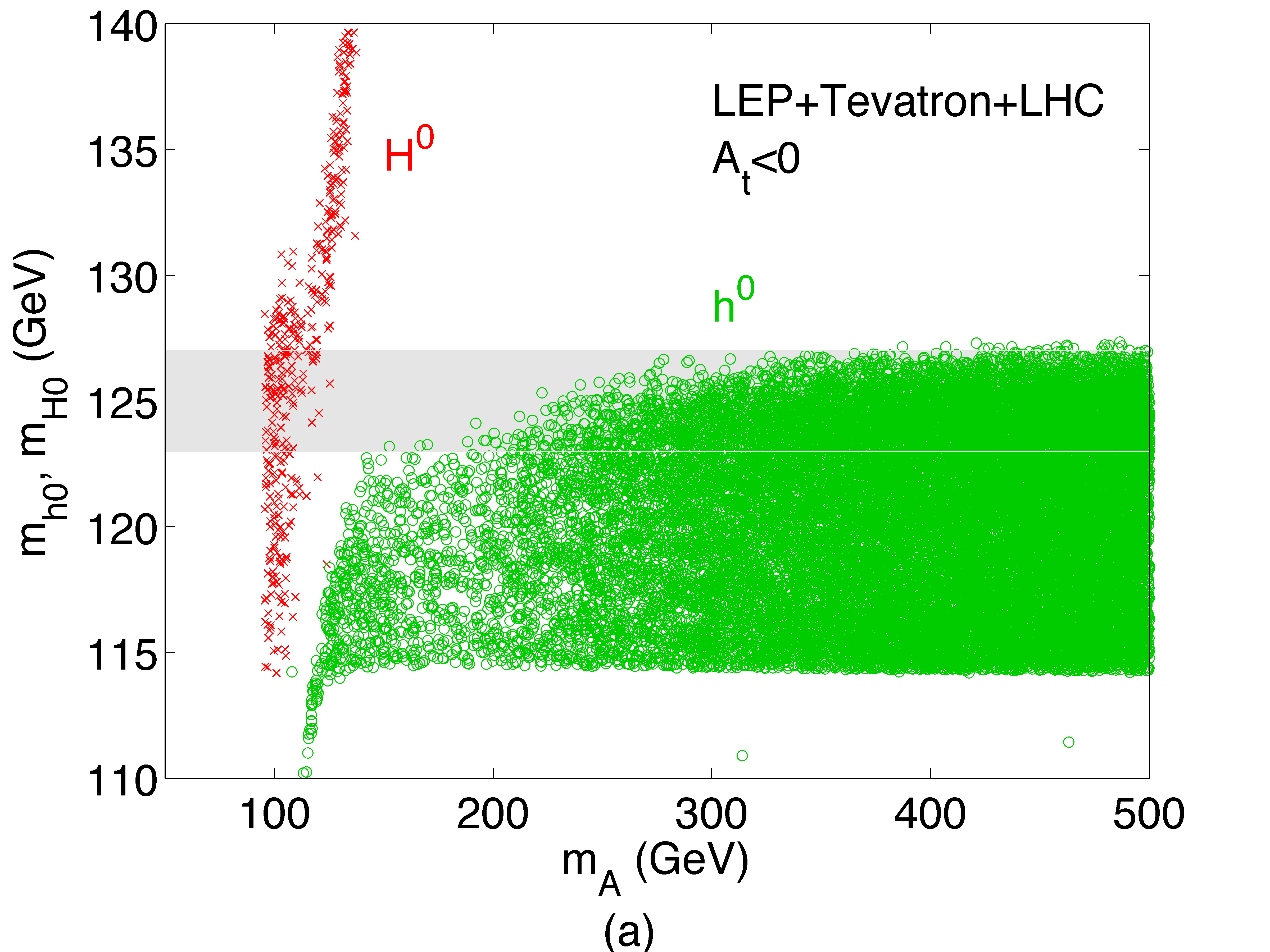}
\includegraphics[scale=1,width=8cm]{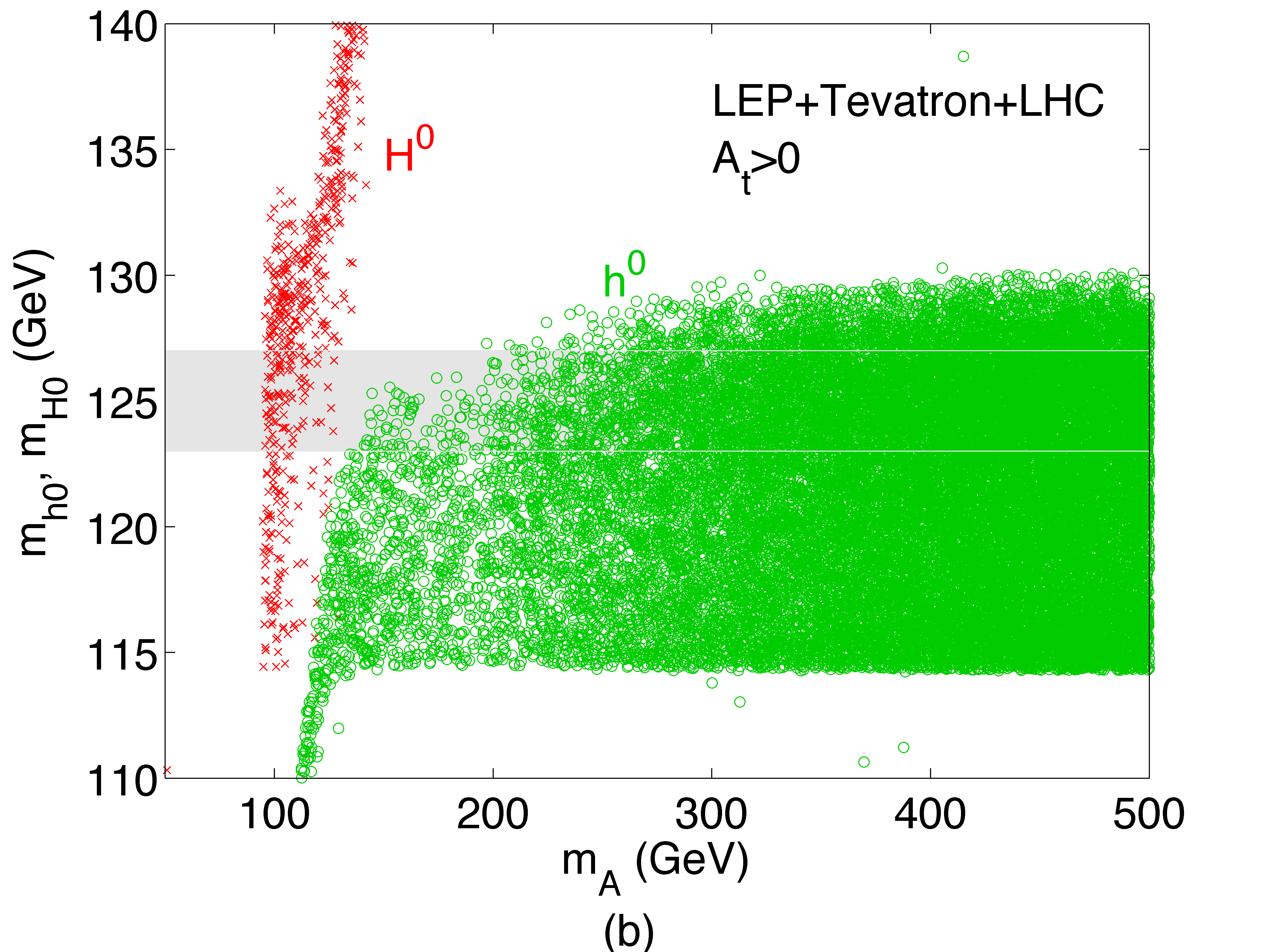}
\caption{
A zoom-in plot of Fig.~\ref{fig:mass}(b) for the light CP even $h^0$ (green circles), 
the heavy CP even $H^0$ (red crosses), 
 including the bounds from the LEP2, the Tevatron and the LHC
for (a) $A_t<0$ and (b) $A_t>0$. 
The horizontal lines mark the mass range of Eq.~(\ref{eq:mh1}).
}
\label{fig:masscloseup}
\end{figure}

\subsection{Allowed Regions for Higgs Boson Masses} 

We first reexamine the Higgs boson masses for $h^0$, $H^0$, $H^\pm$ subject to various current constraints from the direct searches. In Fig.~\ref{fig:mass}(a), we present the scanning output which satisfies the LEP2 \cite{LEP2H} bounds. The band widths reflect the scanning of the other SUSY parameters. 
The LEP2 bound sharply cuts off the allowed masses at a little above $90$~GeV near the kinematic limit for $ZH$ or $AH$.  
Figure \ref{fig:mass}(b) further includes the Tevatron \cite{CDFD0} and 
 the most recent LHC bounds\footnote{We have implemented the ATLAS Higgs search update 
 presented at Moriond meeting for individual channels at $95\%$ C.L. bounds.} \cite{ATLASHnew,CMSHnew,CMS-tautau,CMSA0,HpmCMS,HpmATLAS} with the search for the light Higgs boson and $H^{0},A^{0}\to \tau\tau$.  
We find that although many of the points that passed LEP2 are no longer allowed, the result is qualitatively the same in terms of the $m_{A}$ coverage.  
Figure \ref{fig:masscloseup}(a) and (b) show the allowed mass values of the CP-even Higgs bosons in the close-up region as in Fig.~\ref{fig:mass}(b) with $A_t<0$ and $A_t>0$. The horizontal lines mark the mass range of Eq.~(\ref{eq:mh1}). We see the subtle difference between the signs of $A_{t}$, for which $A_t>0$ yields more accessible solutions especially for a heavier $m_{h^0}$ due to two-loop radiative corrections.

\begin{figure}[tb]
\includegraphics[scale=1,width=8cm]{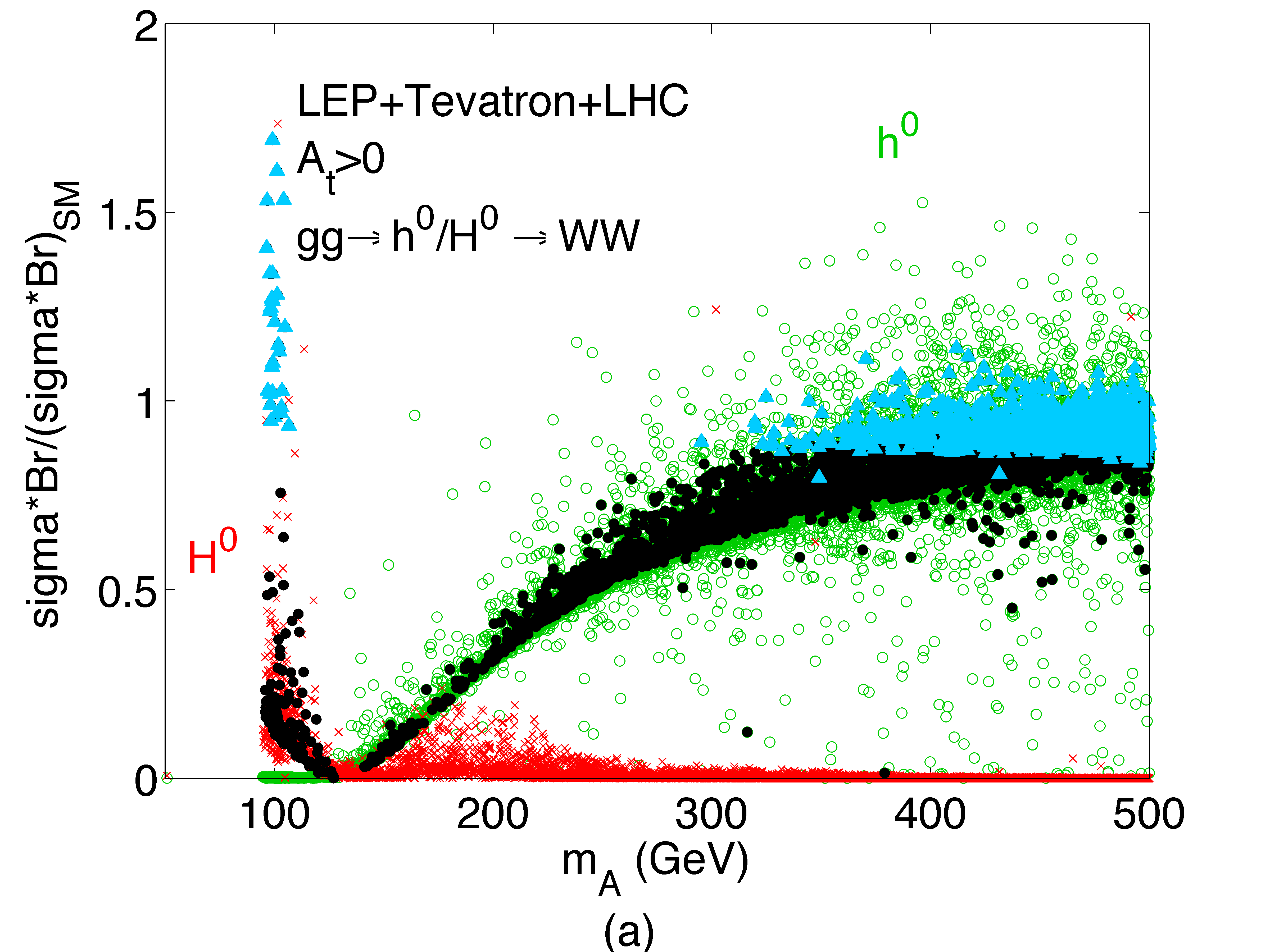}
\includegraphics[scale=1,width=8cm]{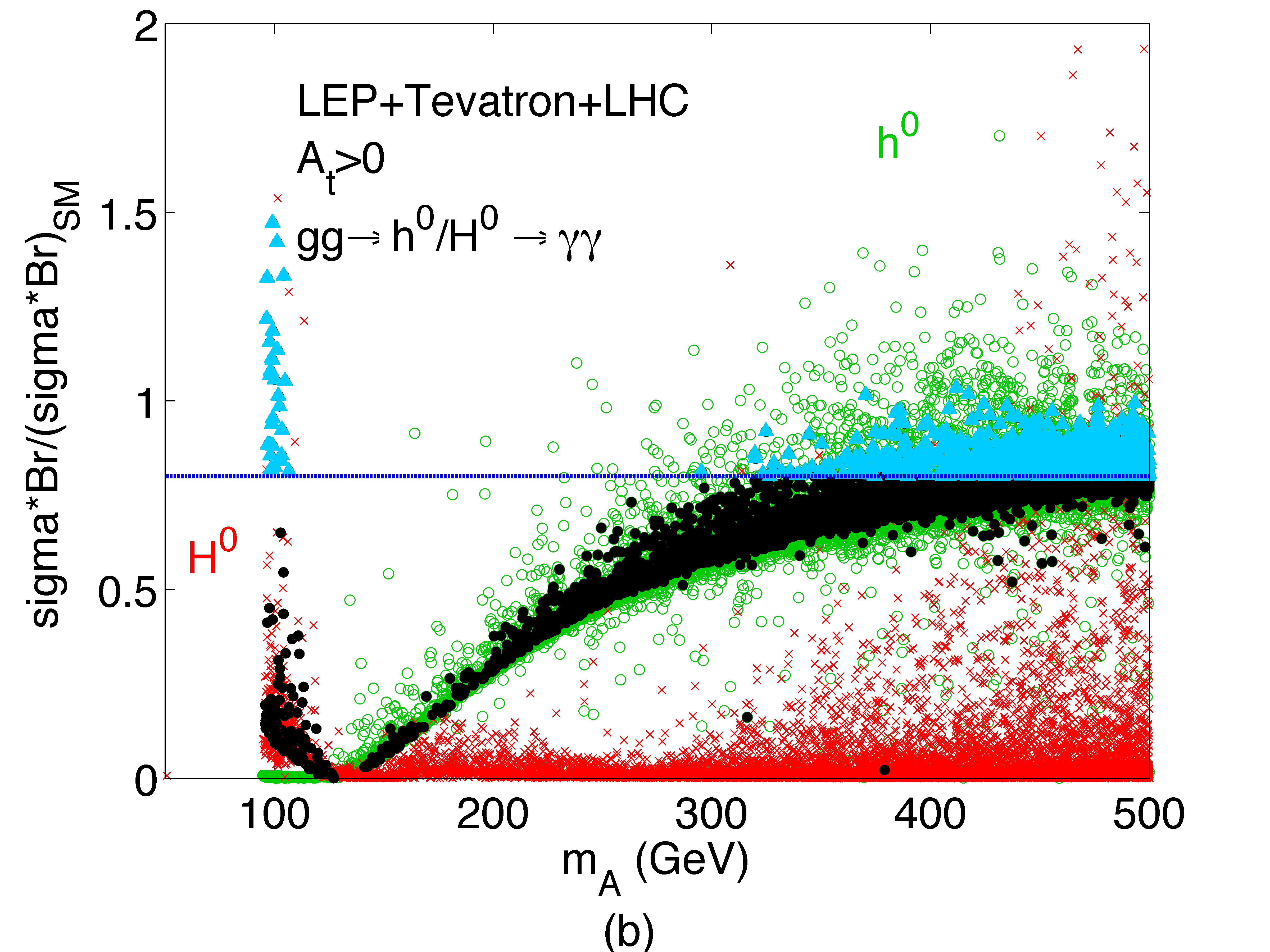}
\includegraphics[scale=1,width=8cm]{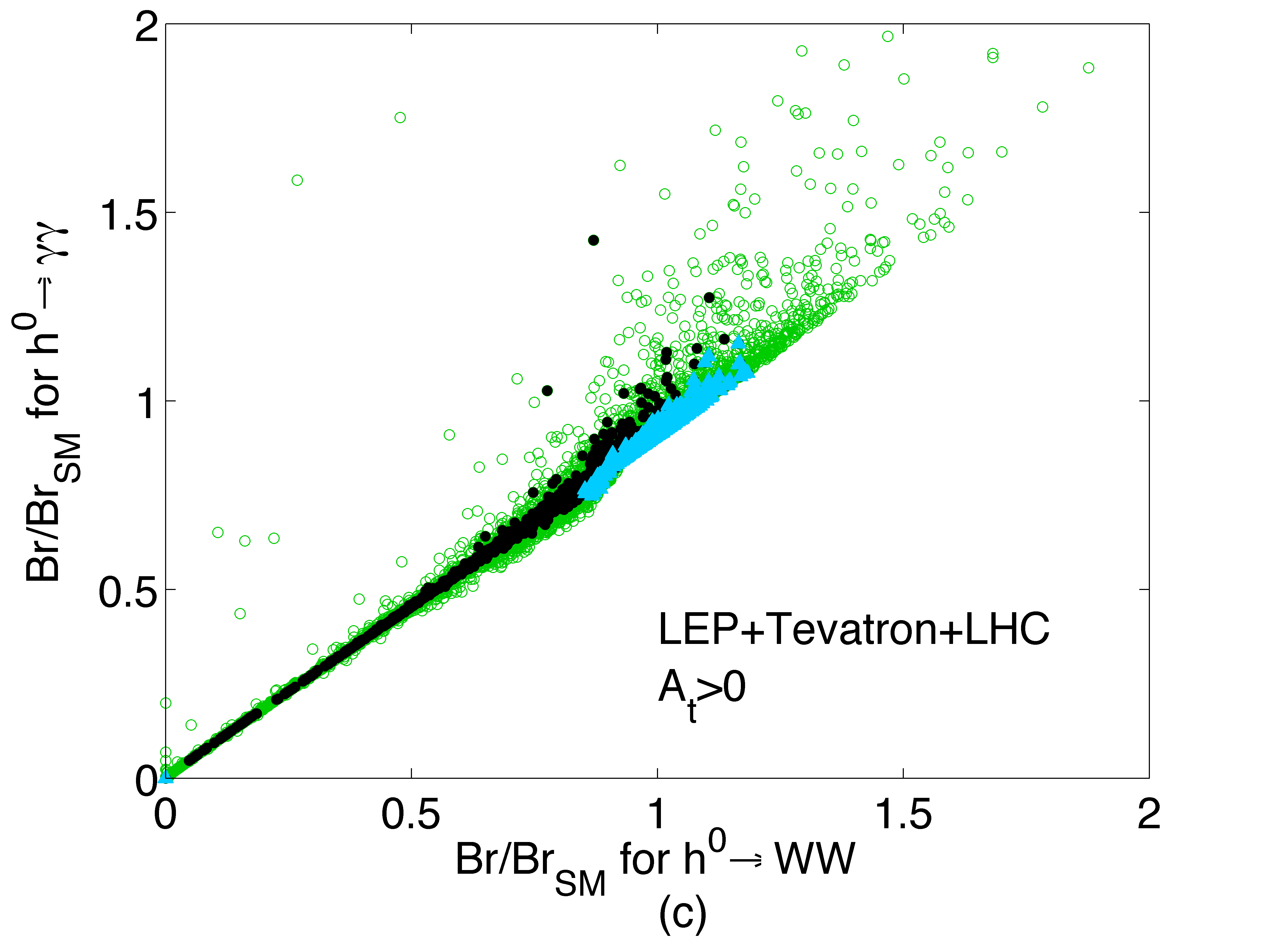}
\includegraphics[scale=1,width=8cm]{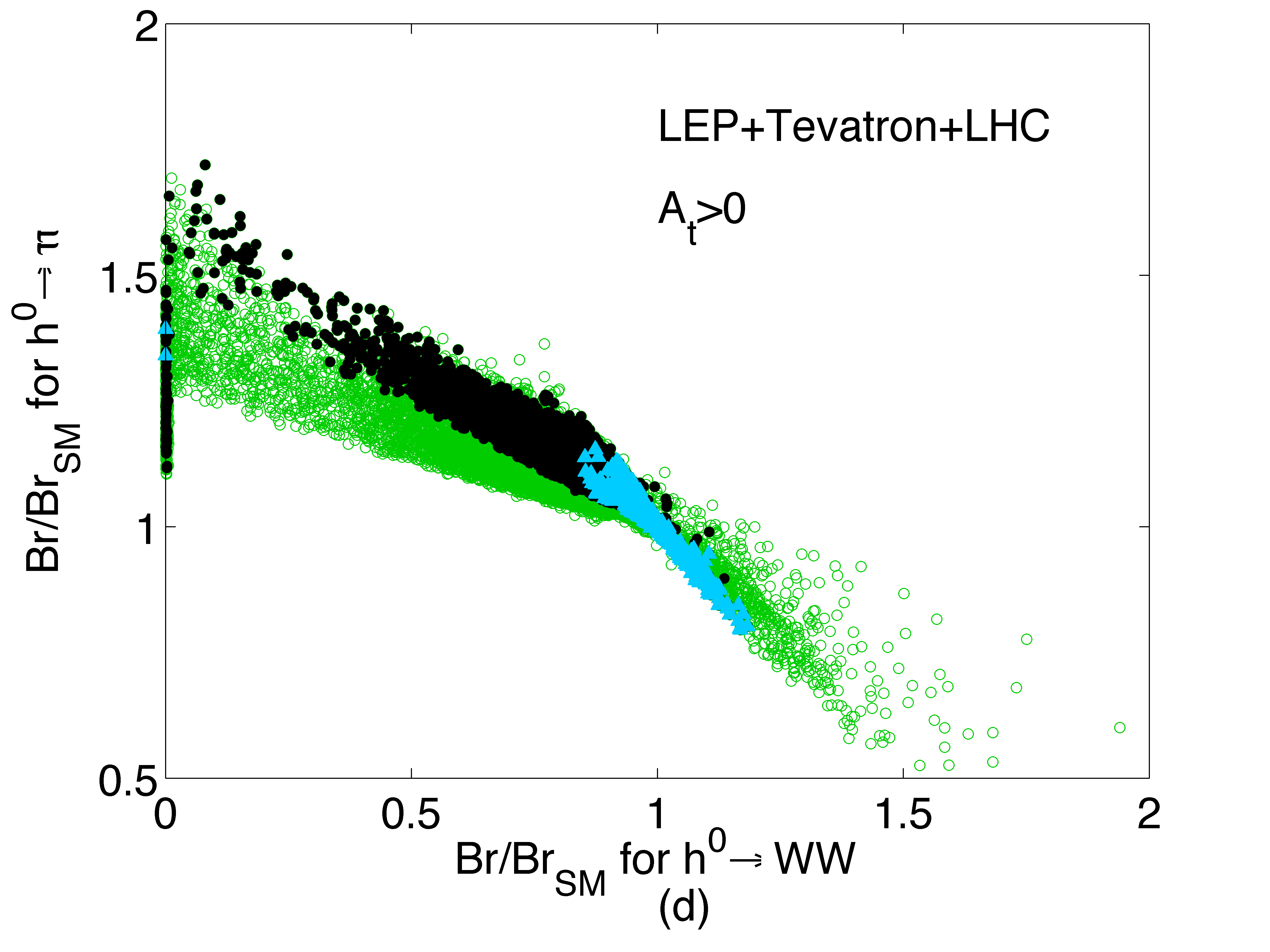}
\caption{Signal cross section ratios $\sigma / \sigma_{SM}$ versus $\ma$ for (a) the $\ww $ final state 
with $h^{0}$ (green circles) and $H^{0}$ (red crosses), (b) the $\gaga$ final state, and the branching fraction  correlation (${\rm Br} / {\rm Br}_{SM}$) for (c) $h^{0}\to \gaga$ versus $h^{0} \to \ww$ and for (d) $h^{0}\to \tautau$ versus $h^{0} \to \ww$. All the LEP2 and hadron collider direct search bounds are imposed.
The black dots in all the panels represent those satisfying the narrower Higgs mass window in Eq.~(\ref{eq:mh1}). 
The light blue triangles are those satisfying the cross section requirement Eq.~(\ref{eq:sigma}). 
Other parameters are scanned over the range in Eq.~(\ref{eq:para}) with $A_{t}>0$.  
}
\label{fig:csR}
\end{figure}

\begin{figure}[tb]
\includegraphics[scale=1,width=8cm]{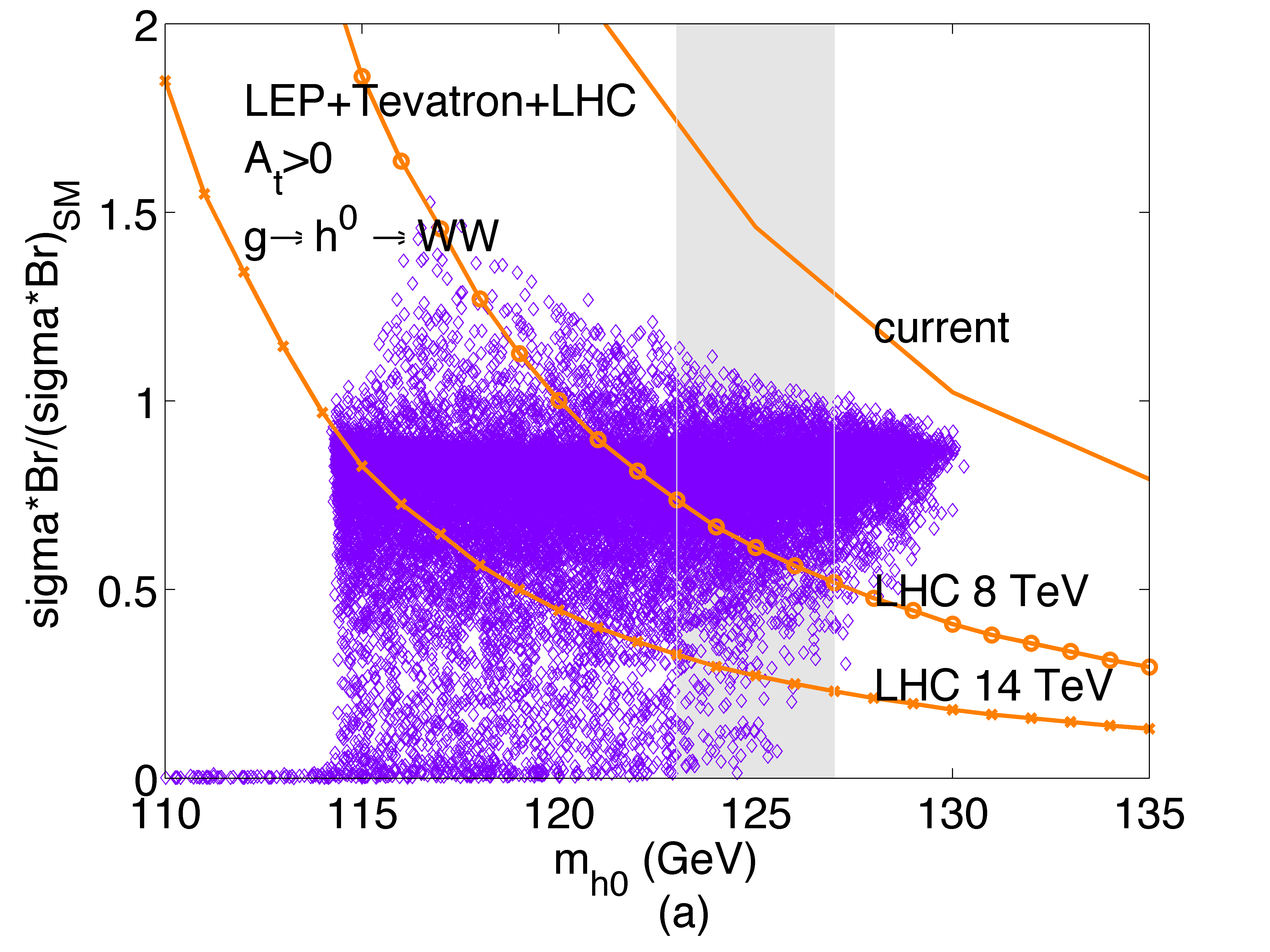}
\includegraphics[scale=1,width=8cm]{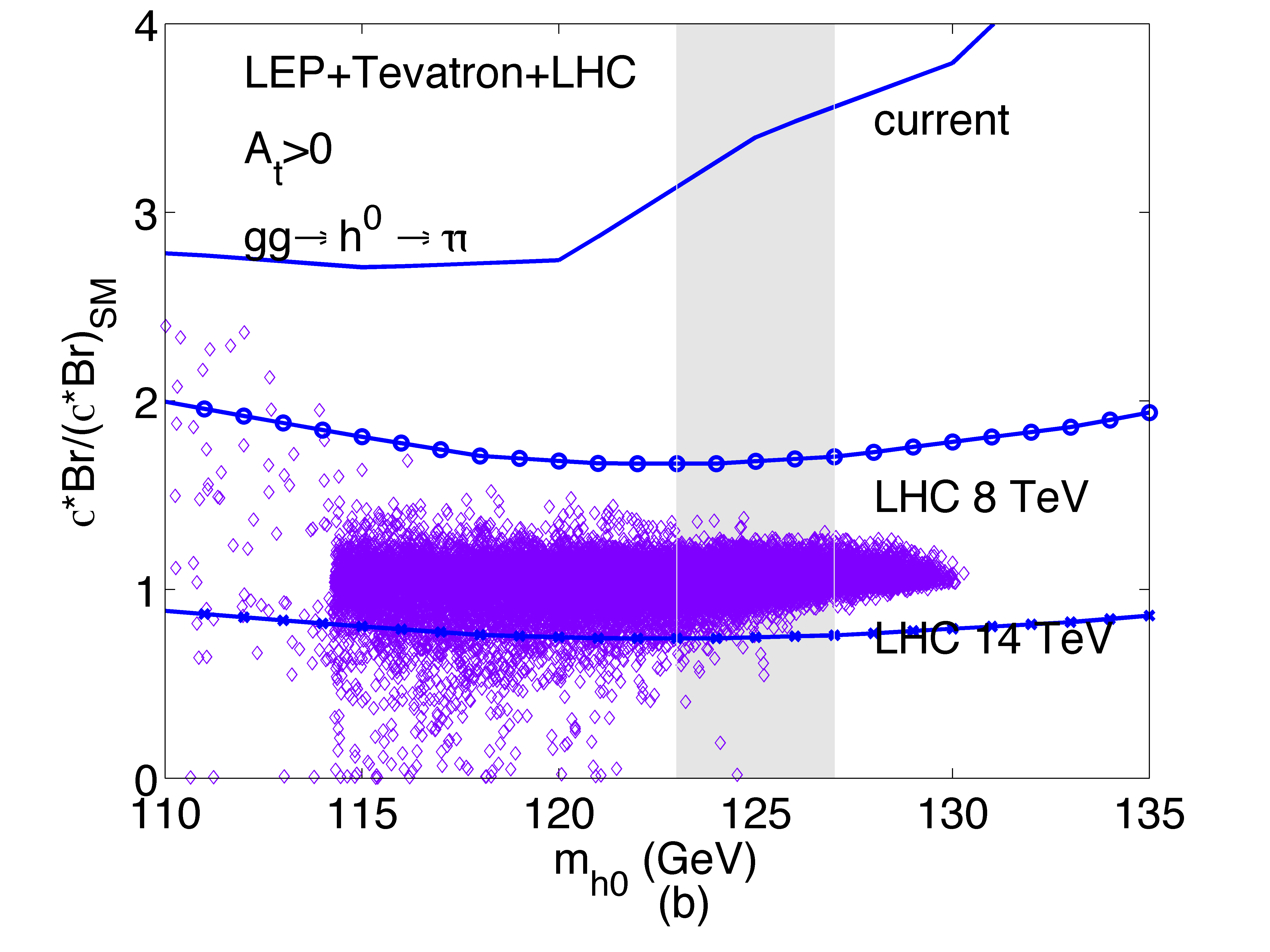}
\caption{Signal cross section ratios $\sigma / \sigma_{SM}$ versus 
%the SM-like Higgs mass 
$m_{h^0}$ for 
(a) $\ww$ and (b) $\tautau$ final state,  with the LEP2 and hadron collider direct search bounds except the latest bounds on $W^+W^-$ and $\tau^+\tau^-$ channel from ATLAS searches \cite{ATLASHnew}.
The upper solid curve in each panel is from the current $95\%$ C.L.~bound \cite{ATLASHnew}. 
The two lower curves indicate the estimated improvements at a 8 TeV and 14 TeV LHC 
(see Sec.~\ref{sec:Future}). The vertical bands indicate the narrow mass window in Eq.~(\ref{eq:mh1}).
Other parameters are scanned over the range in Eq.~(\ref{eq:para}) with $A_{t}>0$.  
}
\label{fig:csRmh}
\end{figure}

We next calculate the CP-even SUSY Higgs production cross section for the channels
\begin{eqnarray}
gg \rightarrow h^{0},\ H^{0} \to  \gamma\gamma,\ \ww,\ ZZ.
\label{eq:ggzz}
\end{eqnarray}
Let us consider the CP-even Higgs bosons $h^{0}$ and $H^0$ after passing both the LEP2 and the hadron collider bounds. 
Figure \ref{fig:csR} presents the ratios of the MSSM cross sections to the SM values versus $\ma$ for (a) 
$\ww$ and (b) $\gaga$ final states, with green circles for $h^{0}$ and red crosses for $H^{0}$.
The result for the $ZZ$ channel is very similar to the $\ww$ channel due to the SU(2) symmetry.

For a SM-like Higgs boson, the Higgs-$WW$ coupling is the main source for both the $WW$ and $\gamma\gamma$ decay channels. In the SM, the ratio at $\mh=125$ GeV is fixed as 
Br$({\ww})_{SM} : $Br$({\gamma\gamma})_{SM} \approx15\% : 2.2\times 10^{-3}$. 
In the MSSM even with our broad parameter scan, there is a strong correlation. This is shown in Fig.~\ref{fig:csR}(c) for Br$(\gaga)$ versus Br$(\ww)$. We see an empirical linear relation
\begin{equation}
{ {\rm Br}(\gamma\gamma) \over {\rm Br}(\gamma\gamma)_{SM} } \approx 0.9\ { {\rm Br}(\ww) \over {\rm Br}(\ww)_{SM} }.
\label{eq:r}
\end{equation}
The smaller-than-unity prefactor is due to some level of cancellation in the loops of $h^{0} \to \gaga$. 
In Fig.~\ref{fig:csR}(d), we show another correlation for the channels of $\tautau$ and $\ww$. The SM prediction is at a value Br$({\ww})_{SM} : $Br$({\tautau})_{SM} \approx15\% : 7\%$ at 125 GeV. It is interesting to note that they are ``anticorrelated''. Thus a consistency check of the predicted correlations as shown in Fig.~\ref{fig:csR} could provide crucial information regarding the underlying theory. 

In Fig.~\ref{fig:csRmh}, we show the cross section ratios $\sigma / \sigma_{SM}$ versus 
%SM-like Higgs mass 
$m_{h^0}$ for 
(a) $\ww$ and (b) $\tautau$ final states, with the LEP2 and hadron collider direct search bounds  except the latest bounds on $W^+W^-$ and $\tau^+\tau^-$ channel from ATLAS searches \cite{ATLASHnew}. 
The solid curve in each panel is from the current $95\%$ C.L.~bound \cite{ATLASHnew}. The vertical bands indicate the narrow mass window in Eq.~(\ref{eq:mh1}). We see that the recent ATLAS bounds from those channels alone are not strong enough to have a direct impact on the existing bounds, leaving solutions with a factor of 1.5 larger than the SM predictions.

Given the tantalizing hint for the $\gaga$ events near 125 GeV, we
take an important step to assume the existence of a CP-even Higgs boson 
\begin{eqnarray}
h^{0}\ {\rm or}\ H^0 \ {\rm in\ the\ mass\ range\ of}\ 123\gev -127\gev, 
\label{eq:mass} \\
\label{eq:sigma}
 \sigma\times{\rm Br}(gg\rightarrow h^0, H^0 \rightarrow \gamma\gamma)_{MSSM} \ge 80\% (\sigma\times{\rm Br})_{SM}.
\end{eqnarray}
The mass window requirement in Eq.~(\ref{eq:mass}) yields a very selective parameter region as indicated by the black dots in the panels in Fig.~\ref{fig:csR}. The simultaneous requirement of the sizable cross section for the $\gaga$ mode forces $\ma$ into two distinct and separate regions, as seen from the light blue triangles above the dashed horizontal line in Fig.~\ref{fig:csR} (b).
The bulk region of the allowed parameter space is pushed to heavy $\ma$ (roughly $\ma > 300$ GeV), the ``decoupling region'' with the light CP-even Higgs being SM-like.
There is, however, a small region at lower $\ma$ that survives in the ``non-decoupling'' region (roughly $95\gev < \ma < 130 \gev$) with the heavy CP-even Higgs being SM-like \cite{Heinemeyer:2011aa}.
The non-decoupling region,
which satisfies both the mass  and the cross section requirement as in  Eqs.~(\ref{eq:mass}) and (\ref{eq:sigma}),
occurs mainly for $A_t>0$.
%(a factor of $\sim50$ times fewer points were found for $A_t<0$) 
This is because a suppression of $H^0 \rightarrow b\bar b$ is needed in order for $gg\rightarrow H^0 \rightarrow \gamma\gamma$ to be above 0.8 of the SM value.    Such a suppression could be due to a large positive radiative correction to the bottom Yukawa,  $\Delta m_b$ (as defined in Eq.~(\ref{eq:deltamb})), as well as a small $\cos\alpha_{\rm eff}$, where $\alpha_{\rm eff}$ is the CP-even Higgs mixing parameter $\alpha$ with radiative corrections. Both could be realized in the positive $A_t$ case, where  $\Delta m_b$ is always positive, and $\cos\alpha_{\rm eff}$ could be as small as zero, while keeping $A_t$ large enough to satisfy the mass region in 
Eq.~(\ref{eq:mass}).   For negative $A_t$, due to the cancellation between the sbottom-gluino loop ($\propto M_3 \mu$) and the stop-Higgsino loop ($\propto A_t \mu$), a small $|A_t|$ is preferred to obtain a positive $\Delta m_b$.  In addition, to get a small value for $\cos\alpha_{\rm eff}$ also requires a relatively small $|A_t|$.
The radiative correction to the Higgs mass, however, is small for such a small value of $|A_t|$, leading to a strong tension between the Higgs mass requirement in Eq.~(\ref{eq:mass}) and the cross section requirement in Eq.~(\ref{eq:sigma}) for $A_t<0$. 

We summarize these two distinctive regions  as
\begin{eqnarray}
\label{eq:d}
&& {\rm Decoupling\ region:\ } h^{0}\ {\rm SM-like,}\ \mH\sim \mHpm \sim \ma \gtrsim 300 \gev; \\
&&  {\rm Non-decoupling\ region:\ }H^{0}\ {\rm SM-like,}\ \mh \sim \ma,\ \ \mH \sim \mHpm.
\label{eq:nond}
\end{eqnarray}
The non-decoupling region is of great interest both in terms of the theoretical implication and the LHC searches.

\begin{figure}[tb]
\includegraphics[width=8cm]{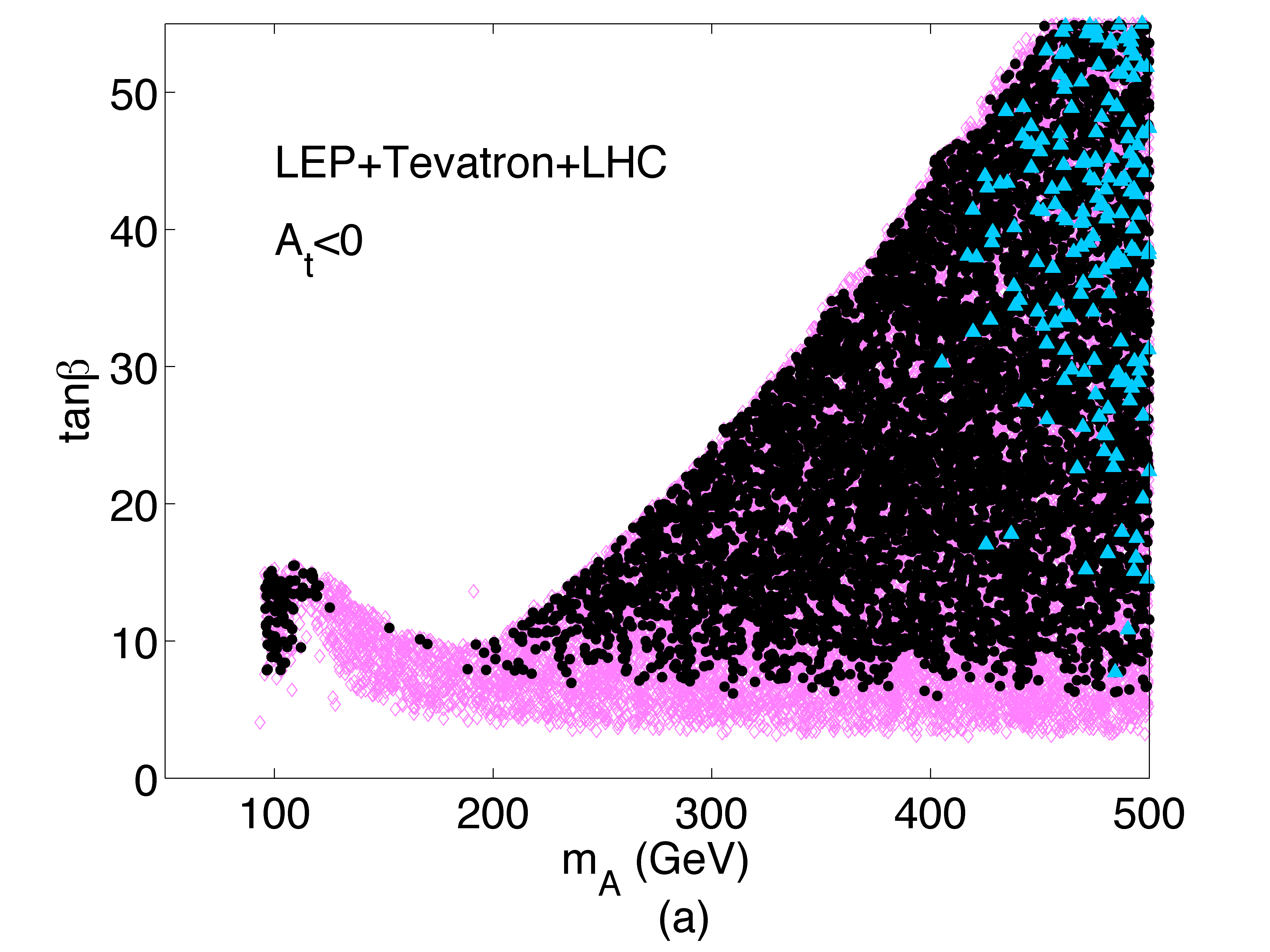}
\includegraphics[width=8cm]{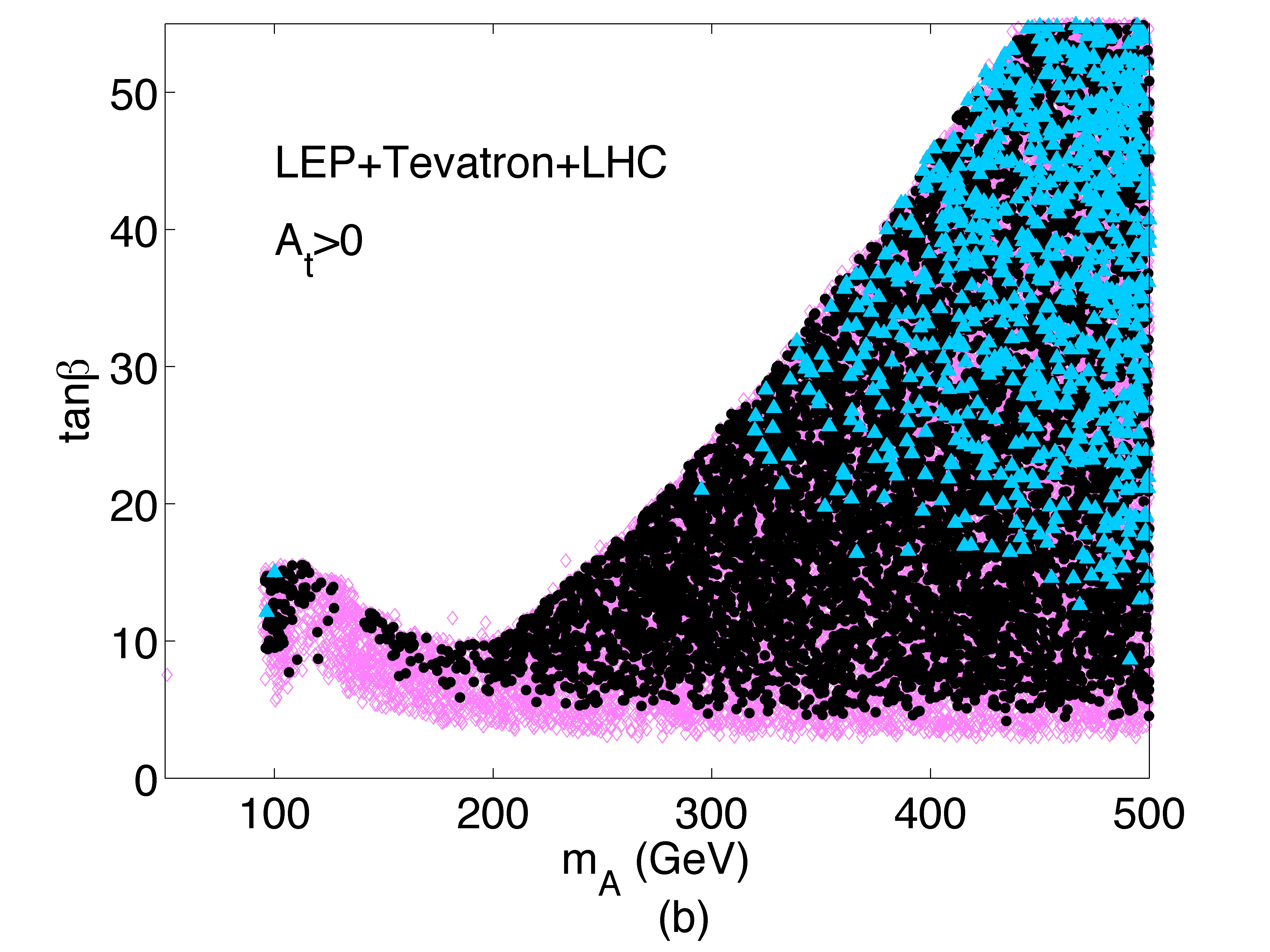}
\caption{Allowed region of $\tan\beta$ versus $\ma$ for (a) $A_{t}<0$ and for (b) $A_{t}>0$, respectively.
The region with purple diamonds satisfies all the LEP2, Tevatron and the LHC direct search constraints. 
The black dots represent those in the narrow mass window in Eq.~(\ref{eq:mass}).
The light blue triangles are those satisfying the cross section requirement Eq.~(\ref{eq:sigma}). 
Other parameters are scanned over the range in Eq.~(\ref{eq:para}).
}
\label{fig:tanb}
\end{figure}

\subsection{Allowed Regions for Other SUSY Parameters} 

It turns out that the above constraints have significant implication for the other SUSY parameters associated with the Higgs sector.
% as given in Eqs.~(\ref{eq:deltamh}) and (\ref{eq:At}). 

\subsubsection{$\tan\beta$ versus $\ma$}
 
We first examine the allowed region of $\tan\beta$ versus $\ma$.  
We present the region for $A_t<0$ in Fig.~\ref{fig:tanb}(a), and for $A_t>0$  in Fig.~\ref{fig:tanb}(b). Not shown in the figures are the regions allowed by LEP2 alone, which are uniformly from $\ma\approx 90$ GeV and on.
The bounds from the hadron colliders (purple diamonds) remove the region of low $m_A$ and high $\tan\beta$.
This is largely due to the searches for $h^0, H^0, A^0 \rightarrow \tau\tau$ \cite{ CMSA0}, as well as $t\to b H^\pm$  \cite{CDFD0, HpmCMS, HpmATLAS}.
The final requirements for the existence of a SM-like Higgs as in Eqs.~(\ref{eq:mass})  and (\ref{eq:sigma}) 
once again highly limit the parameter space (black dots and light blue triangles, respectively).   Requiring the existence of a SM-like Higgs in the mass range of 123 $-$ 127 GeV results in $m_A \gtrsim 400$ GeV for $A_t<0$ and $m_A \gtrsim 300$ GeV for $A_t>0$. 

\begin{figure}[tb]
\includegraphics[scale=1,width=8cm]{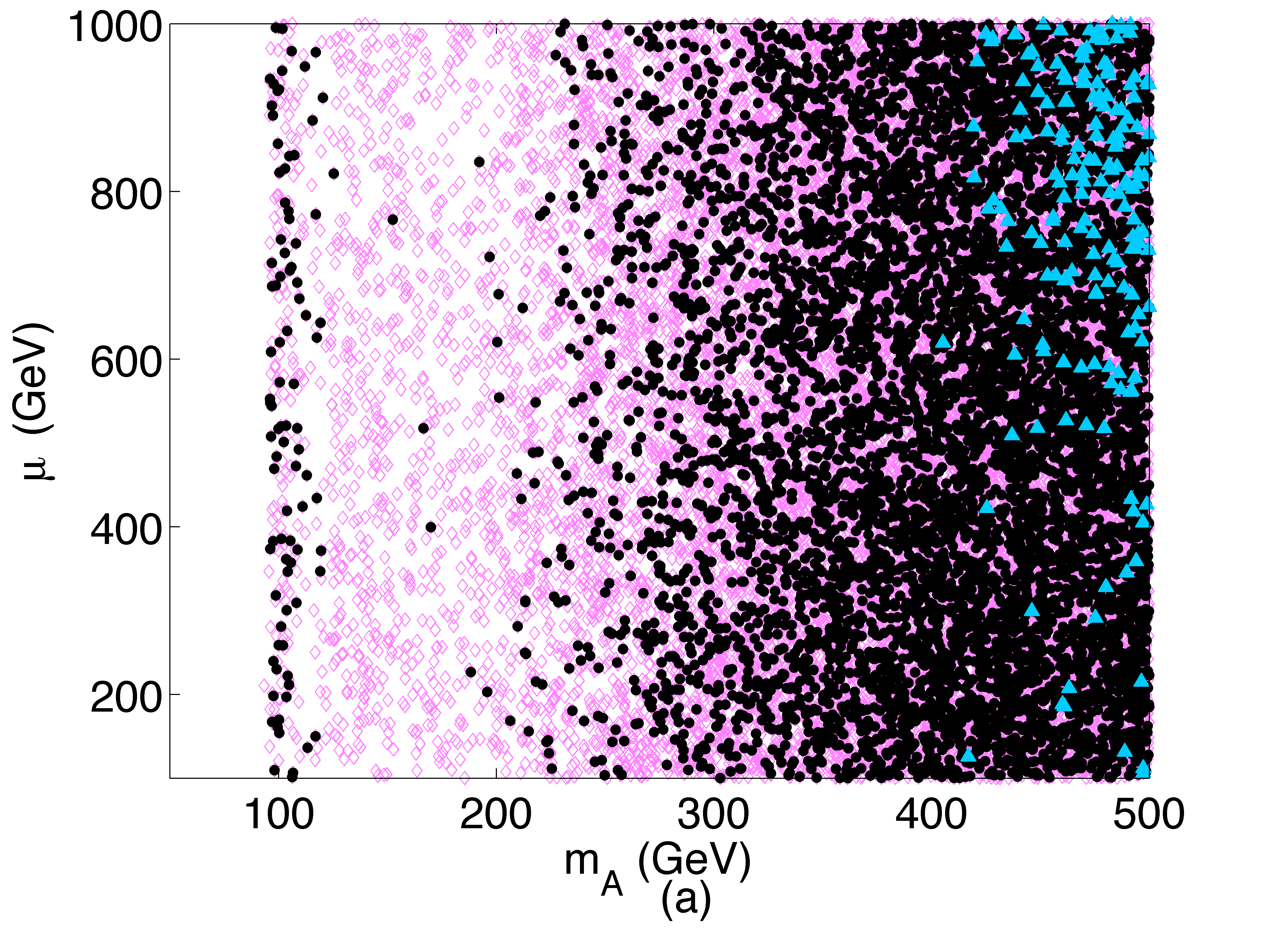}
\includegraphics[scale=1,width=8cm]{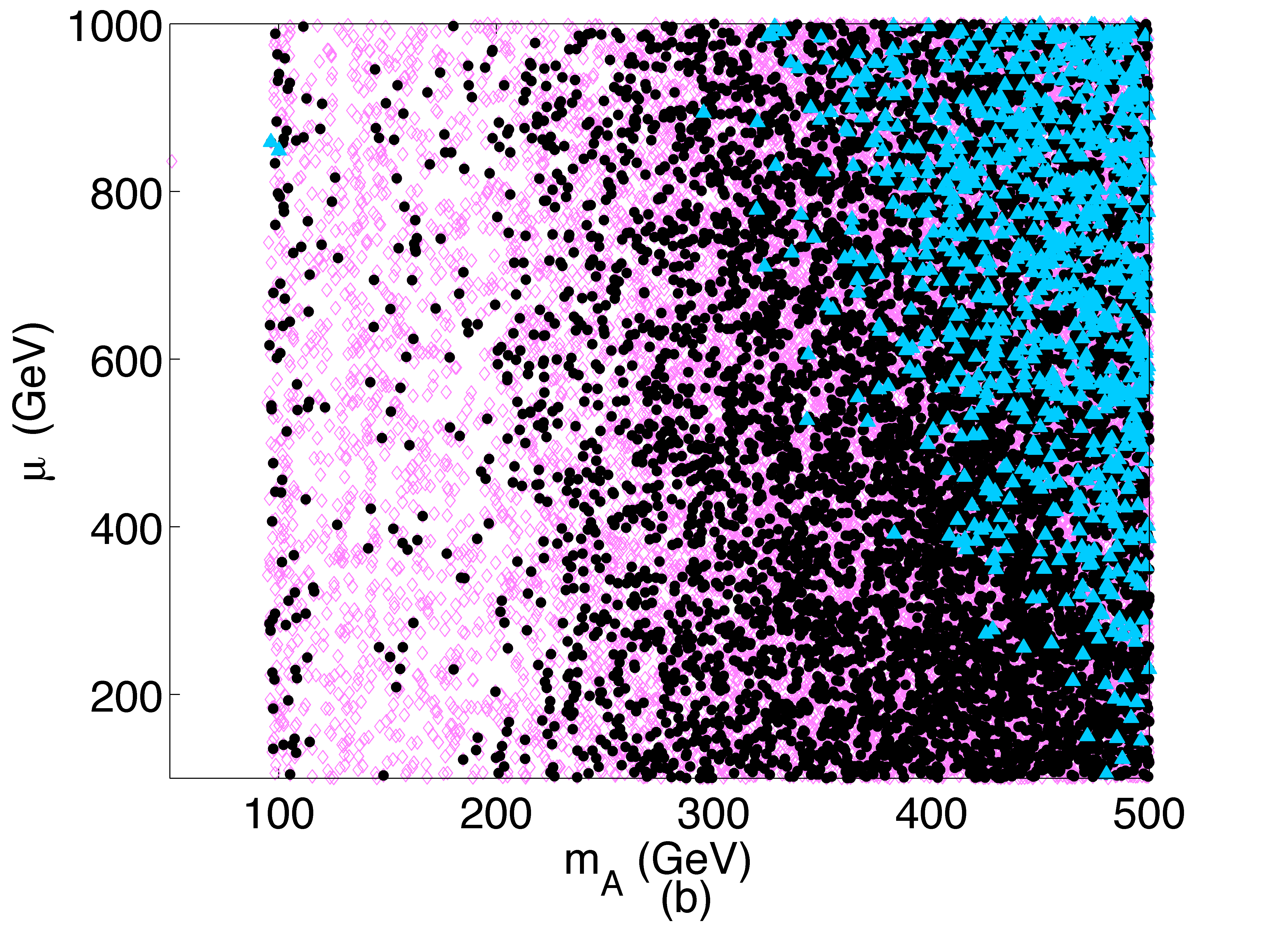}
\caption{Allowed region for the SUSY Higgs mixing parameter $\mu$ versus $\ma$. 
The legends are the same as in Fig.~\ref{fig:tanb}.
}
\label{fig:mu}
\end{figure}

\subsubsection{$\mu-\ma$ correlation}

The Higgs mixing parameter $\mu$ plays an important role for radiative corrections to the Higgs production and decay channels and we vary it in the range of Eq.~(\ref{eq:para}). 
We show the impact on this parameter in Fig.~\ref{fig:mu}, where the legends are the same as in Fig.~\ref{fig:tanb}.
We note the interesting correlation in the decoupling region for $A_{t}>0$  once we impose the cross section requirement as in Eq.~(\ref{eq:sigma}) (regions indicated by light blue triangles)
that a lower value of $\ma$ results in a higher $\mu$. This is because a smaller $\mu$ leads to a suppressed $gg\rightarrow h^0 \rightarrow \gamma\gamma$  and is, therefore, disfavored \cite{Carena:1999bh, Carena:1998gk}. 

\begin{figure}
\includegraphics[scale=1,width=8cm]{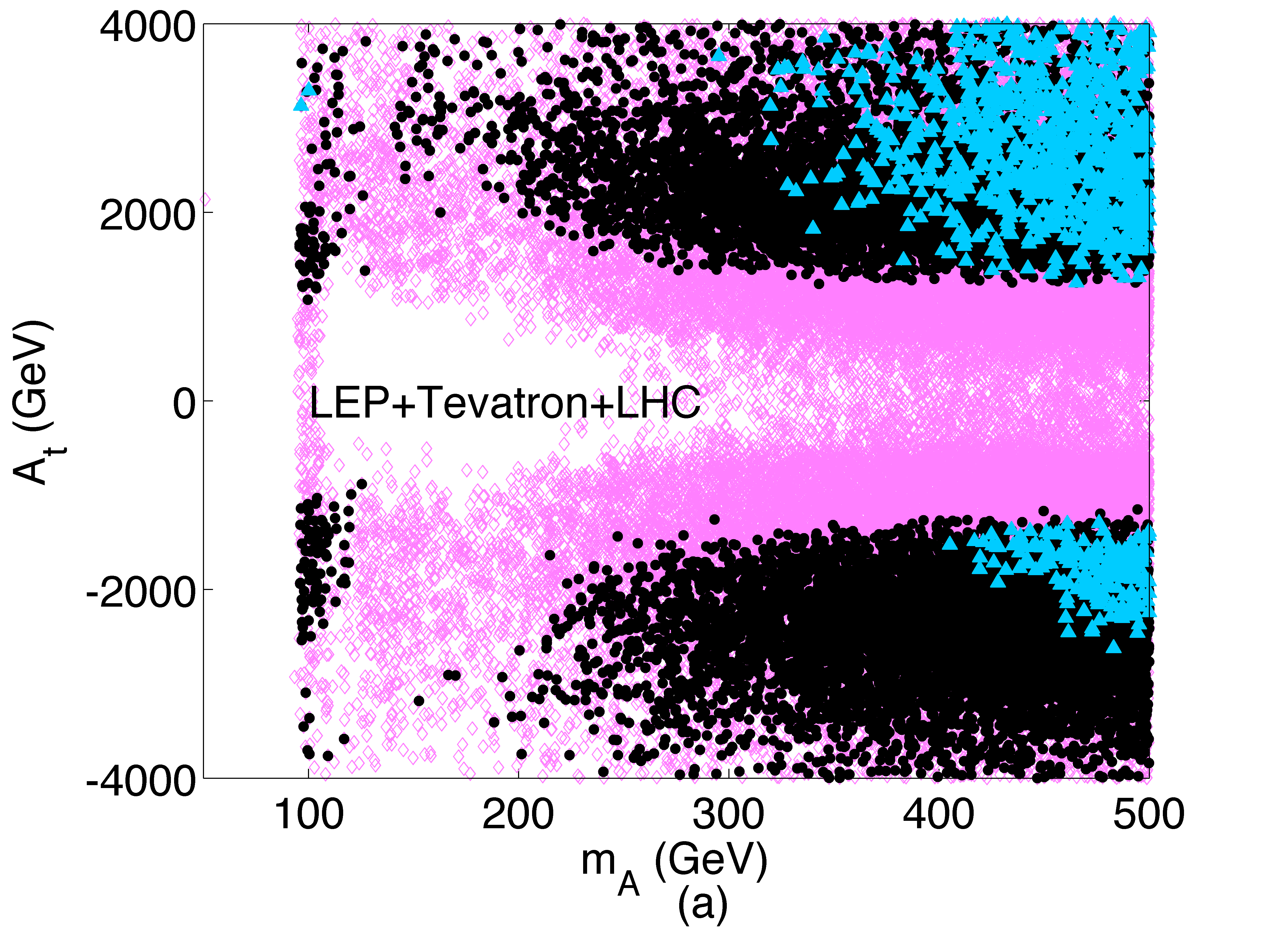}
\includegraphics[scale=1,width=8cm]{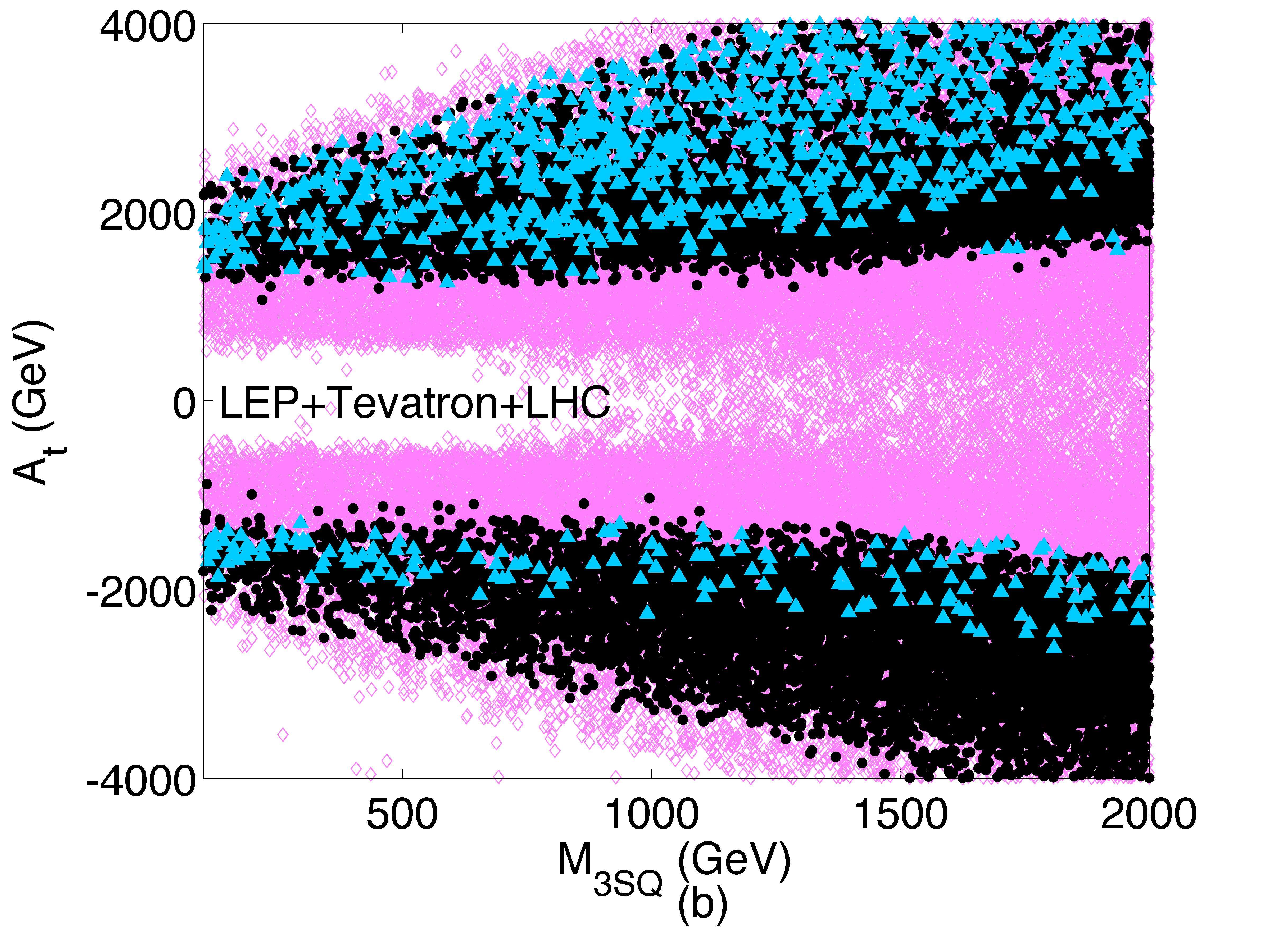}
\caption{Allowed region for (a) the SUSY stop-quark mixing parameter $A_{t}$  versus $\ma$ and (b) $A_{t}$  versus $M_{3SQ}$. The legends are the same as in Fig.~\ref{fig:tanb}.
}
\label{fig:at}
\end{figure}

\subsubsection{$A_{t}-\ma$ correlation}

The next SUSY parameter relevant to the Higgs sector is $A_{t}$ (see Eq.~(\ref{eq:At})) and we vary it in the range of Eq.~(\ref{eq:para}). 
We show the effect on this parameter in Fig.~\ref{fig:at}(a), with the legends the same as in  Fig.~\ref{fig:tanb}. 
The smaller $\left|A_t\right|$ region is disfavored due to the smallness of $m_{h^0}$,  while the large $\left|A_t\right|$ region is removed by demanding sizable $gg\rightarrow h^0 \rightarrow \gamma\gamma, \ww$ 
cross sections \cite{Carena:1999bh, Carena:1998gk}.  Such correlation of $A_t$ with $m_A$ is more pronounced for the negative $A_t$ case.
Similar effects were already observed earlier in \cite{Low:2009nj}.

\subsubsection{SUSY breaking scale $M_{3SQ}$}
 
In Fig.~\ref{fig:at}(b), we present the allowed region in the plane of the soft SUSY breaking scale $M_{3SQ}$ and $A_t$, with the legends the same as in Fig.~\ref{fig:tanb}. The behavior for $M_{3SU}$ is very similar.  
An approximate $m_h^{\rm max}$ relation of $\tilde{A}_t \sim \sqrt{6} M_{3SQ}, \sqrt{6}M_{3SU}$ and/or large $M_{3SQ}$, $M_{3SU}$ are needed to have a relatively heavy Higgs mass in the range of 123 to 127 GeV \cite{Heinemeyer:2011aa,  Arbey:2011ab, Arbey:2011aa, Carena:2011aa, Arvanitaki:2011ck,Cao:2012fz}.  Imposing the cross section requirement of Eq.~(\ref{eq:sigma}) further narrows down the range of $A_t$.  In particular, 
for the negative $A_t$ case, $A_t$ is typically in the narrow range from  $-2500$ to $-1000$ GeV, 
while for the positive $A_t$ case, the allowed region is much broader, from 1000 GeV and higher.
The difference between positive and negative $A_t$ is mainly due to the difference in the radiative correction to $\Delta m_b$  from the stop sector \cite{deltahb}.  
 
 \subsubsection{Non-decoupling region}

\begin{figure}
\includegraphics[scale=1,width=8cm]{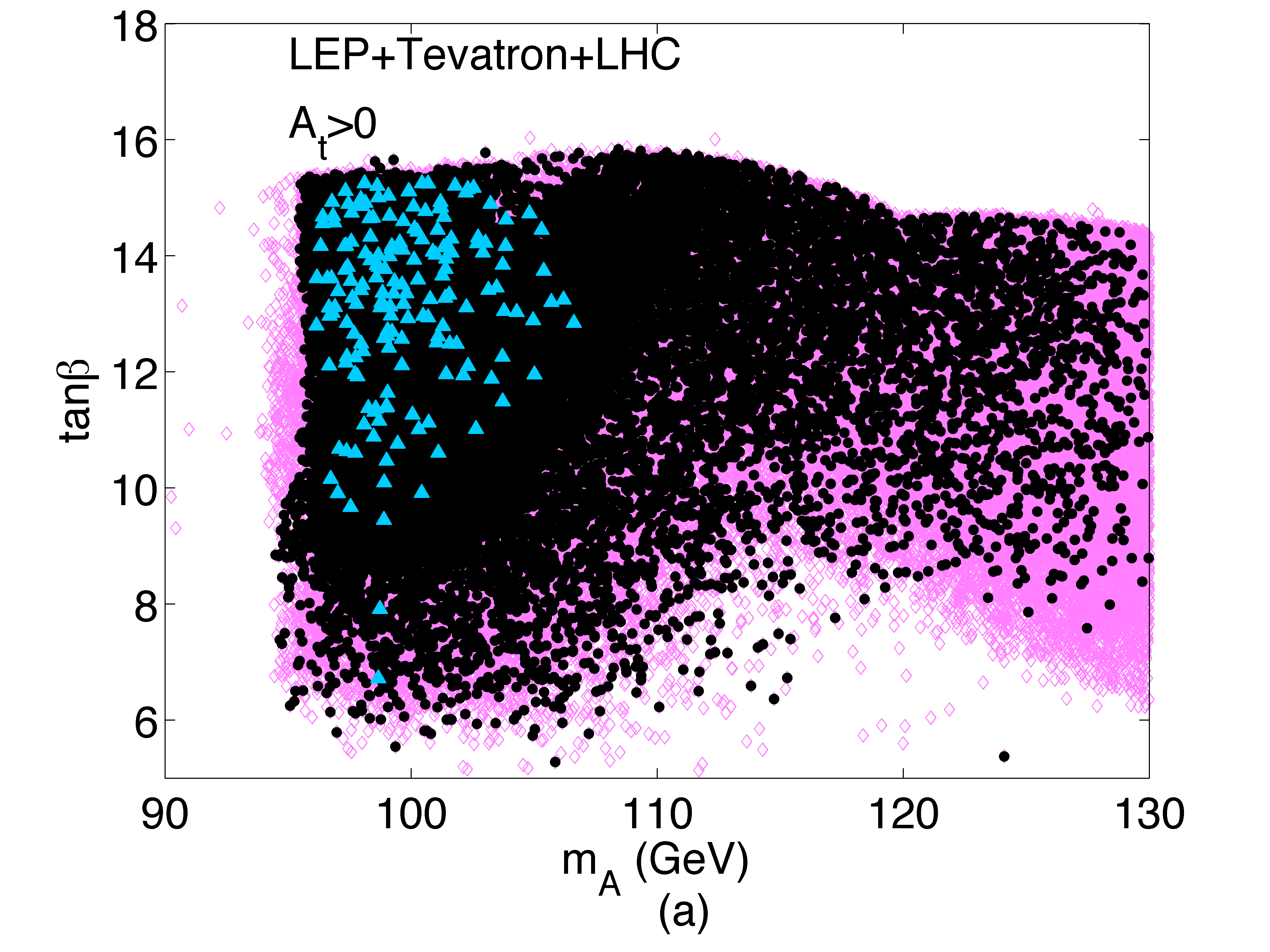}
\includegraphics[scale=1,width=8cm]{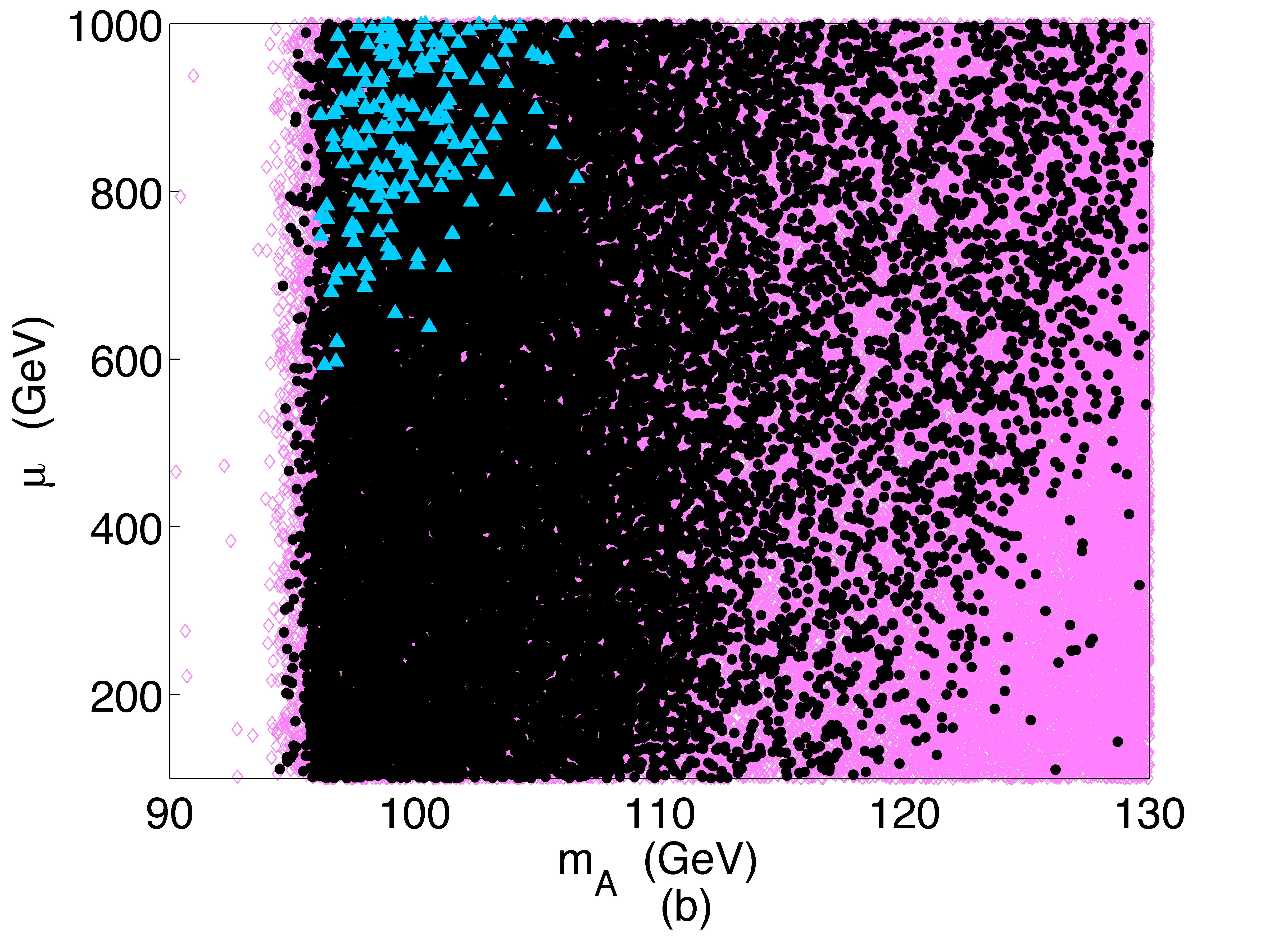}
\caption{Allowed region for (a) $\tan\beta$  versus $\ma$ and (b) $\mu$ versus $\ma$ in the non-decoupling region for $A_t>0$.  The legends are the same as in Fig.~\ref{fig:tanb}.
}
\label{fig:non-decoupling}
\end{figure}

As discussed earlier, the non-decoupling region mainly appears when $A_t>0$.
In Fig.~\ref{fig:non-decoupling}, we zoom into the non-decoupling region and impose the mass and cross section requirements as in Eqs.~(\ref{eq:mass}) and (\ref{eq:sigma}).  
Panel (a) shows that only a narrow region of 
\begin{equation}
95\ {\rm GeV} < \ma < 110 \ {\rm GeV}, \ \ \ 6 < \tan\beta < 16 
\end{equation}
can accommodate a SM-like heavy CP-even Higgs  in the mass range of 123 $-$ 127 GeV \cite{Heinemeyer:2011aa}.   Panel (b) shows that a higher value of $\mu$ is preferred  for larger $m_A$ after imposing the cross section requirement.    This is because a large $\mu$ leads to a larger positive $\Delta m_b$, resulting in a more suppressed $H^0 \rightarrow b \bar b$ and a more enhanced $H^0 \rightarrow \gamma\gamma$.
The surviving region in $A_t$ versus $\ma$ and $A_t$ versus $M_{3SQ}$ are similar to the decoupling case.
 
%%%%%%%%%%%%%%%%%%%%%%

\subsection{Extended Discussions}
\label{sec:Discussion}
  % Ref.\cite{Heinemeyer:2011aa, Arbey:2011ab, Arbey:2011aa, Draper:2011aa, Carena:2011aa, Arvanitaki:2011ck}
  
  In our study, we scanned over the six parameters, $m_A$, $\tan\beta$, $\mu$, $M_{3SQ}$, $M_{3SU}$ and $A_t$, which are the parameters most relevant to the Higgs sector phenomenology.  The other MSSM sectors, {\it i.e.}, sbottoms, staus etc., could also contribute to the Higgs sector, radiatively, as we briefly summarize below.  Most of our discussion  applies to the SM-like Higgs boson being either $h^0$ in the decoupling region or $H^0$ in the non-decoupling region.     
 
 \subsubsection{Higgs mass corrections}
 
\begin{figure}
\includegraphics[scale=1,width=8cm]{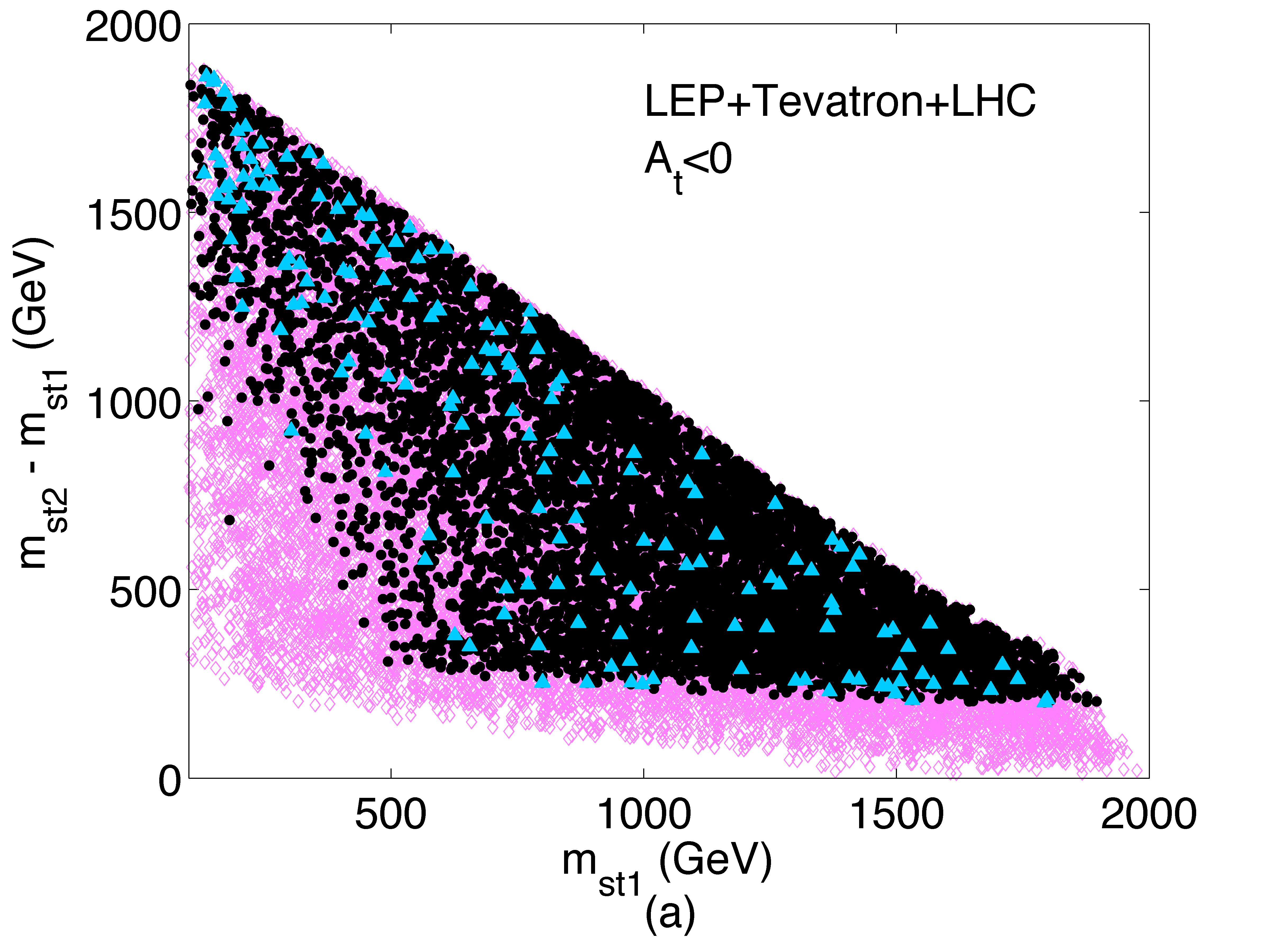}
\includegraphics[scale=1,width=8cm]{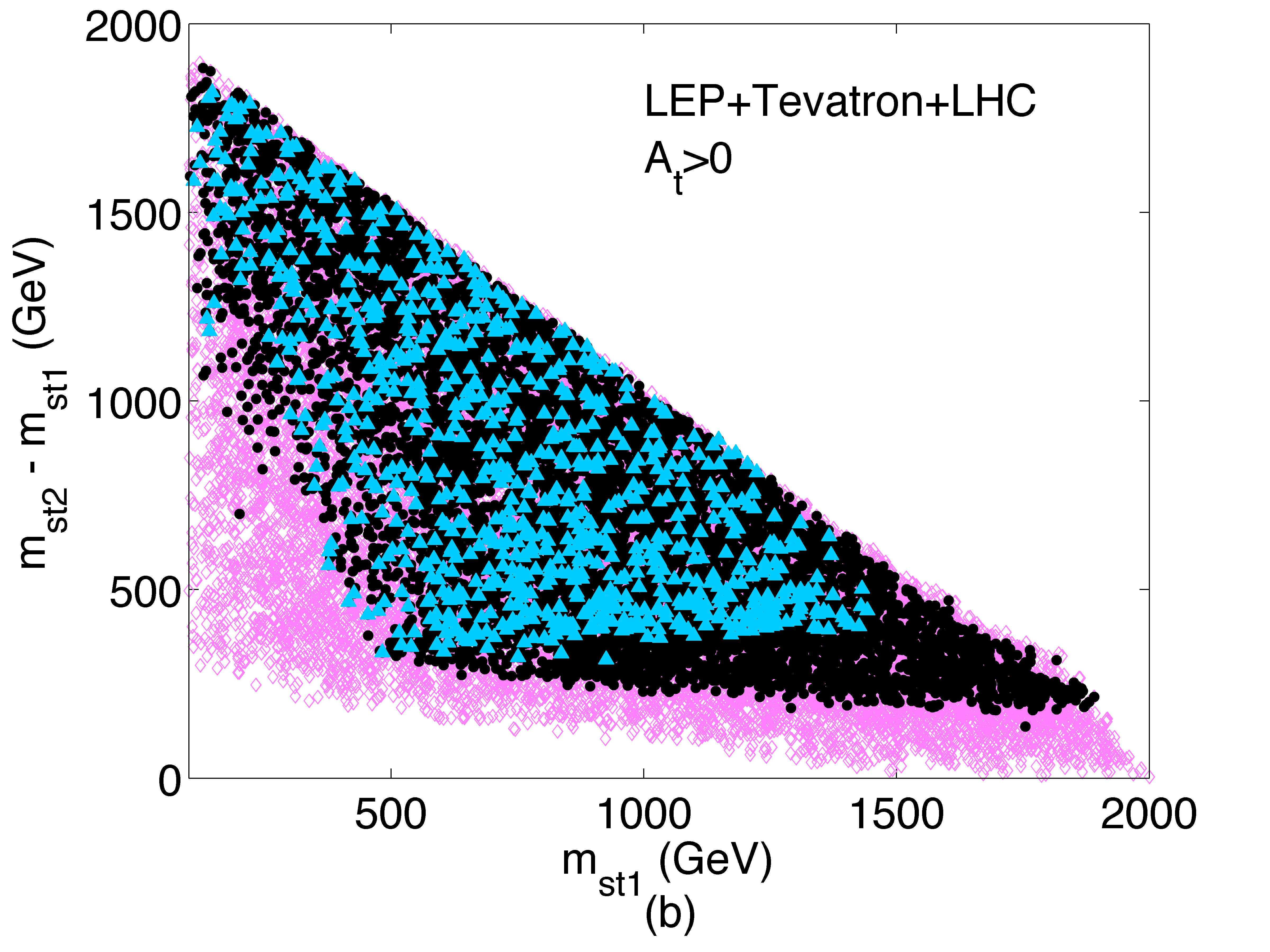}
\caption{Allowed region in $\Delta m_{\tilde{t}}=m_{\tilde{t}_2}-m_{\tilde{t}_1}$ versus $m_{\tilde{t}_1}$ 
for (a) $A_t<0$ and (b) $A_t>0$. The legends are the same as in Fig.~\ref{fig:tanb}.
}
\label{fig:stopmass}
\end{figure}

 As seen in Eq.~(\ref{eq:deltamh}), the stop sector provides substantial radiative corrections to the Higgs mass.  Large contributions from the stop sector need a relatively large $A_t$ term and at least  one of the stop mass parameters ($M_{3SQ}$ or $M_{3SU}$)  to be large.  In particular, when we restrict the Higgs mass to the narrow window of $125 \pm 2$ GeV, the mass splitting between the two stop mass eigenstates is found to be at least 200 GeV (300 GeV) for $A_{t}<0\ (A_{t}>0$). Although one of the stops can still be as light as 100 $-$ 200 GeV \cite{Arbey:2011aa, Carena:2011aa,Arvanitaki:2011ck}, the lighter the stop mass is, the larger the mass split would have to be, as seen from Fig.~\ref{fig:stopmass}. The collider phenomenology of the light stop $\tilde{t}_1$ (as well as $\tilde{t}_2$ when it is within collider reach) depends on the stop mixing angle $\theta_{\tilde{t}}$ and on the spectrum of gauginos, which is under current investigation \cite{MSSMStop_Su}.

Another way to reach a large positive correction to $m_h$ is to allow extremely heavy stop masses, which we did not explore. Even when  the stop masses are pushed up to 5$-$10 TeV, we could barely obtain a SM-like Higgs boson with a mass of around 125 GeV for $\tilde{A}_t\sim 0$.  Such a heavy stop mass would suffer from a severe fine-tuning problem, unless we envision the focus point scenario \cite{Feng:2011ew}.

Note that there could be negative contributions to the Higgs mass from the sbottom and stau sectors when those states are light. The mixing parameter takes the form $\tilde A_{b,\tau} = A_{b,\tau} - \mu \tan\beta$. Sizable corrections could be obtained for large $\tan\beta$ and large $\mu$, with  $\mu M_3<0$ (for sbottom contribution) and $\mu M_2<0$ (for stau contribution) \cite{Carena:1995wu, Carena:1995bx,Degrassi:2002fi,Heinemeyer:1998np}.
  
\subsubsection{${\rm Br}(h^0 \rightarrow \gamma\gamma, \ww, ZZ$)}

Observation of the processes $h^0,\ H^{0} \rightarrow \gamma\gamma, \ww, ZZ$ is of the utmost importance for the discovery and determination of the properties of the Higgs boson.
As we discussed above, in the decoupling region ($m_A > 300$ GeV) $h^0$ is SM-like and all the partial widths 
$h^{0} \to gg,\ \gaga,\ \ww,$ and $ZZ$ are typically slightly suppressed compared to the SM values, while they are highly correlated in the generic MSSM sector. However, there are certain MSSM parameter regions where  Br$(h^0 \rightarrow \gamma\gamma,\ \ww,\ ZZ)$ are not suppressed and even enhanced, and the predicted correlation is modified. 

Given the dominant decay of $h^0 \rightarrow b \bar b$, a suppression of the $h^0b \bar b$ coupling leads to the enhancement of the decay branching fractions of all three channels. There are two ways to suppress the $h^0b \bar b$ coupling, either through the Higgs mixing effects in the CP-even Higgs sector, or through the suppression of the bottom Yukawa coupling via SUSY radiative corrections.  The former is referred to as the ``small $\alpha_{\rm eff}$ region" in the literature \cite{Carena:2002qg}.  When the loop correction from the stop, sbottom, or stau sector to  $(M_H^2)_{12}$ is large and positive, the Higgs mixing angle $\alpha_{\rm eff}$ is small, leading to a suppressed $h^0b \bar b$ coupling, which is proportional to $\sin\alpha_{\rm eff}/\cos\beta$.  Such a region typically appears for moderate to large $\tan\beta$, small to moderate $m_A$, light stop, sbottom, stau masses, as well as large  $A_t$, $A_b$, $A_\tau$ and $\mu$ \cite{Carena:1999bh}.
The bottom Yukawa  could also receive large radiative corrections in the MSSM, which can either be enhanced or suppressed compared to its tree-level value \cite{Carena:1998gk}. In particular, strong suppression of the bottom Yukawa could be achieved for a large and positive value of $\mu M_3$ \cite{deltahb}. 

While the partial decay width for $h^0\rightarrow \ww, ZZ$ ( which are $\propto \sin^2(\beta-\alpha_{\rm eff})$) are typically suppressed in the MSSM compared to the SM values,  loop induced decay of $h\rightarrow \gamma \gamma$, on the other hand, could be enhanced with stop, sbottom, or stau contributions with large left-right mixing and small sparticle masses.   For light stop and light sbottom, however, the simultaneous suppression of the production channel $gg\rightarrow h^0$ results in an overall suppression of $gg\rightarrow h^0 \rightarrow \gamma\gamma$.  Stau, on the other hand, does not lead to the suppression of $gg\rightarrow h^0$.  For 
stau mass around 300 GeV with large $\tan\beta$ and $A_\tau$, an enhancement of $gg\rightarrow h^0 \rightarrow \gamma\gamma$ as large as a factor of 2 is possible \cite{Carena:2011aa}.

As noted above, the stop left-right mixing $A_{t}$ is of critical importance since  it has multiple roles here.  First, it affects the correction to the Higgs mass with positive $A_t$ and gives a larger correction compared to the case of negative $A_t$, due to a two-loop contribution with gluino and stops. 
 $A_t$ could also affect the mixing in the CP-even Higgs sector,  bottom Yukawa, Higgs coupling to $\gamma\gamma$, as well as the production of $gg\rightarrow h^0$. Third, the sign of $A_t$ also changes the sign of the chargino contribution to $b\rightarrow  s \gamma$, as discussed below.
 
 Note that similar effects could also occur in the non-decoupling region with $H^0$ being the SM-like Higgs.  Our discussion above is still valid with the substitution of $h^0$ by $H^0$, $\sin(\beta-\alpha_{\rm eff})$ by $\cos(\beta-\alpha_{\rm eff})$ and $\sin\alpha_{\rm eff}/\cos\beta$ by $\cos\alpha_{\rm eff}/\cos\beta$.

\subsubsection{$b\rightarrow s \gamma$}
The dominant indirect constraints on a light Higgs sector comes from $b\rightarrow s \gamma$. The current observed value is  
${\rm Br} (b \rightarrow s\gamma)_{\rm exp} = (3.55 \pm 0.24 \pm 0.09) \times 10^{-4}$ \cite{Asner:2010qj} 
and  the Next-to-Next-to-Leading Order QCD correction gives
${\rm Br} (b \rightarrow s \gamma)_{\rm SM} = (3.15 \pm 0.23) \times 10^{-4}$ \cite{Misiak:2006zs, Misiak:2006ab}.
There are two dominant MSSM contributions, namely, charged Higgs-top loop corrections and chargino-stop loop corrections.
While charged Higgs loops always give positive contributions,   contributions from the chargino loops   
depend on the signs of $M_2$, $\mu$, and $A_t$ \cite{Barbieri:1993av}.  In particular, the contribution from the Higgsino-stop loop that is proportional to the left-right mixing in the stop sector gives a negative contribution for $\mu A_t<0$ and a positive contribution for $\mu A_t>0$.  For our choice of $M_2>0$, $\mu>0$, the rest of the chargino loop contributions typically provides a negative correction to $b\rightarrow s \gamma$.
In the nondecoupling region with small $m_A$, significant negative contributions from chargino loops are needed to cancel the charged Higgs contribution, which typically requires a small $M_2$.    
In the decoupling region where the charged Higgs contribution is negligible, given that the current SM prediction is lower than the experimental value, 
a positive $A_t$ is slightly preferred so that the MSSM corrections do not make the deviation worse. 
  
 %%%%%%%%%%%%%%%%%%%%

\section{Future Expectation with the search for the SM-like Higgs boson}
\label{sec:Future}

\begin{figure}[tb]
\includegraphics[scale=1,width=8cm]{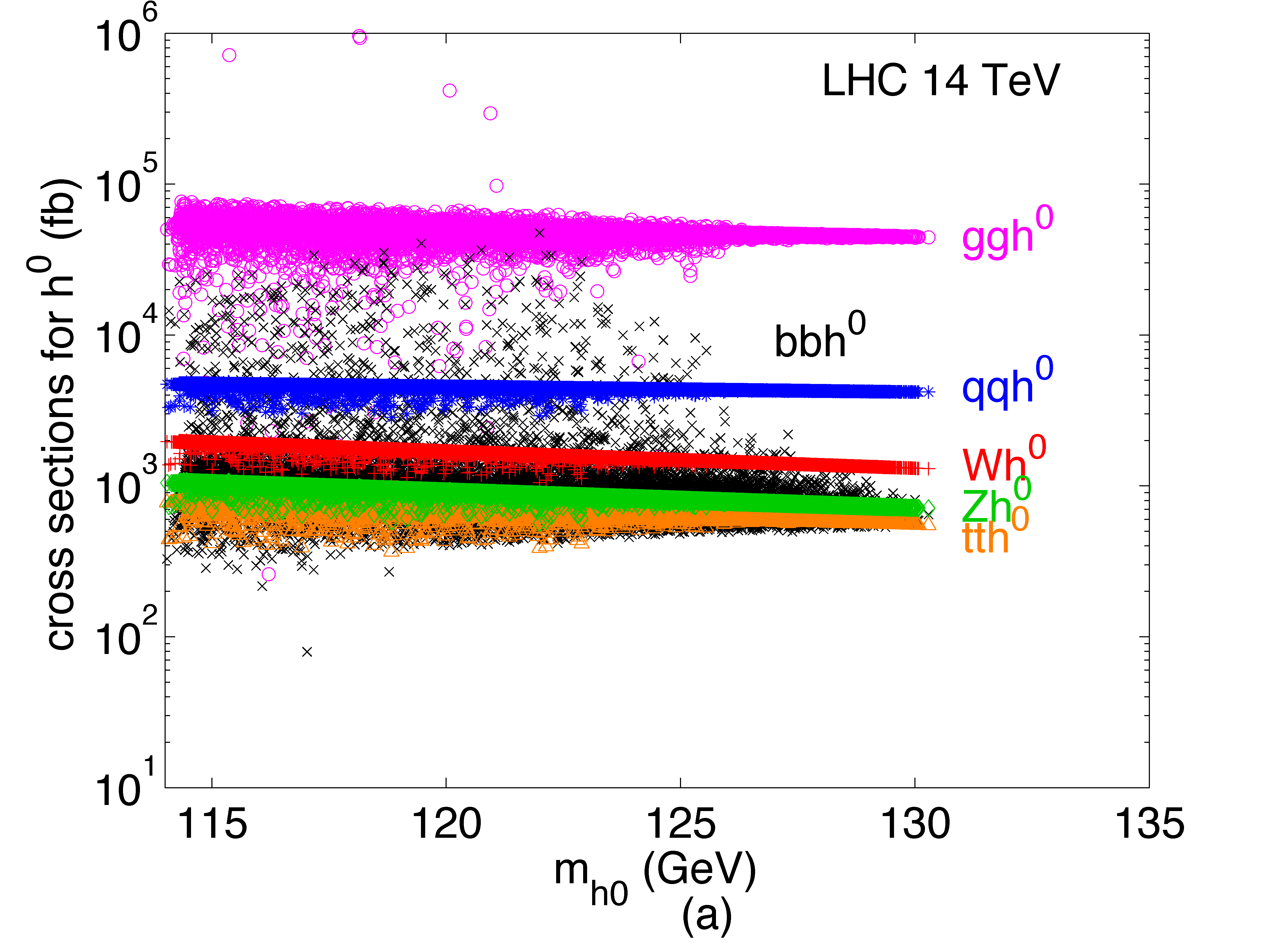}
\includegraphics[scale=1,width=8cm]{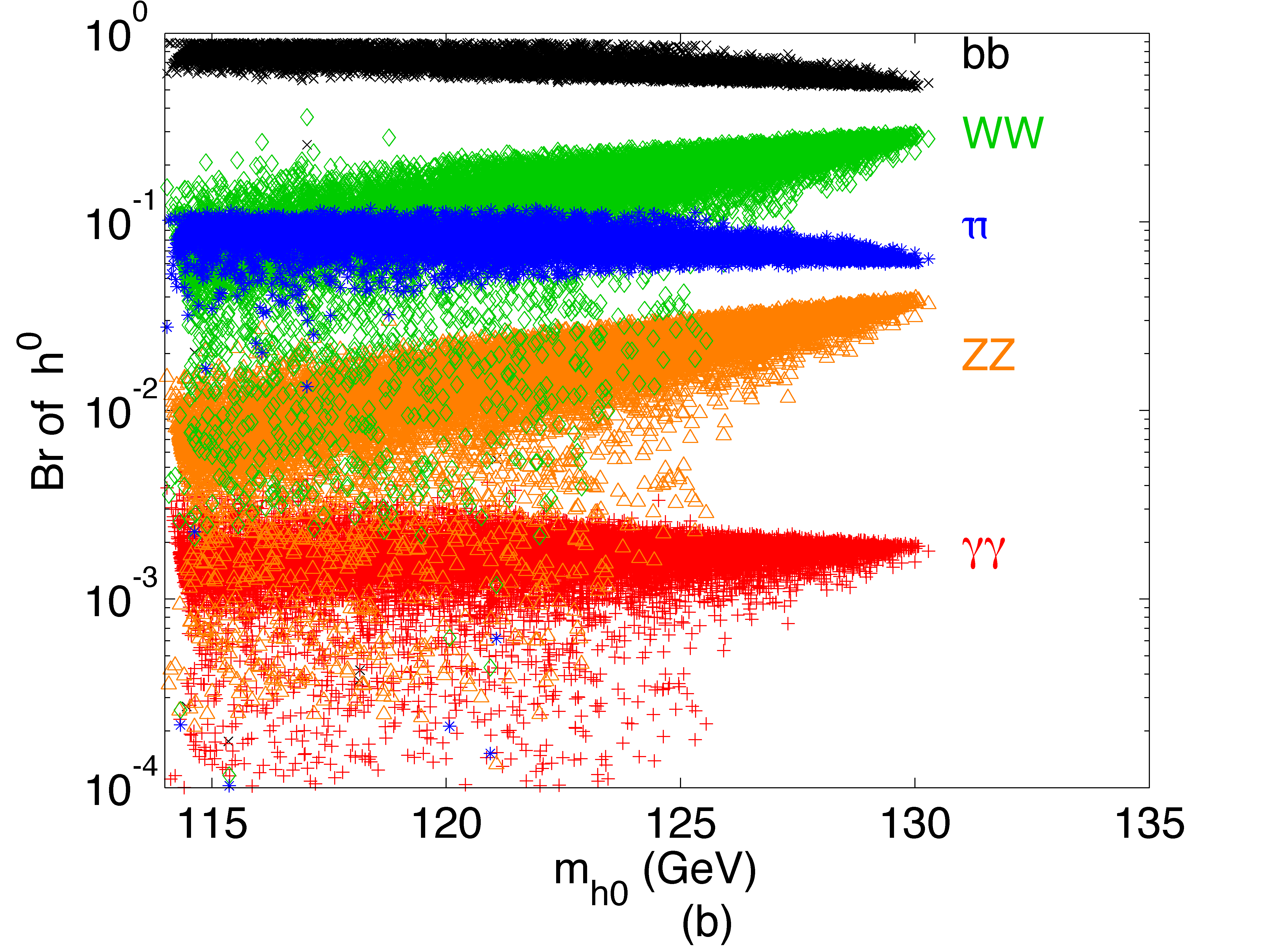}
\includegraphics[scale=1,width=8cm]{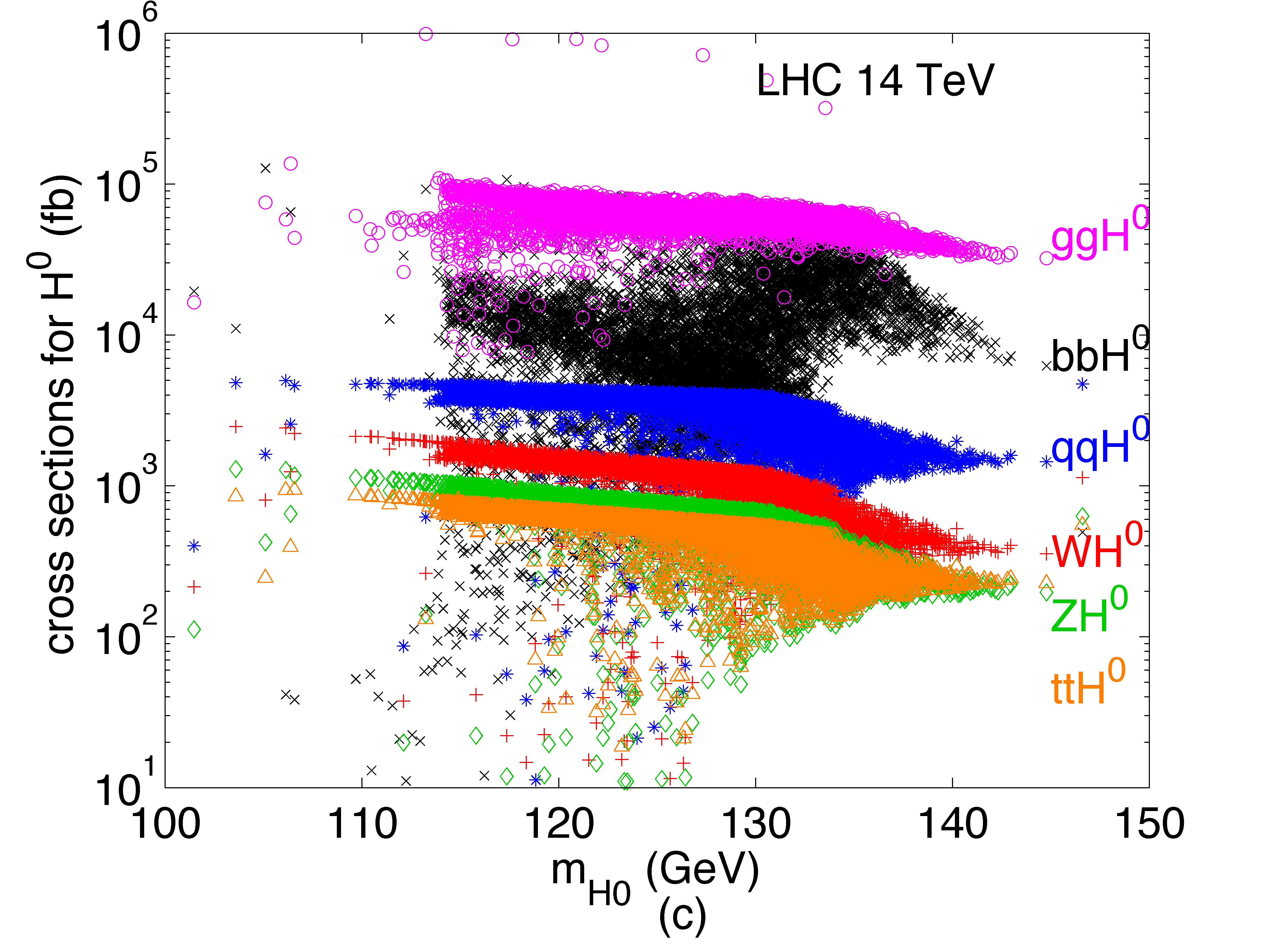}
\includegraphics[scale=1,width=8cm]{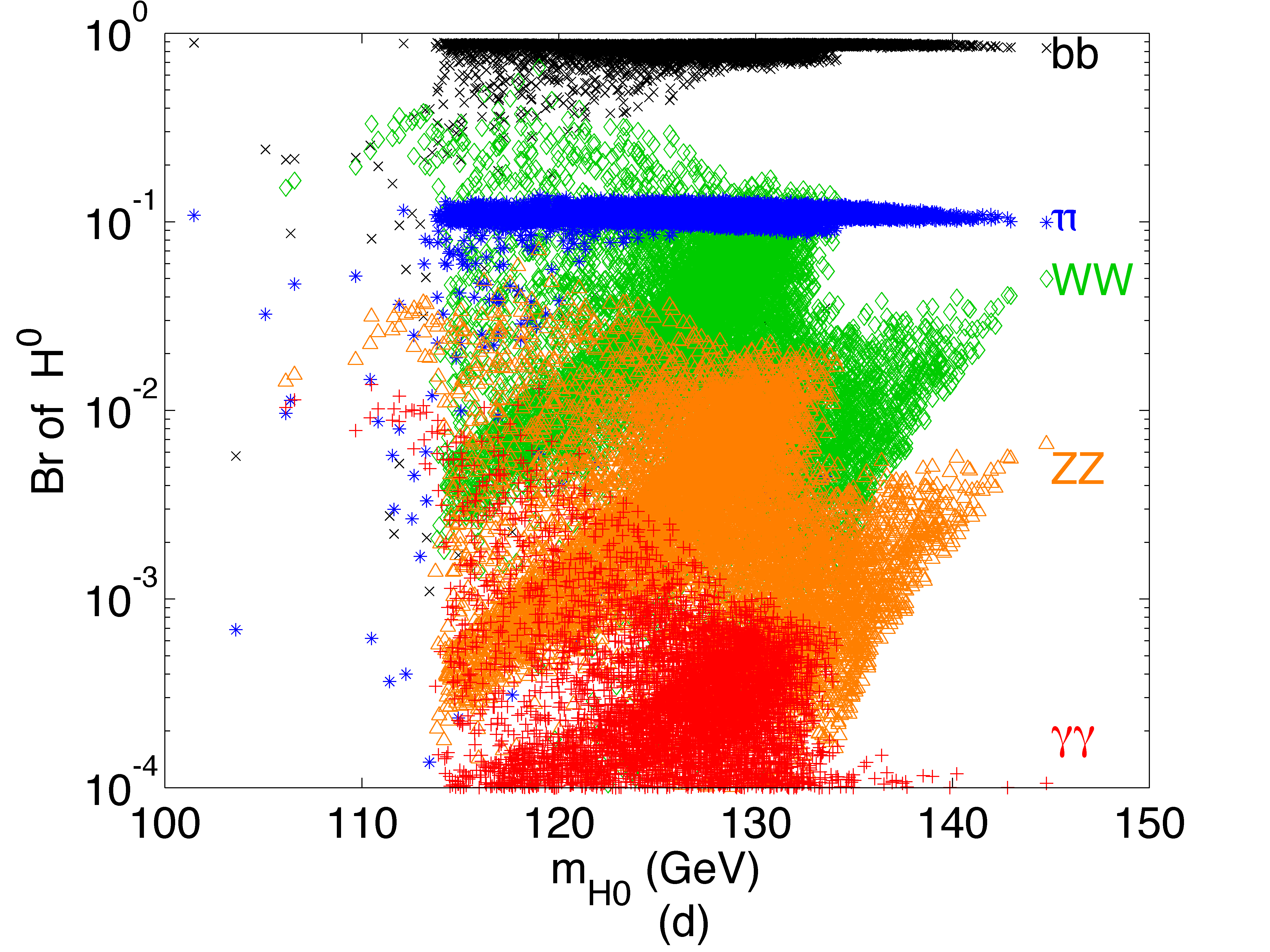}
\caption{Cross sections (left panels) at the 14 TeV LHC and branching fractions (right panels) for the SM-like Higgs boson. (a) and (b) are for the decoupling region for $h^{0}$, (c) and (d) in the non-decoupling region for $H^{0}$. }
\label{fig:cs_h}
\end{figure}

In anticipation of the successful operation at the energies of 8 and 14 TeV, the LHC will deliver a large amount of quality data in the years to come. If a signal for a SM-like Higgs boson is confirmed, then the task would be to determine its basic properties to good precision \cite{Duhrssen:2004cv}. 
On the other hand, if the signal for a SM-like Higgs boson continues to be elusive, it would provide further important information about the MSSM Higgs sector. 

We first reiterate the production and decay of a SM-like Higgs boson in the MSSM at the LHC.
For the convenience of future discussions, we divide the $\ma$ mass parameter into two regions\footnote{This division is not meant to be a rigorous definition, rather for the purpose of numerical illustration.},
\begin{eqnarray}
\nonumber
&&{\rm Non-decoupling\ region:}\quad 90 \gev < \ma < 130\gev ; \\
&&{\rm Decoupling\ region:}\quad 130 \gev < \ma  .
% < 500\gev .
\label{eq:division}
\end{eqnarray}
In Fig.~\ref{fig:cs_h}, the total cross sections (left panels) at 14 TeV and decay branching fractions (right panels) for the leading channels of the SM-like Higgs boson are shown 
after passing all the constraints, (a)-(b) in the decoupling region for the SM-like $h^{0}$, and (c)-(d) in the non-decoupling region for the SM-like $H^{0}$. 
As before, other parameters in the MSSM are scanned over the range in Eq.~(\ref{eq:para}).  
The leading production channel is via $gg$ fusion and of a rate at the order of 50 pb \cite{xsection}
\begin{equation}
gg \to h^{0} \ (H^{0}).
\end{equation}
The $b\bar b$ initial process is known to be small in the SM at the order of 0.6 pb for a 125 GeV mass at 14 TeV \cite{Maltoni:2003pn}, but it could be significantly enhanced in certain SUSY parameter region especially at large $\tan^{2}\beta$ \cite{DiazCruz:1998qc}. This is seen in the plot by the large spread in Figs.~\ref{fig:cs_h}(a) and (c). The electroweak processes of vector-boson-fusion and Higgs-strahlung are the next important sources for the SM-like Higgs boson production
\begin{equation}
q q' \to q q' h^{0} \ (H^{0}),\quad 
q\bar q' \to W h^{0} \ (H^{0}),\quad  Z h^{0} \ (H^{0}), \quad 
\label{eq:EW}
\end{equation}
which are roughly in  the range of $0.5-5$ pb. 
For those production channels that do not involve heavy quarks, the cross section rates are well predicted as seen from the narrow bands. 
%The channels $gg$ and especially $b\bar b$ are more spread out due to the variation of $\tan\beta$, as well as other relevant SUSY parameters associated with the $b$ quark.
For the branching fractions, the $b\bar b,\ \tau\tau$ modes are stable due to the cancellation of a common factor 
$\tan^{2}\beta$ in the ratios, while all  other modes result in a large spread.  Because of the nature of the SM-like Higgs boson, in either the decoupling or the non-decoupling region, the cross sections of $h^{0}$ or $H^{0}$ behave similarly. 

\begin{figure}[tb]
\includegraphics[scale=1,width=8cm]{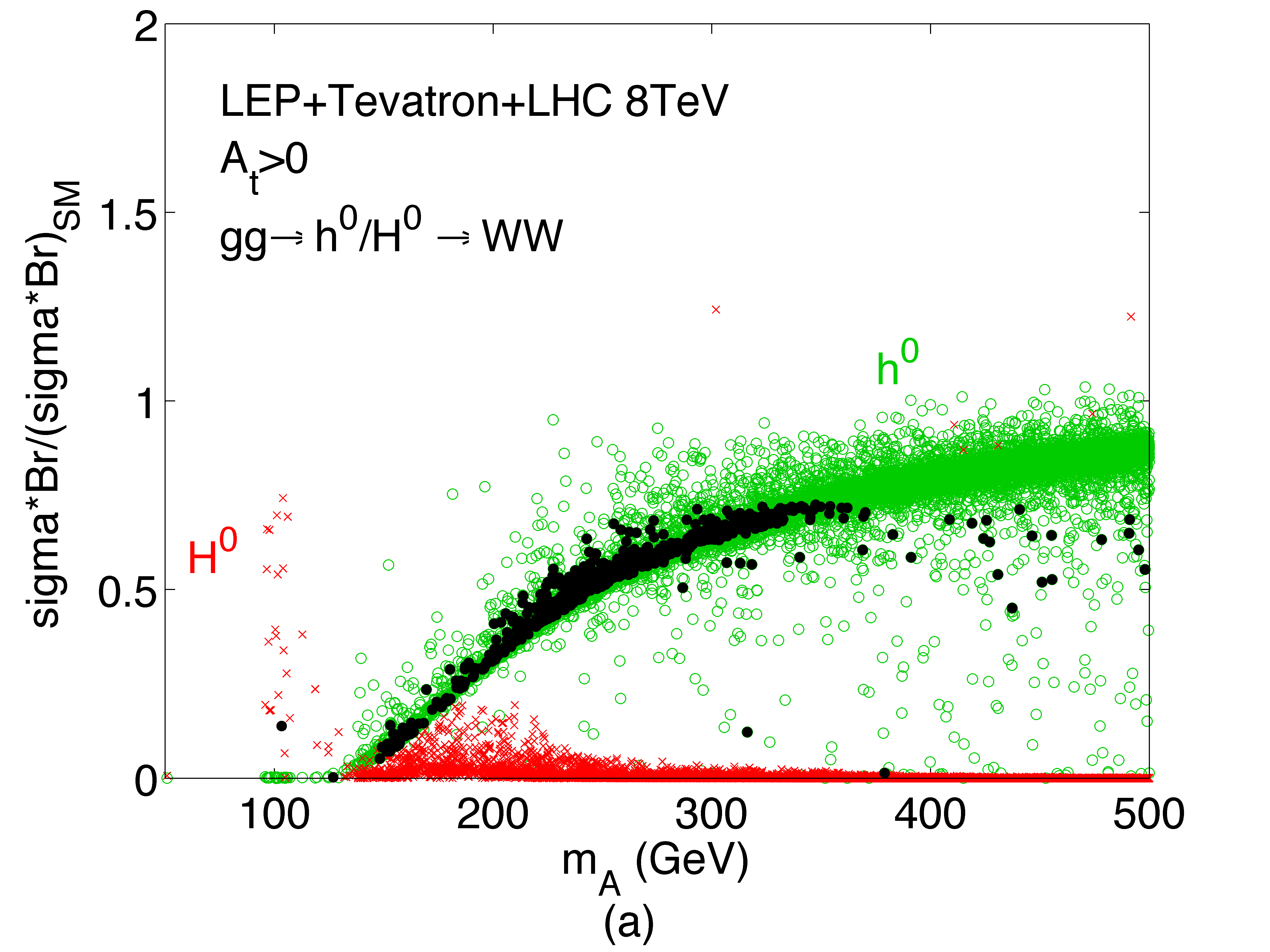}
\includegraphics[scale=1,width=8cm]{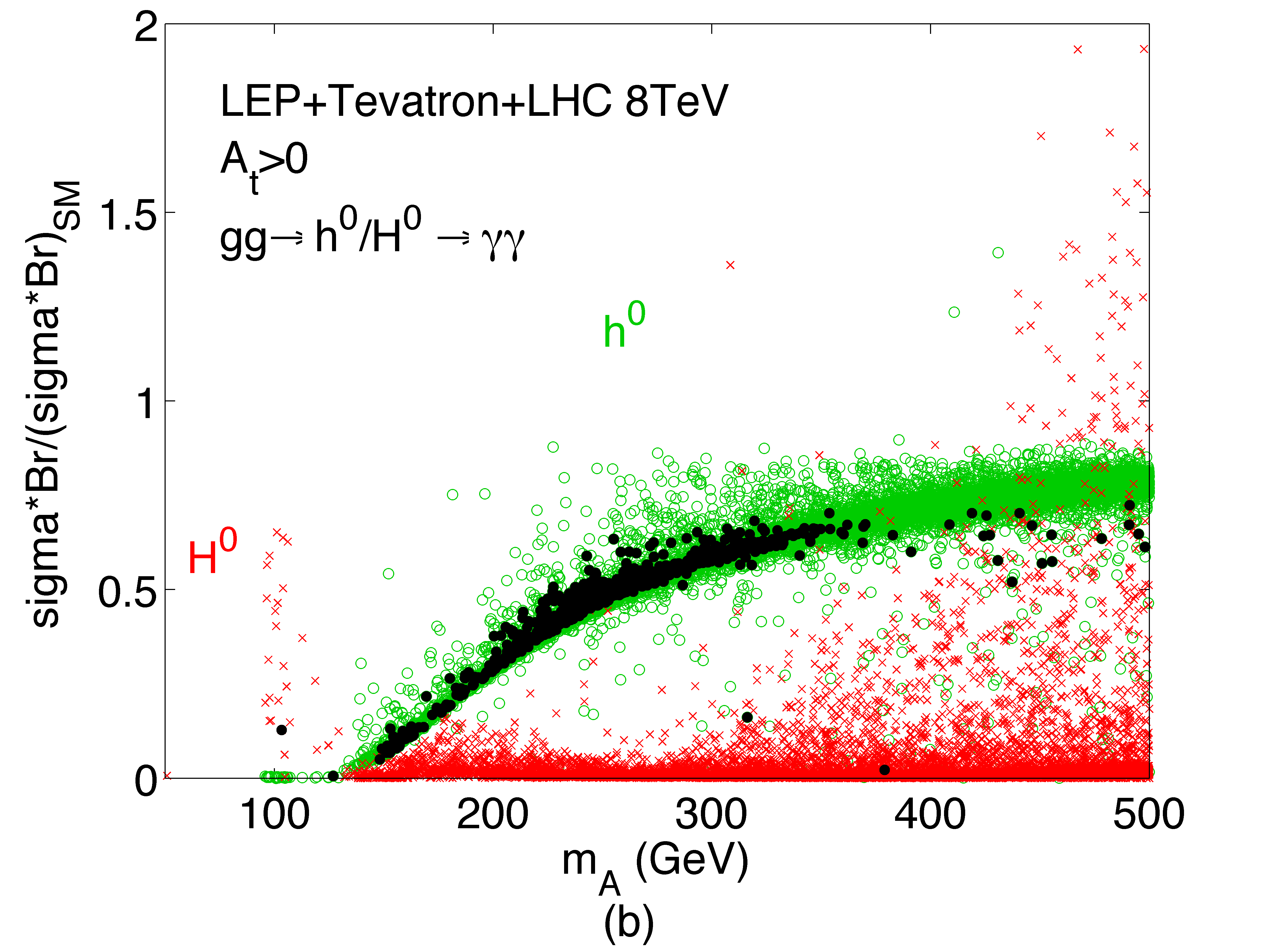}
\caption{(a) Signal cross section ratio $\sigma / \sigma_{SM}$ versus $\ma$ with the 8 TeV improvement of sensitivity at the LHC for (a) the $W^+W^-$ channel, and for (b) the $\gaga$ channel.
The legends are the same as in Fig.~\ref{fig:csR}. Other parameters are scanned over the range in Eq.~(\ref{eq:para}). 
}
\label{fig:BoundWW}
\end{figure}
 
We now consider improved measurements for the search for the SM-like Higgs boson and see the implication for the Higgs sector of the MSSM. 
Without going through detailed signal and background simulations, we simply assume the future data collection  as in Table \ref{tab:Improve}. The signal sensitivity improvements are scaled with $\sqrt{\sigma_{signal}\times L}$ where $\sigma_{signal}$ is the total cross section (We use $\mh=125 \gev $ as an illustration.) for SM Higgs boson production \cite{xsection}, and $L$ is the integrated luminosity.
\begin{table}[h]
\begin{tabular}{|c|c|c|c|}
\hline
C.M.~Energy & 7 TeV & 8 TeV & 14 TeV \\ \hline
Integrated luminosity & 5 fb$^{-1}$ & 15 fb$^{-1}$ & 30 fb$^{-1}$ \\ \hline
Cross section $gg\to h$ & 15.3 pb & 19.5 pb & 51.4 pb \\ \hline
Signal statistical improvement  &1 & 2 & 4.5 \\ \hline
\end{tabular}
\caption{Statistical improvement factors for the SM-like Higgs boson search with $\mh =125 \gev $ at the different energies of the LHC and with different luminosity assumption. }
\label{tab:Improve}
\end{table}

Estimated improvements could have already been seen in Fig.~\ref{fig:csRmh} by the two lower curves both for $\ww$ and $\tautau$ channels. As expected, the $\ww$ channel has stronger experimental sensitivity. With this channel alone, a Higgs boson in the MSSM with SM-like couplings could be excluded at $95\%$ C.L. at the LHC, giving the allowed mass ranges
\begin{equation}
\ww:\quad \mh <120\gev  \ \ {\rm at\  8\ TeV}, \quad \mh <115 \gev \ \ {\rm at\ 14\ TeV.}
\end{equation}
These upper bounds could be relaxed if the coupling to $W^{\pm}$ is weaker than that of the SM.

\begin{figure}
\includegraphics[scale=1,width=8cm]{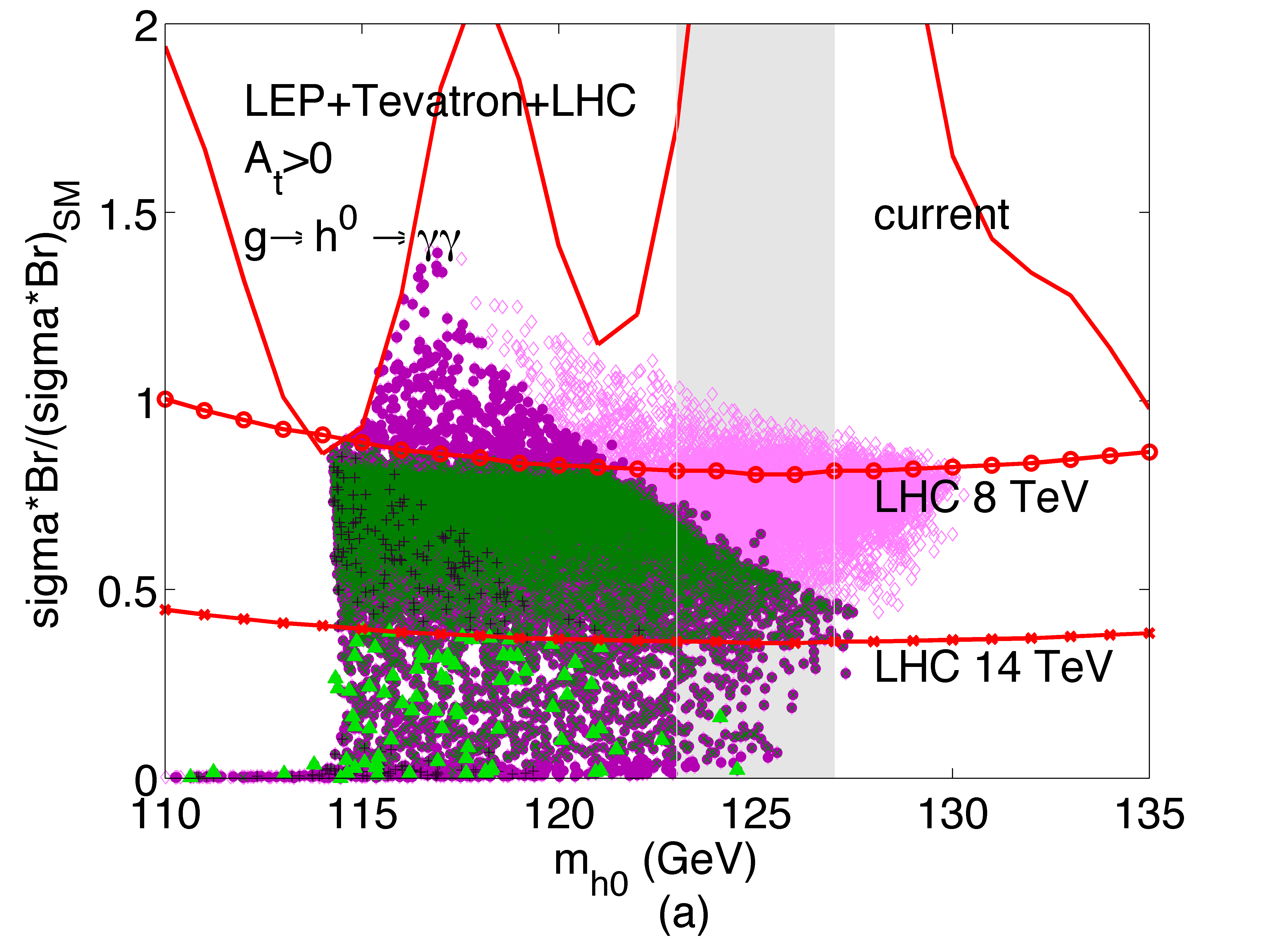}
\includegraphics[scale=1,width=8cm]{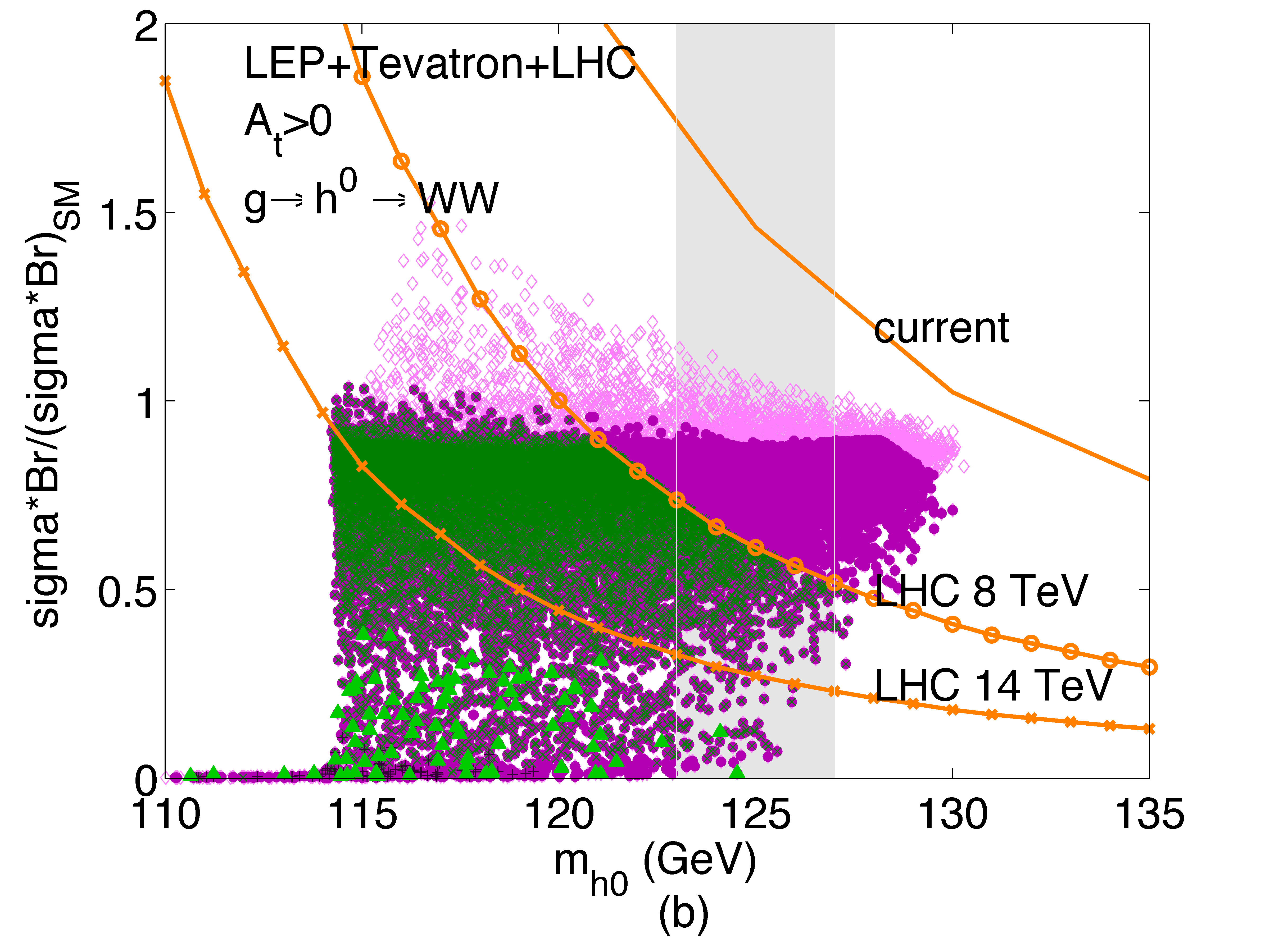}
\includegraphics[scale=1,width=8cm]{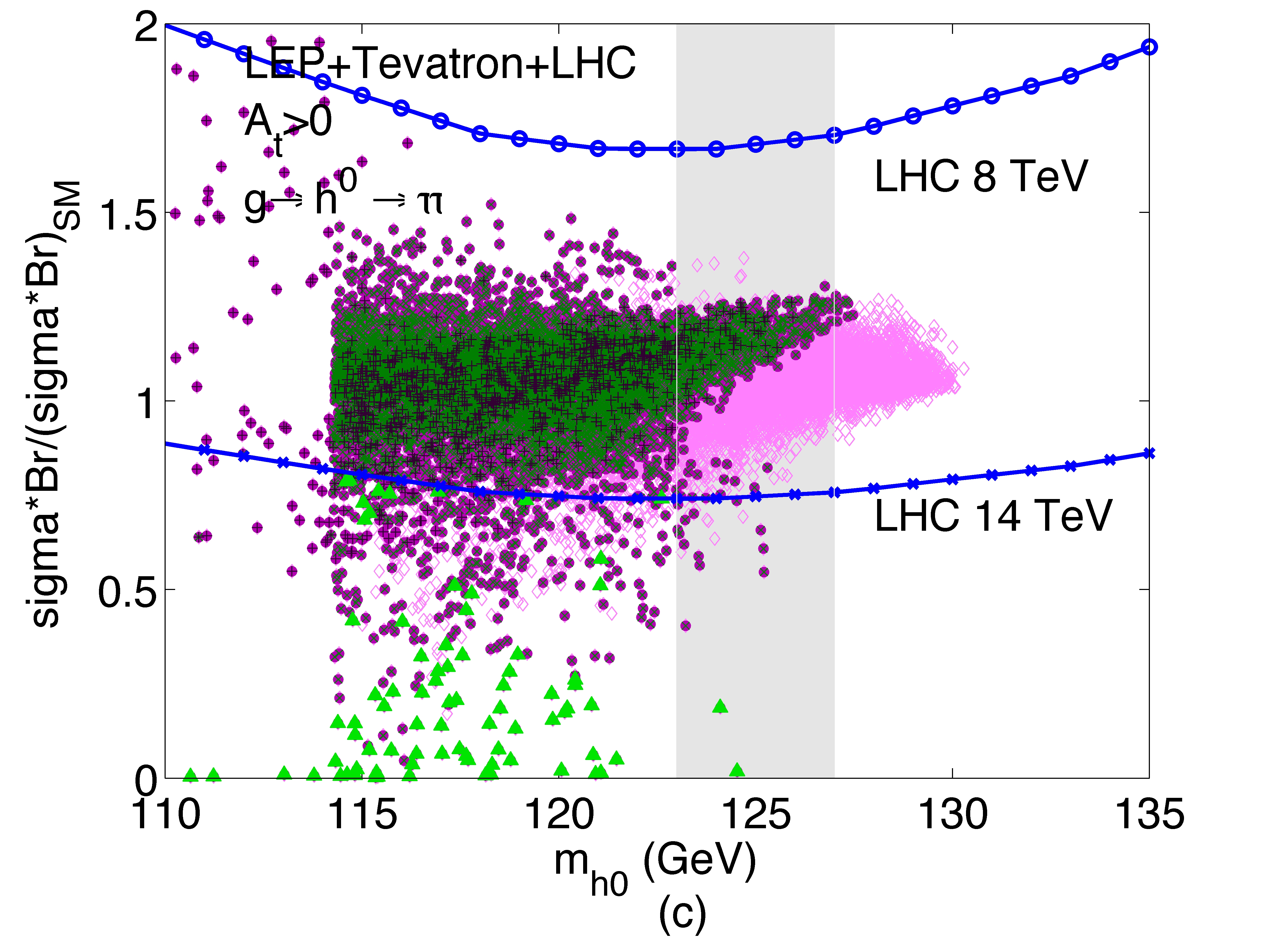}
\includegraphics[scale=1,width=8cm]{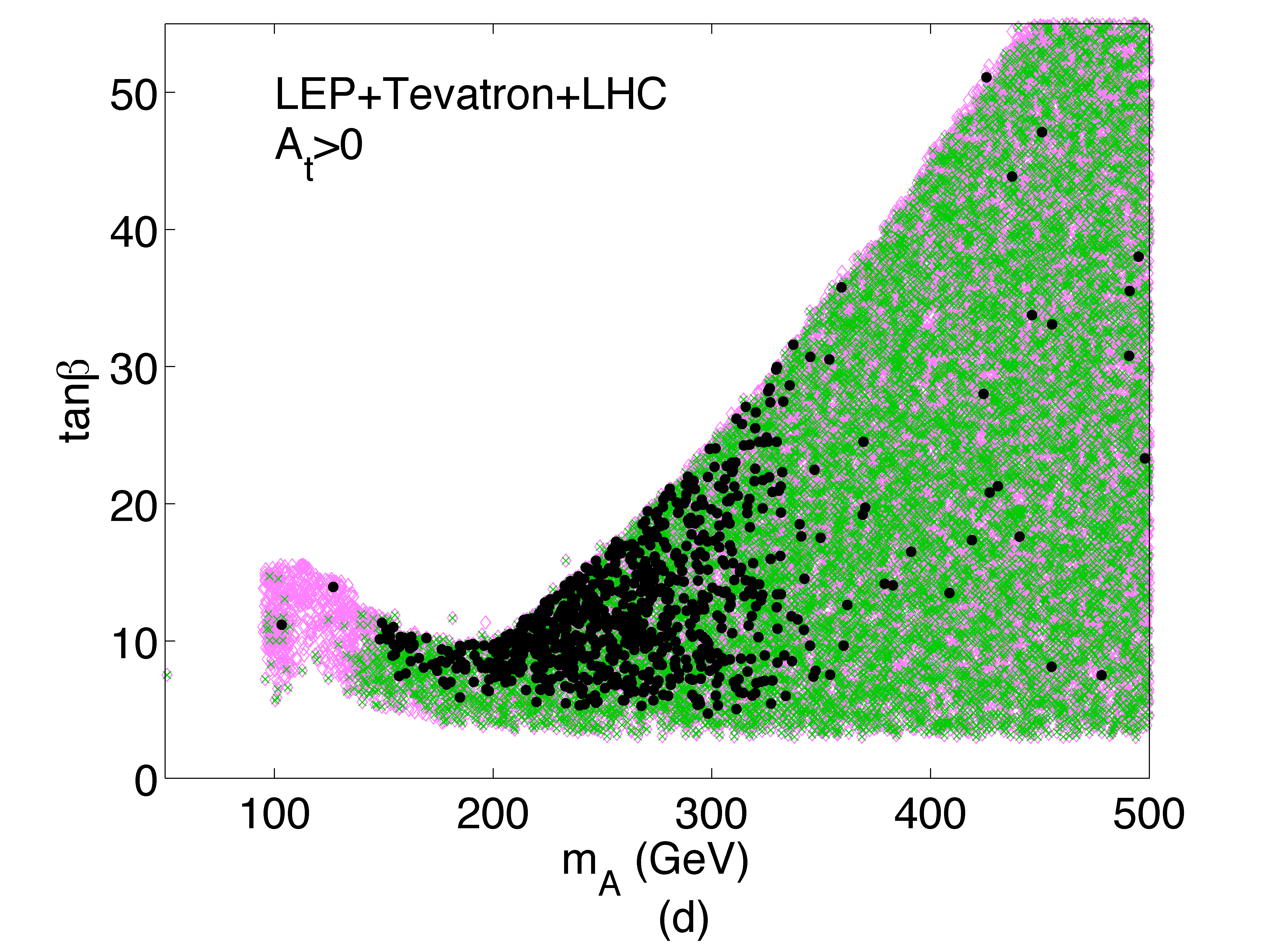}
\caption{Signal cross section ratio $\sigma / \sigma_{SM}$ versus $\mh$. (a) For the $\gaga$ channel:
the large light purple region is from the current LEP2, Tevatron and LHC bounds, the middle medium purple region is with the 8 TeV improvement including $\ww,\ ZZ,\ \tautau$ channels, and the medium green includes all four channels. The dark purple and light green are the same as the medium purple and medium green above, but with the 14 TeV improvement.
(b) and (c) For the $\ww$ channel and $\tautau$ channel, with the switch of $\gaga \leftrightarrow \ww$, and $\gaga \leftrightarrow \tautau$, respectively.
The two lower curves are the expected improvements from the individual channels (a) for $\gaga$, (b) for $\ww$ and (c) for $\tautau$. For comparison, the current observed $95\%$ bounds at ATLAS for $\gaga$  and $\ww$ are shown in solid curves as well.  Vertical bands indicate the 123 $-$127 GeV mass window for $h^0$.    Other parameters are scanned over the range in Eq.~(\ref{eq:para}).    Panel (d) shows the
constrained region in the parameter space of $\tan\beta - \ma$ for 8 TeV with 
 $A_t>0$.  The light purple shows the current LEP2, Tevatron and LHC bounds. The   green
 is the 8 TeV expected improvement including all $\ww,\ ZZ,\ \gaga$ and $\tautau$ channels. The
black dots includes the requirement of the mass  window in Eq.~(\ref{eq:mass}).
}
\label{fig:Boundgaga}
\end{figure} 

We consider the potential improvement by combining the $\ww, \ ZZ, \ \gaga$ and $\tautau$ channels. Although theoretically correlated as discussed earlier, these channels are experimentally complementary since they are sensitive to a Higgs signal in different mass regions. We thus scale the ATLAS expected curves by the sensitivity factors in Table \ref{tab:Improve} and estimate the expected improvements at the 8 TeV and 14 TeV LHC. 
In Fig.~\ref{fig:BoundWW}, we present the reduced regions for the cross sections versus $\ma$ for (a) the $\ww$ channel and (b) the $\gaga$ channel. We note that, similar to the case in Fig.~\ref{fig:csR}, a narrow mass window would further force the CP-even Higgs boson to have weaker couplings to the electroweak gauge boson, and thus less SM-like. The related consequence would be to drag $\ma$ lower, away from the decoupling region. We also note from Fig.~\ref{fig:BoundWW}(b), that the cross section spread for the $\gaga$ channel, especially for $H^{0}$ is significantly larger than that for $\ww$, due to the other SUSY parameter effects in the loop for $H^{0} \to \gaga$.

 To gain more intuition with respect to the experimental observables, we now examine the signal cross section ratio as a function of the SM-like Higgs boson mass $\mh$ with the progressive steps in Fig.~\ref{fig:Boundgaga}(a) for the $\gaga$ channel, (b) for the $\ww$ channel and  (c)  for the $\tautau$ channel.
In Fig.~\ref{fig:Boundgaga}(a), 
the large light purple region is from the current LEP2, Tevatron and LHC bounds. 
The middle medium purple region is with the 8 TeV expected improvement including $\ww,\ ZZ,\ \tautau$ channels. The medium green   includes the $\gaga$ channel in addition. The lower dark purple and light green  are the same as the medium purple and medium green above, but with the 14 TeV expected improvement. For Fig.~\ref{fig:Boundgaga}(b) and (c), we simply switch $\gaga \leftrightarrow \ww$, and $\gaga \leftrightarrow \tautau$, respectively.
The two lower curves are the expected improvements from the individual channels for $\gaga$ in Fig.~\ref{fig:Boundgaga}(a), for $\ww$ in (b) and for $\tautau$ in (c).  For comparison, the current observed $95\%$ C.L. bound at ATLAS for $\gaga$ and $\ww$ are shown in Fig.~\ref{fig:Boundgaga}(a) and (b)  as the red and orange curve at the top. 

Figure \ref{fig:Boundgaga} contains essential results and several remarks are thus in order. First, as seen from the $\gaga$ channel, the expected improvements look impressive.
The 8 TeV expected improvement will already be able to cover the full MSSM mass range with a SM coupling strength. The 14 TeV expected improvement will be able to probe a weaker coupling down to about a half of the SM cross section. 
Second, the $\gaga$ channel and the $\ww,\ ZZ$ channels are complementary, with the former more sensitive in the low-mass region and the latter in the high-mass region. 
Third, due to the correlation of the Higgs decay channels to $\gaga$ and to $\ww,\ ZZ$ as predicted in the MSSM, one would expect their sensitivity curves to move down consistently. If otherwise the signal in the $\gaga$ channel remains as the red curve at the top, while the $\ww$ channel continues to be reduced and break the MSSM correlation, then new physics beyond the MSSM must exist. 
The $\tautau$ channel shown in Fig.~\ref{fig:Boundgaga}(c) is less sensitive than the $\gaga$ mode by about a factor of 2 for the cross section measurement
as expected based on the current ATLAS analysis. The qualitative features in Fig.~\ref{fig:Boundgaga}(c) are similar to (a) and (b)  otherwise.

Finally, we illustrate the expected improvement in constraining the parameters in the $\tan\beta-\ma$ plane
in Fig.~\ref{fig:Boundgaga}(d)  for 8 TeV.  Again the narrow mass window Eq.~(\ref{eq:mass}) is crucial when constraining 
the $\ma$ range as indicated by the black dots. 

\section{The Search for Non-SM-Like Higgs Bosons}
\label{sec:LHCSearch}

The searches for the SM Higgs boson in the LHC experiments have a direct impact on our knowledge of the SM-like Higgs boson in the MSSM Higgs sector, as discussed in the previous sections. However, in order to unambiguously confirm the structure of the Higgs sector in the MSSM, the most crucial next step would be to predict and test the other aspects correlated with SM-like Higgs boson searches. Naturally, the other Higgs bosons in the MSSM are of the highest priority. In this section, we comment on the search strategy for the two parameter regions as defined in Eq.~(\ref{eq:division}).
We use FeynHiggs to calculate the cross section for all the production modes except for the two Higgs modes, which it does not provide.  For these, we use the couplings that FeynHiggs provides and we calculate the cross sections using CalcHEP \cite{Pukhov:1999gg}.  We then multiply the two Higgs cross sections by a K-factor of 1.3 \cite{Dawson:1998py}.

%%%%%%%%%%%%%%%%%%%%%%%%%%%
\subsection{Non-decoupling region: $\mh \sim \ma \sim \mz,\quad \mH \sim \mHpm \sim 125\gev $}
\label{sec:channels_nondecouple}

\begin{figure}[tb]
\includegraphics[scale=1,width=7.6cm]{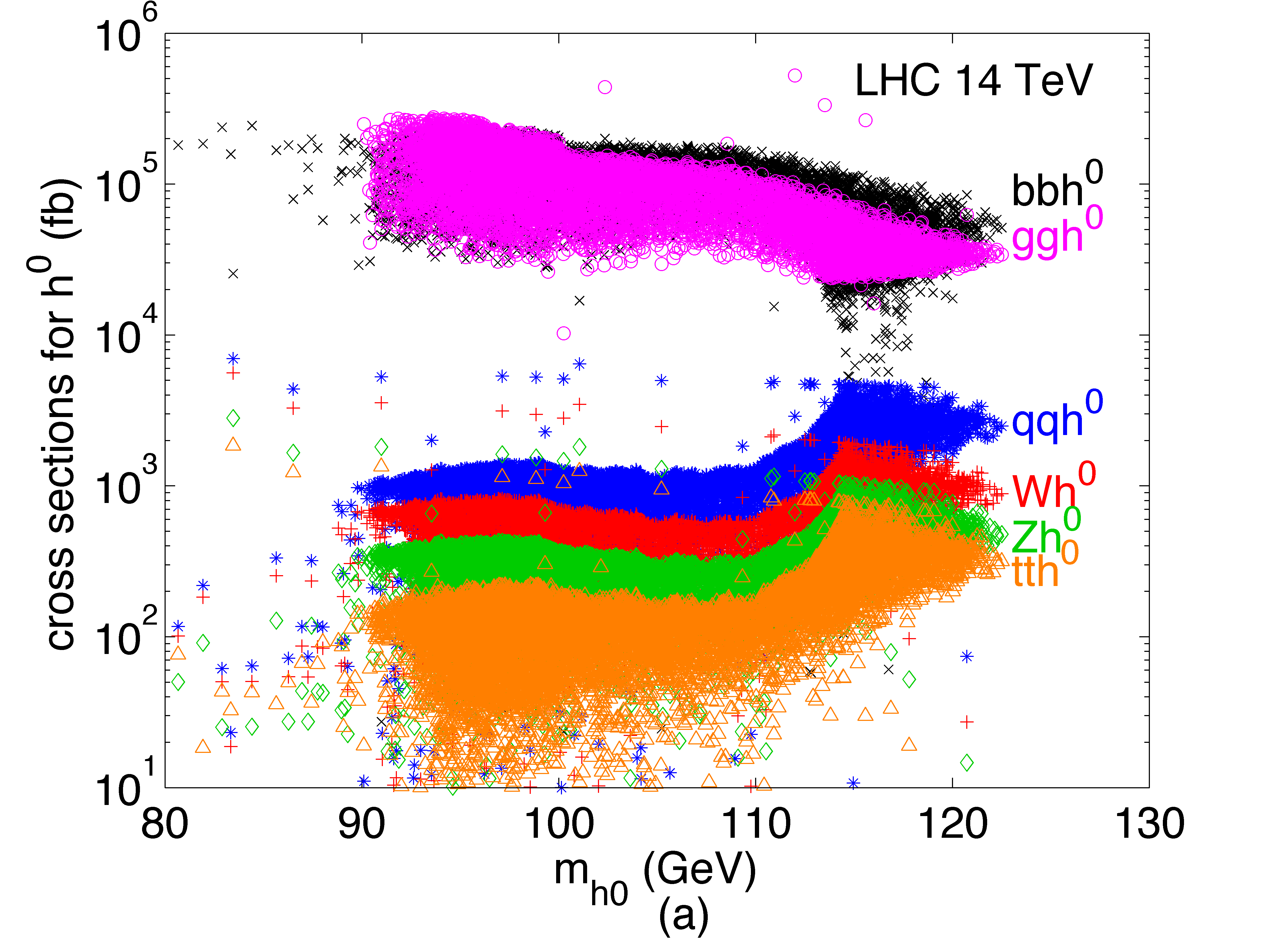}
\includegraphics[scale=1,width=7.6cm]{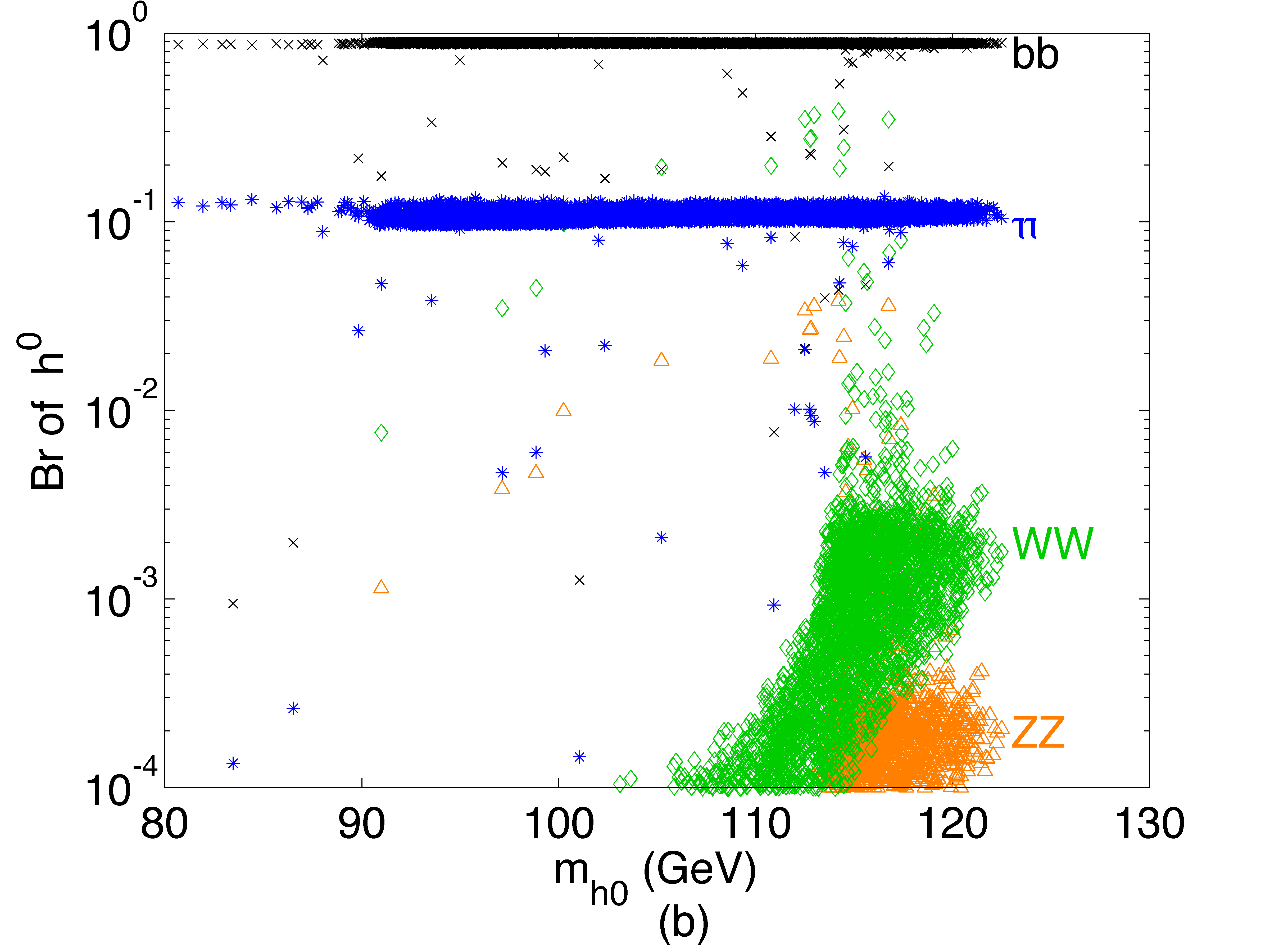}
\includegraphics[scale=1,width=7.6cm]{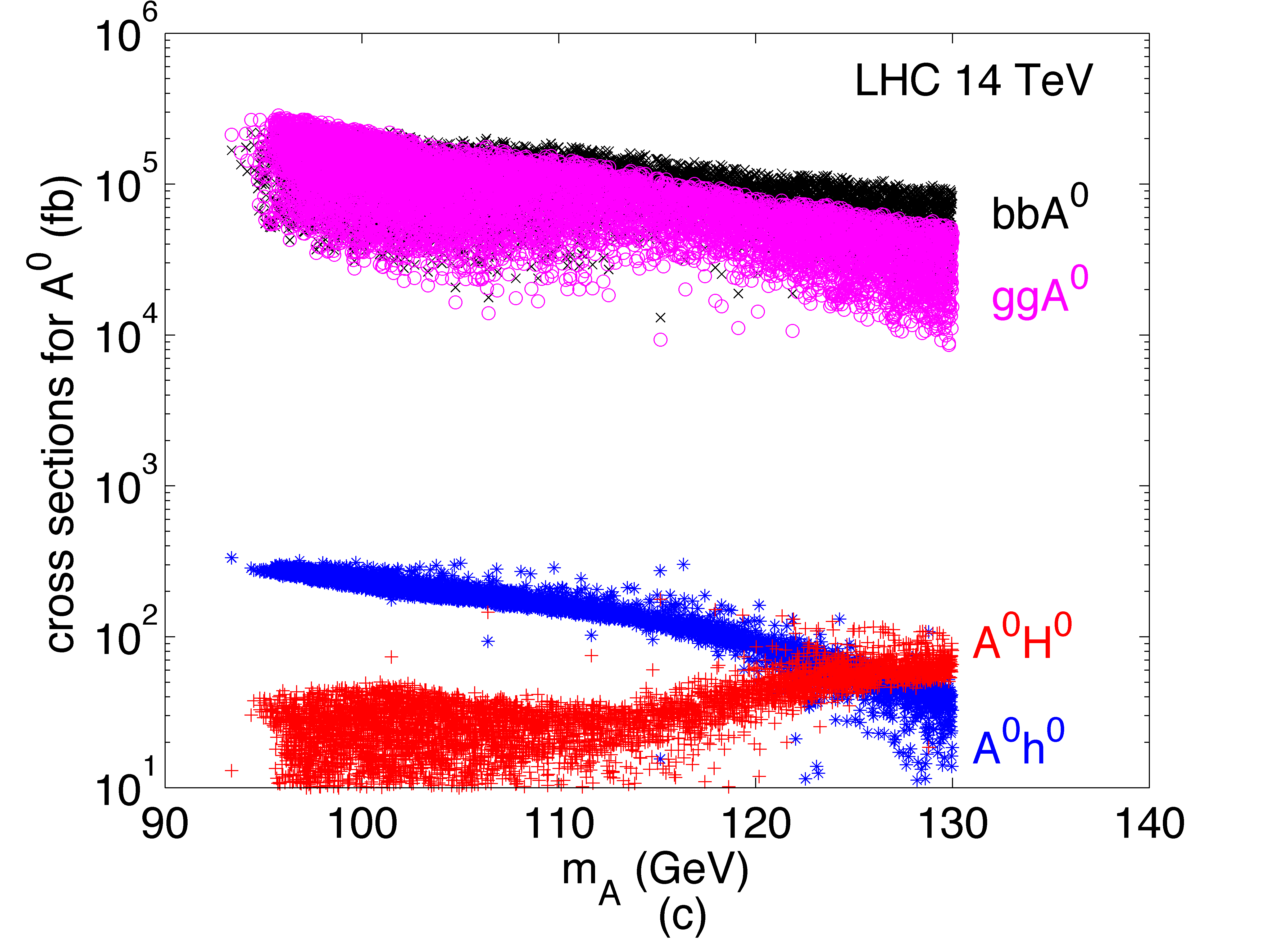}
\includegraphics[scale=1,width=7.6cm]{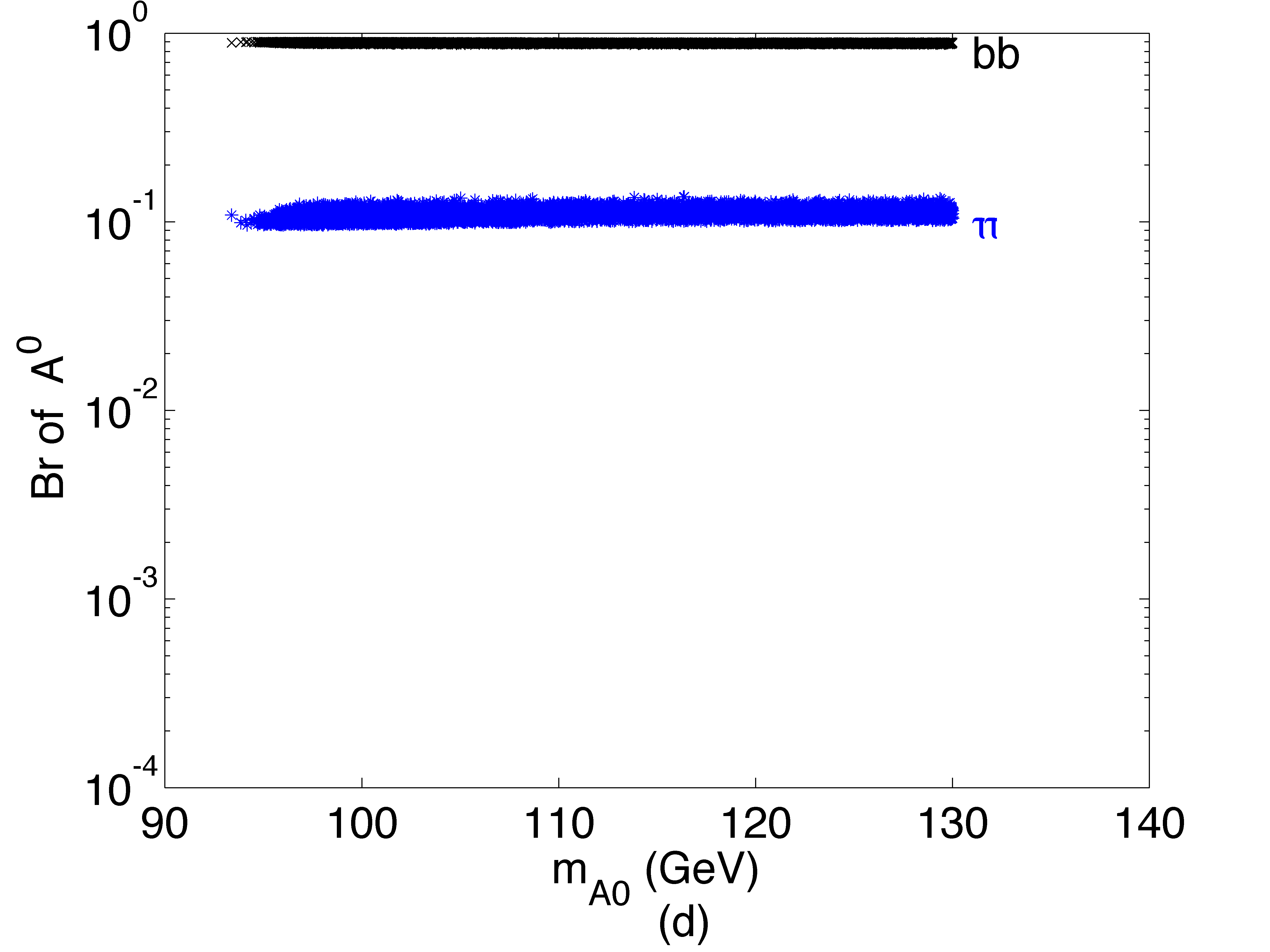}
\includegraphics[scale=1,width=7.6cm]{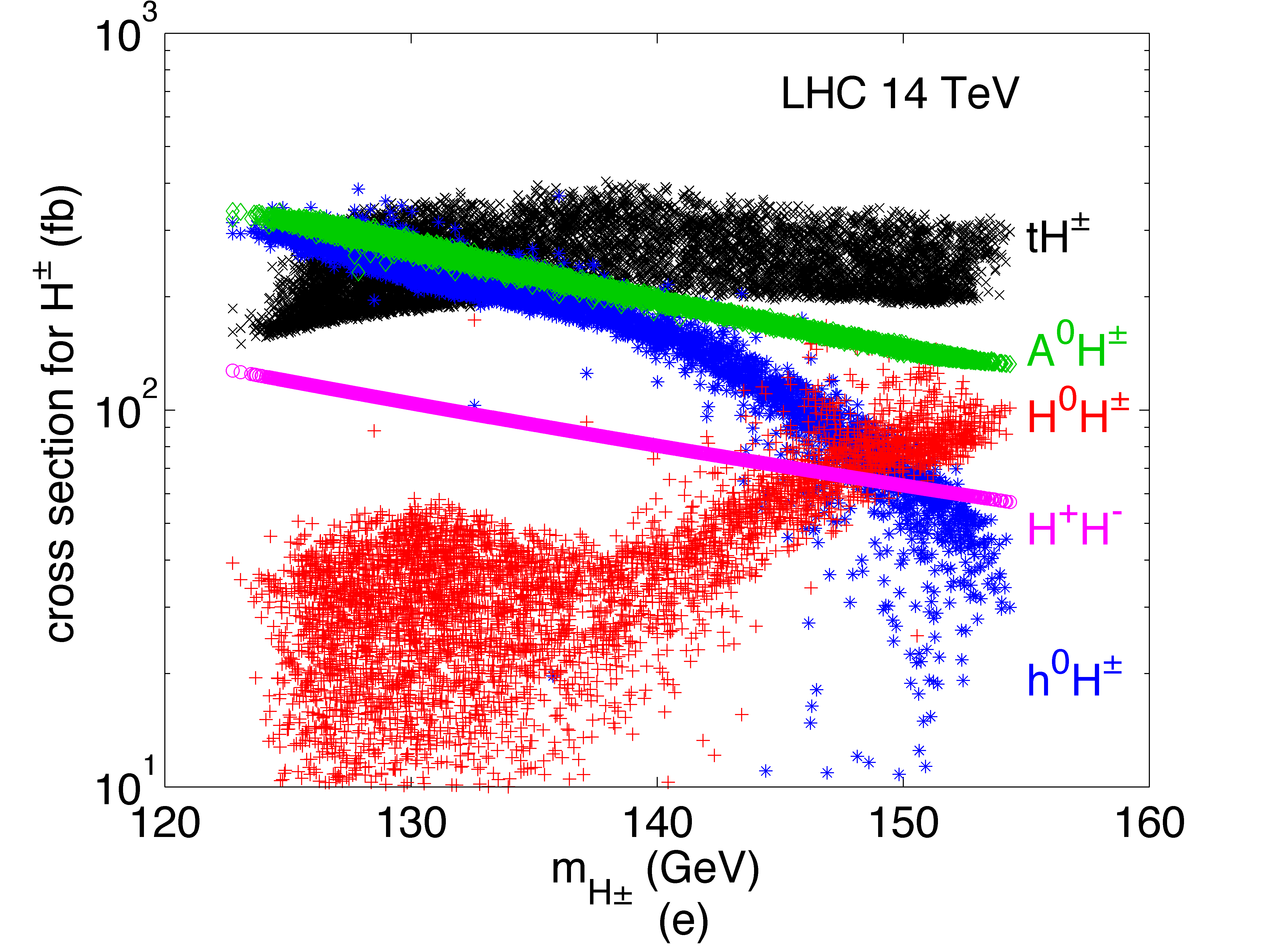}
\includegraphics[scale=1,width=7.6cm]{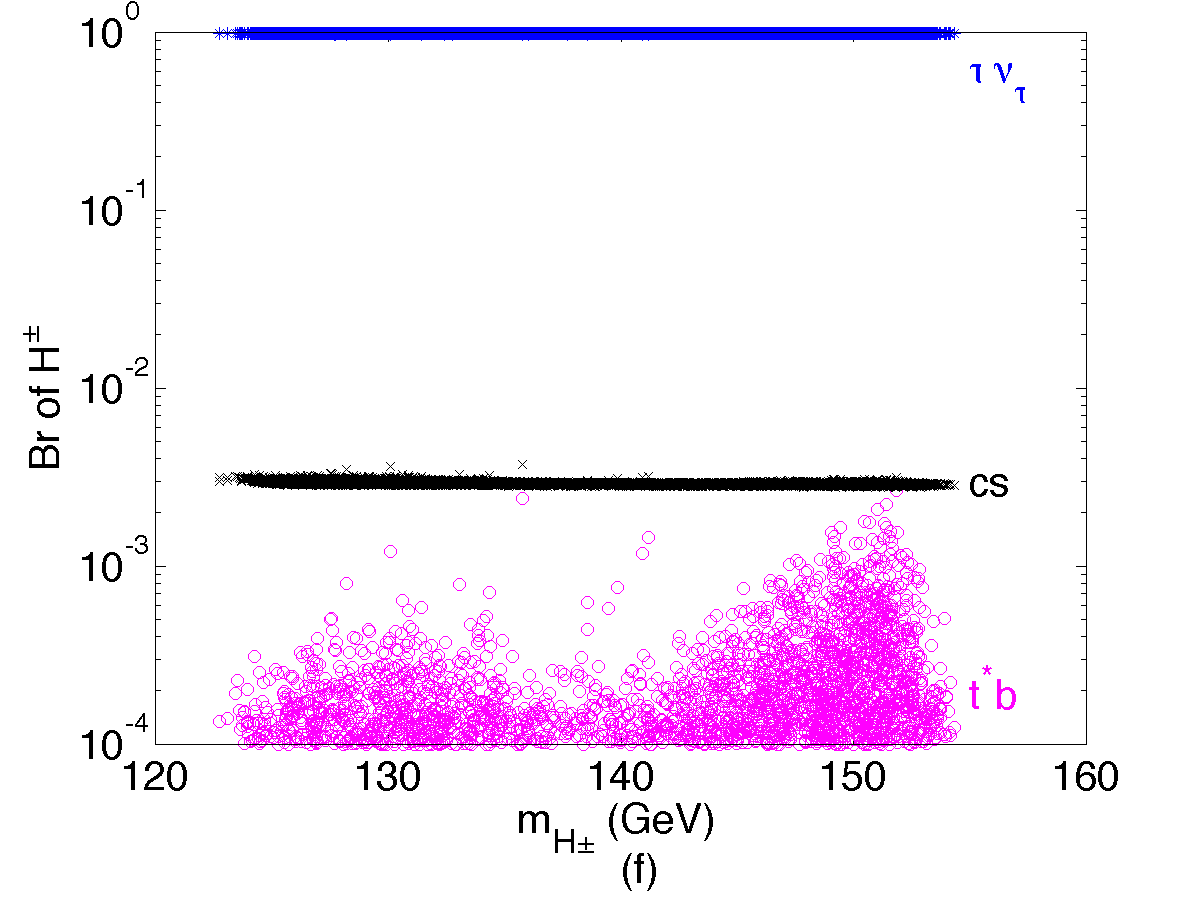}
\caption{Production cross sections at 14 TeV (left panels) and branching fractions (right panels) that satisfy all constraints for the non-SM-like Higgs bosons  in the non-decoupling region, (a) and (b) for $h^{0}$,  (c) and (d) for $A^{0}$,  (e) and (f) for $H^{\pm}$ and associate production.
}
\label{fig:nondec14tev}
\end{figure}

Guided by the results in Fig.~\ref{fig:tanb}(b), a SM-like Higgs boson in the $\gaga$ mode directs us to a possible region with low mass and non-decoupling when $A_{t} >0$. Independently, the lack of $\ww$ signal events indicates a lower cross section for the SM-like Higgs boson and thus prefers lower masses for the non-SM-like Higgs bosons. 
In this parameter region, the SM-like Higgs boson is a heavier one with $\mH\sim \mHpm \sim 125$ GeV, and the other neutral Higgs bosons are all lighter. We show their production cross sections at 14 TeV in Fig.~\ref{fig:nondec14tev} (left panels) along with the branching fractions (right panels). 
Considering the large QCD background to the $b\bar b$ final state, the preferred final state for the Higgs signals are 
$ \tau'$s \cite{CMS-tautau,CMSA0,HpmATLAS,HpmCMS}. It is encouraging that the hadronic mode from both $\tautau$ can be implemented in the search \cite{ATLAS-tautauNew}.
The events may contain one or two accompanying $b$ jets in them. We thus list the leading channels as
\begin{eqnarray}
&& b\bar b \to h^{0},\ A^{0} \to \tau^{+}\tau^{-} + 0,\ 1,\ 2\ b's,\quad gg \to h^{0},\ A^{0} \to \tau^{+}\tau^{-},\\
&& gg\to t\bar t \to H^{\pm}b + W^{\mp} b,\quad  g b \to t H^{\pm} \to Wb+\tau \nu.
%&& g b \to t H^{\pm} \to Wb + \tau \nu, \\
\end{eqnarray}
The cross sections can be quite sizable and are of the order of 100 pb for the $b \bar b$ annihilation channel, largely due to the $\tan^{2}\beta$ enhancement. The next channel is $gg\to h^{0}, A^{0}$, with a comparable cross section. 
 The production rates at the 8 TeV LHC are scaled down by roughly a factor of $2.5-3.5$.
The production cross sections of the neutral Higgs bosons, as well as $t \to H^{\pm} b$ sensitively depend on $\tan\beta$, that could vary by about one order of magnitude.
As for the decay branching fractions, they are all dominated by the heavy fermion channels that are kinematically accessible. They are rather robust with respect to other SUSY parameters. One important exception relevant to the charged Higg search is the decay $t \to H^{\pm} b$, which sensitively depends on $\tan\beta$. For instance, for $\tan\beta < 15$, the branching fraction of the top decay to $H^{+}b$ is only a few percent,

We would like to point out that for low mass, along with the contributions to the SM-like Higgs boson in Eq.~(\ref{eq:EW}), there are several additional electroweak processes that can be competitive
\begin{eqnarray}
\label{eq:gauge}
&& pp \to \gamma / Z^{*} \to H^{+} H^{-} \to \tau\nu\ \tau\nu,\qquad pp \to W^{\pm} \to H^{\pm} A^{0} \to \tau\nu+b\bar b, \\
\label{eq:nongauge}
&& pp \to Z^{*} \to A h^{0} \to \tau\tau+b\bar b , \qquad  pp \to W^{\pm} \to H^{\pm} h^{0} \to \tau\nu+b\bar b .
%&& pp \to Z^{*} \to Z H^{0} \to \ell^{+} \ell^{-}+b\bar b , \qquad  pp \to W^{\pm} \to W^{\pm} H^{0} \to \ell^{\pm} \nu+b\bar b.
\label{eq:semigauge}
\end{eqnarray}
As seen in Fig.~\ref{fig:nondec14tev}(e), the cross sections for these electroweak pair productions \cite{Dawson:1998py} are of the order of 100 fb, at the same order of magnitude as that of the associated production $t H^{\pm}$. We emphasize the potential importance of the electroweak processes of Eq.~(\ref{eq:gauge}) which are independent of the SUSY parameters except for their masses \cite{CPYuan}.
Complementarily, the production cross sections for the other processes of Eq.~(\ref{eq:nongauge}) do depend on the other SUSY parameters \cite{CPYuan2}, which may serve as a discriminator to probe the underlying theory once observed. 
%Of course, the gauge boson and SM-like Higgs boson associate production in Eq.~(\ref{eq:semigauge}) will also be present, as in the standard search, while the  non-SM-like Higgs boson associate production $(Z h^{0},\ W^{\pm} h^{0})$ will be suppressed. 
%
As for the observable signatures, it is imperative that the $\tau$ final state should be adequately identified. In this regard, it has been encouraging to see the outstanding performance by the ATLAS and CMS Collaborations. 

We summarize the leading signals and the unique electroweak processes at the LHC in 
Table~\ref{tab:nondecouple}. 
Some further investigation regarding the signal observability and background suppression is under way.
\begin{table}
\begin{tabular}{|c|c|c|c|}
\hline 
\multirow{2}{*}{Production channels} & \multirow{2}{*}{$\tau$ decay BR ($\%$)} & \multicolumn{2}{|c|}{Signal events/1 fb$^{-1}$}\\
& &  at 8 TeV & at 14 TeV \\ \hline\hline
gg,\ $b\bar b \to h^{0},\ A^{0}\to \tau^{+} \tau^{-}$  & pure leptonic: $12\%$& $480-3850$ & 1450$-$9600 \\
8 TeV:    $4\times (1-8)   \times 10^{4}$ fb$\times 10\%$ & semi leptonic:  $46\%$ & $1850-14700$ & 5200$-$37000 \\
 14 TeV: $4\times (3-20)\times 10^{4}$ fb$\times 10\%$ &   pure hadronic: $42\%$   & $1700-13500$  & $5050-33600$ \\ \hline\hline
$ g g,\ q \bar q \to t \bar t \to  W^{\pm}b\  H^{\mp} b$ & & & \\
8 TeV:     $2\times 2.3 \times 10^{5}$ fb$\times 2\%$ & leptonic: $35\%$& $3200$ & 12600 \\
 14 TeV: $2\times 9 \times 10^{5}$ fb$\times 2\%$ & hadronic: $65\%$  &  6000 &  23400\\ \hline\hline
% t -> H+- b: arxiv:0710.1761 hep-ph, mH 130 GeV, tanb = 10-16
$ g b \to t H^{\pm} \to W^{\pm}b\ \tau^{\mp} \nu$ & & & \\
    8 TeV: $(32-74)$ fb  & leptonic: $35\%$ & $11-26$ & $53-123$ \\
  14 TeV: $(150-350)$ fb & hadronic: $65\%$ & $21-48$ & $98-230$ \\ \hline
\hline
$ q\bar q \to H^{\pm} A^0,\ H^{\pm} h^{0} \to \tau^{\pm}\nu\ b\bar b$ & & & \\ 
8 TeV: $2\times (100-150)$ fb$\times 90\%$  &  leptonic: $35\%$ & $63-95$ &  $126-189$ \\
  14 TeV: $2\times (200-300)$ fb$\times 90\%$  &  hadronic: $65\%$ & $117-176$ & $234-351$ \\ \hline
\hline
$ q\bar q \to H^{+} H^{-} \to \tau^{+}\nu\ \tau^{-}\nu $ & pure leptonic: $12\%$ & $4.8$ & $12$ \\
8 TeV: $40 $ fb  &  semi leptonic:  $46\%$  & $18$ & $46$ \\
  14 TeV: $100$ fb &  pure hadronic: $42\%$ & $17$ & $42$ \\ \hline
\hline
$ q\bar q \to A^{0} h^{0}\to \tautau\ b\bar b$ & pure leptonic: $12\%$ & $2.2-3.2$ &  $4.3-6.5$\\
8 TeV: (100-150) fb$\times 18\%$  &  semi leptonic:  $46\%$  & $8.3-12$ & $17-25$ \\
  14 TeV: (200-300) fb $\times 18\%$  & pure hadronic: $42\%$   & $7.6-11$ & $15-23$ \\ 
\hline
\end{tabular}
\caption{Signal channels and rates at the 8 TeV and 14 TeV LHC for the non-SM-like Higgs bosons in the non-decoupling region with $\ma\approx 100$ GeV and $m_{H^{\pm}}\approx 128$ GeV. The cross section ranges reflect the variation of $\tan\beta \approx 10-15$.
}
\label{tab:nondecouple}
\end{table}

%%%%%%%%%%%%%%%%%%%%%%%%%%%%%%

\begin{figure}[tb]
\includegraphics[scale=1,width=7.6cm]{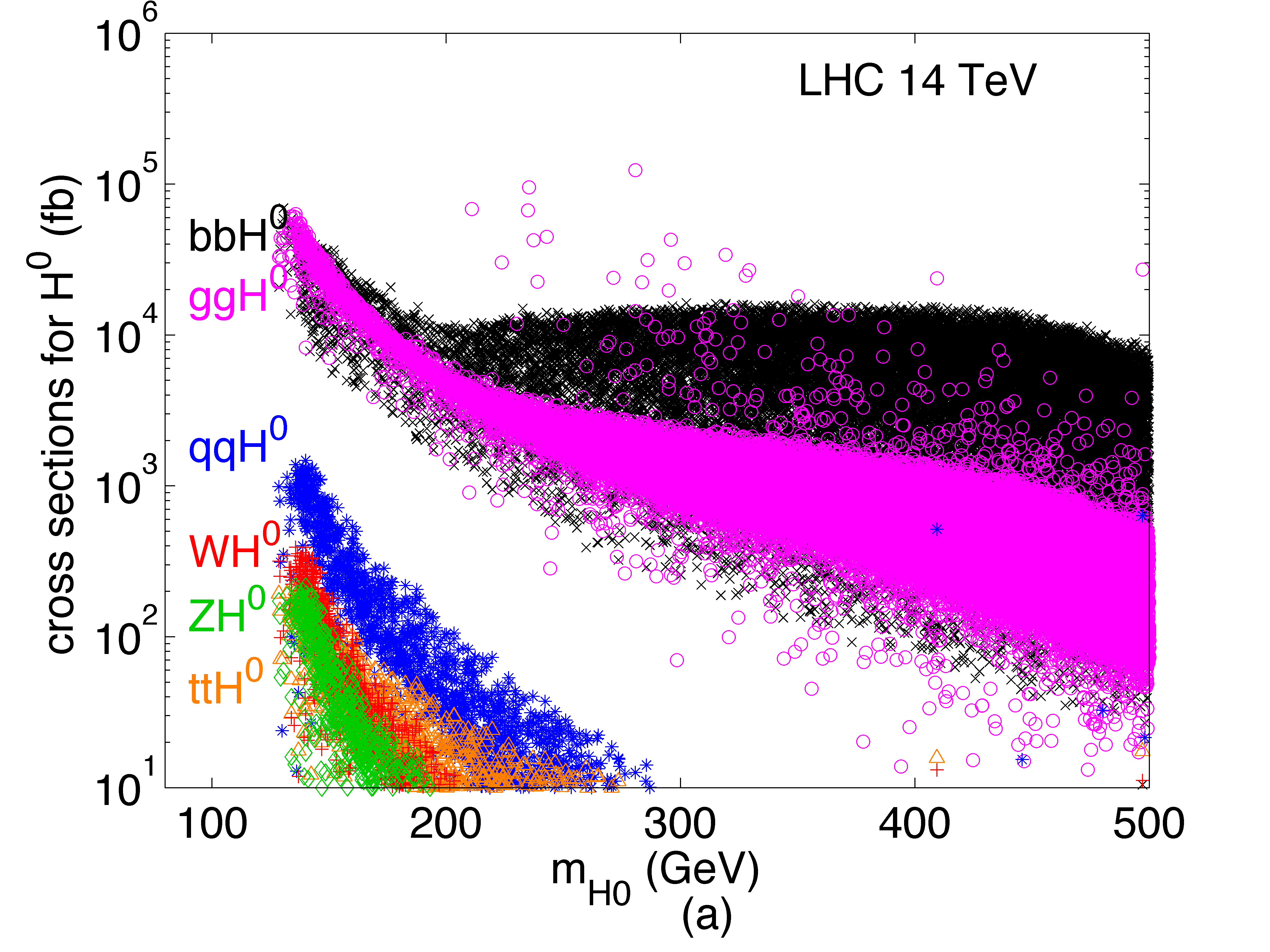}
\includegraphics[scale=1,width=7.6cm]{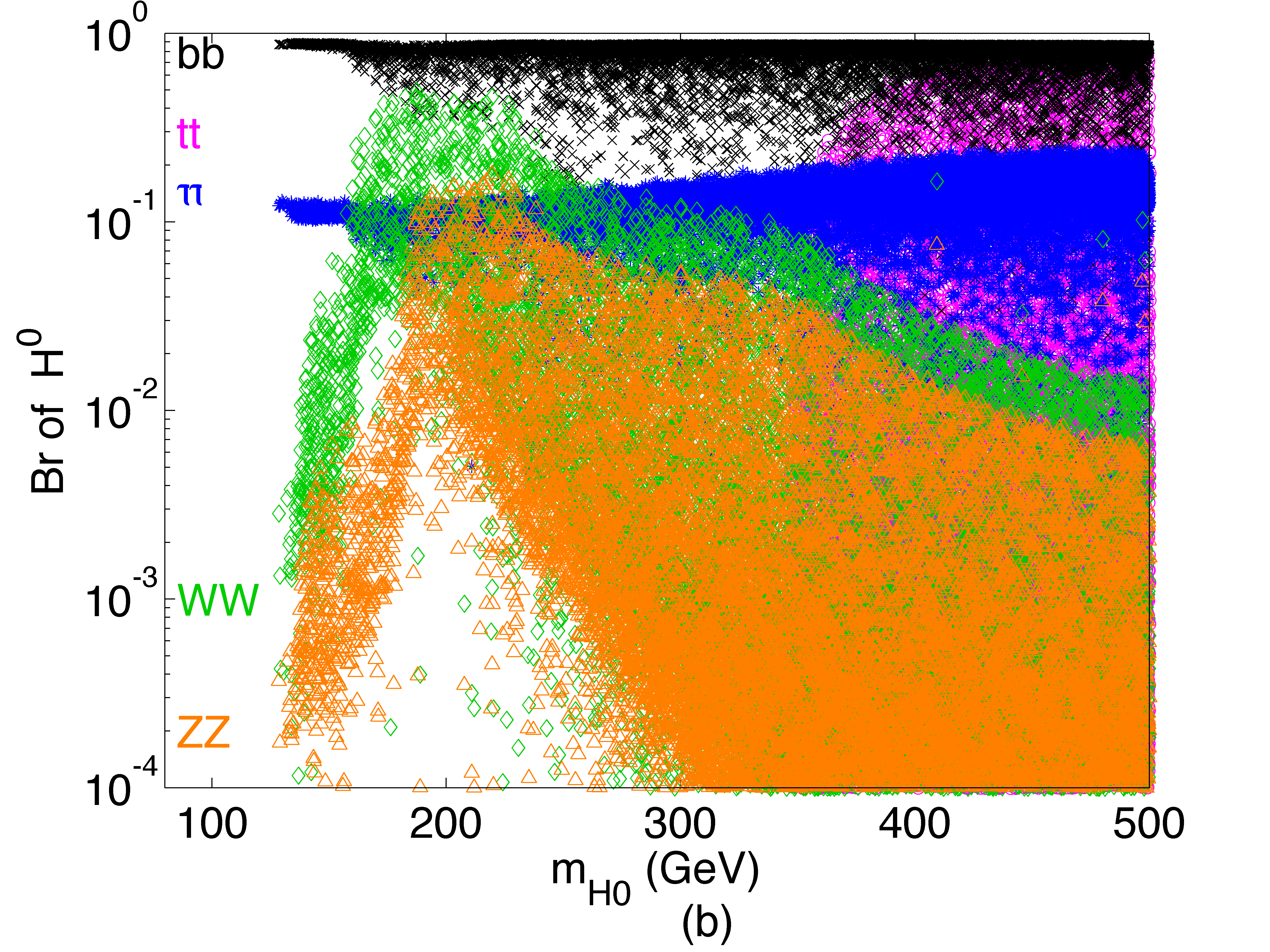}
\includegraphics[scale=1,width=7.6cm]{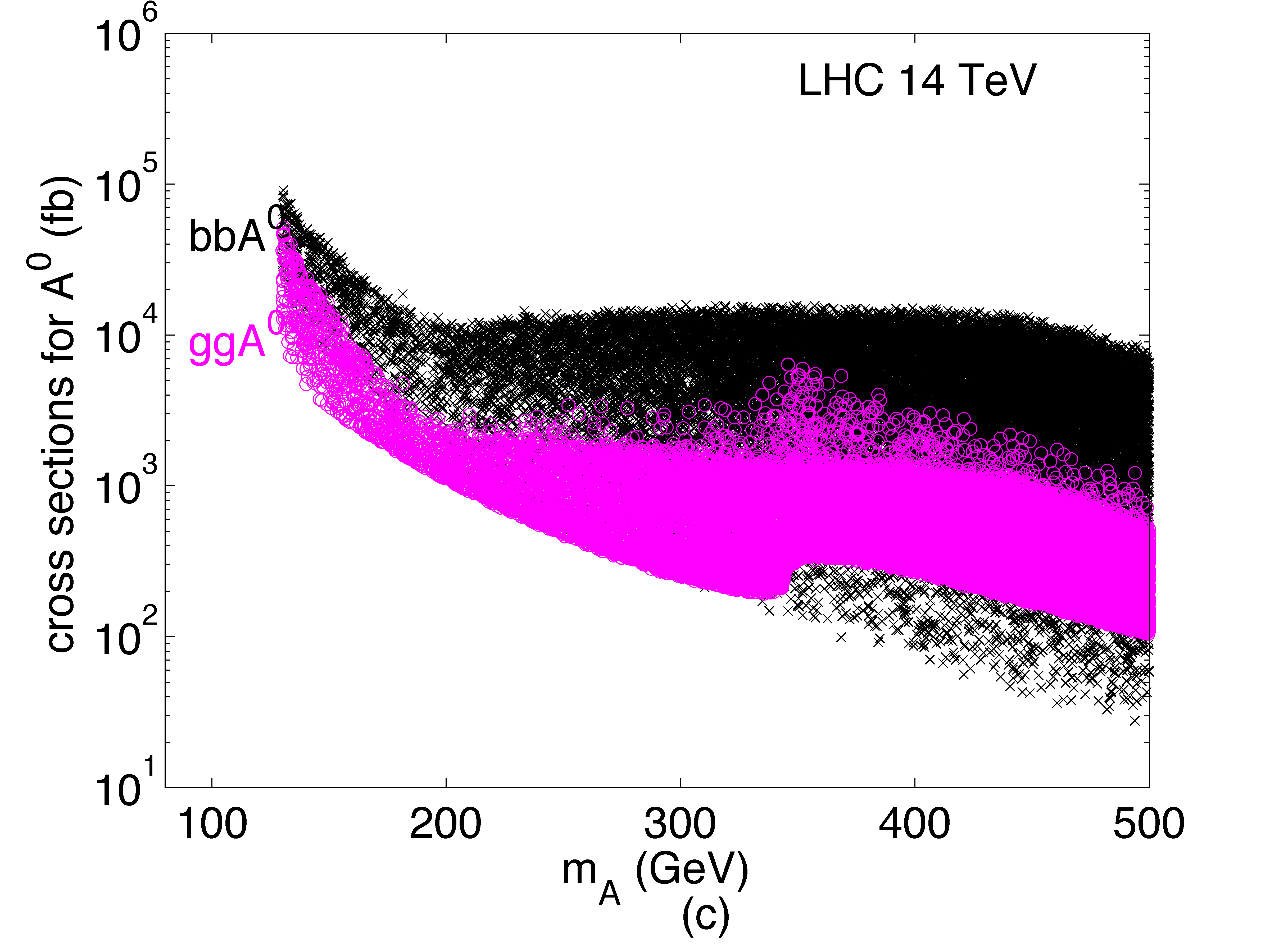}
\includegraphics[scale=1,width=7.6cm]{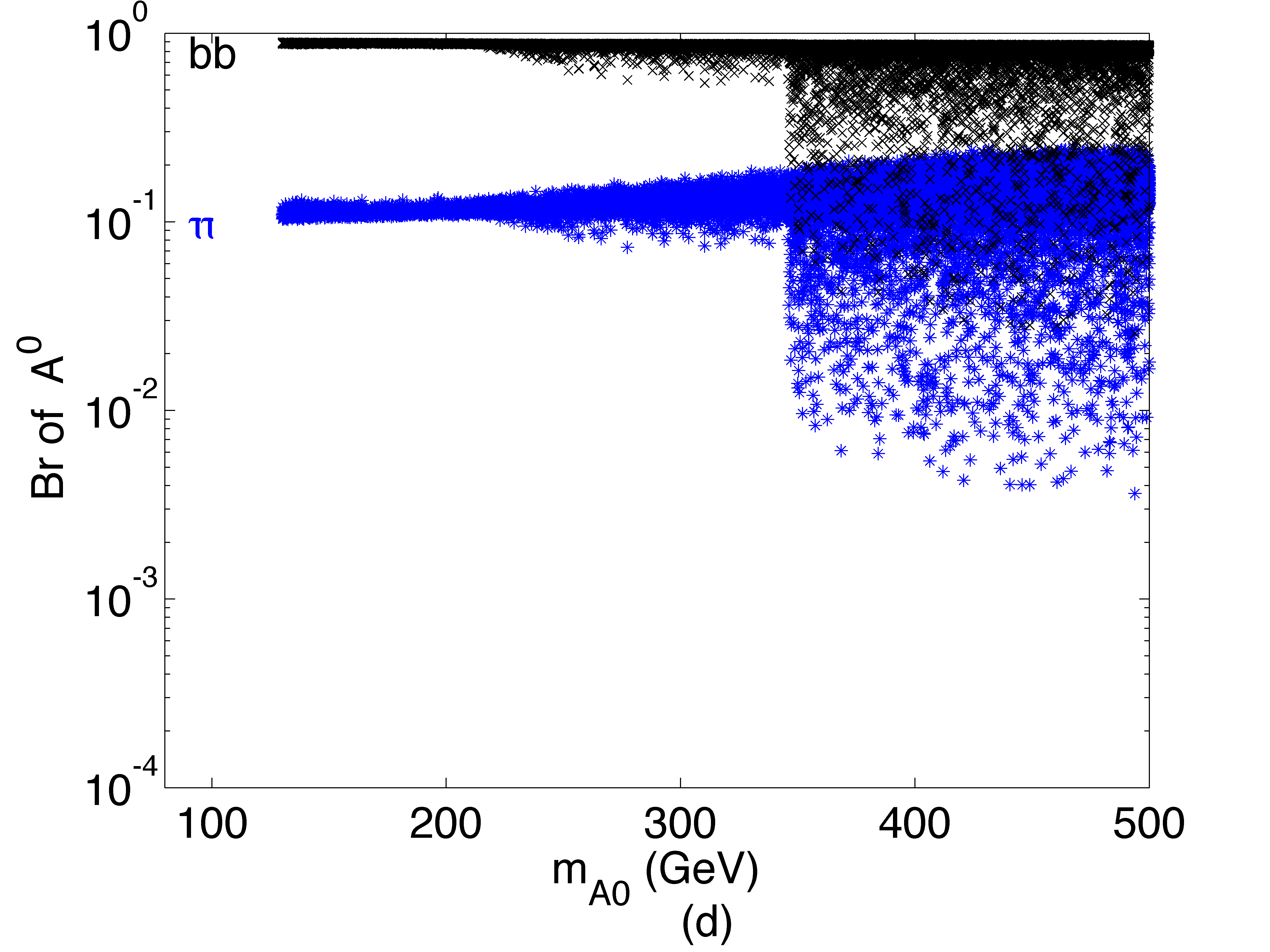}
\includegraphics[scale=1,width=7.6cm]{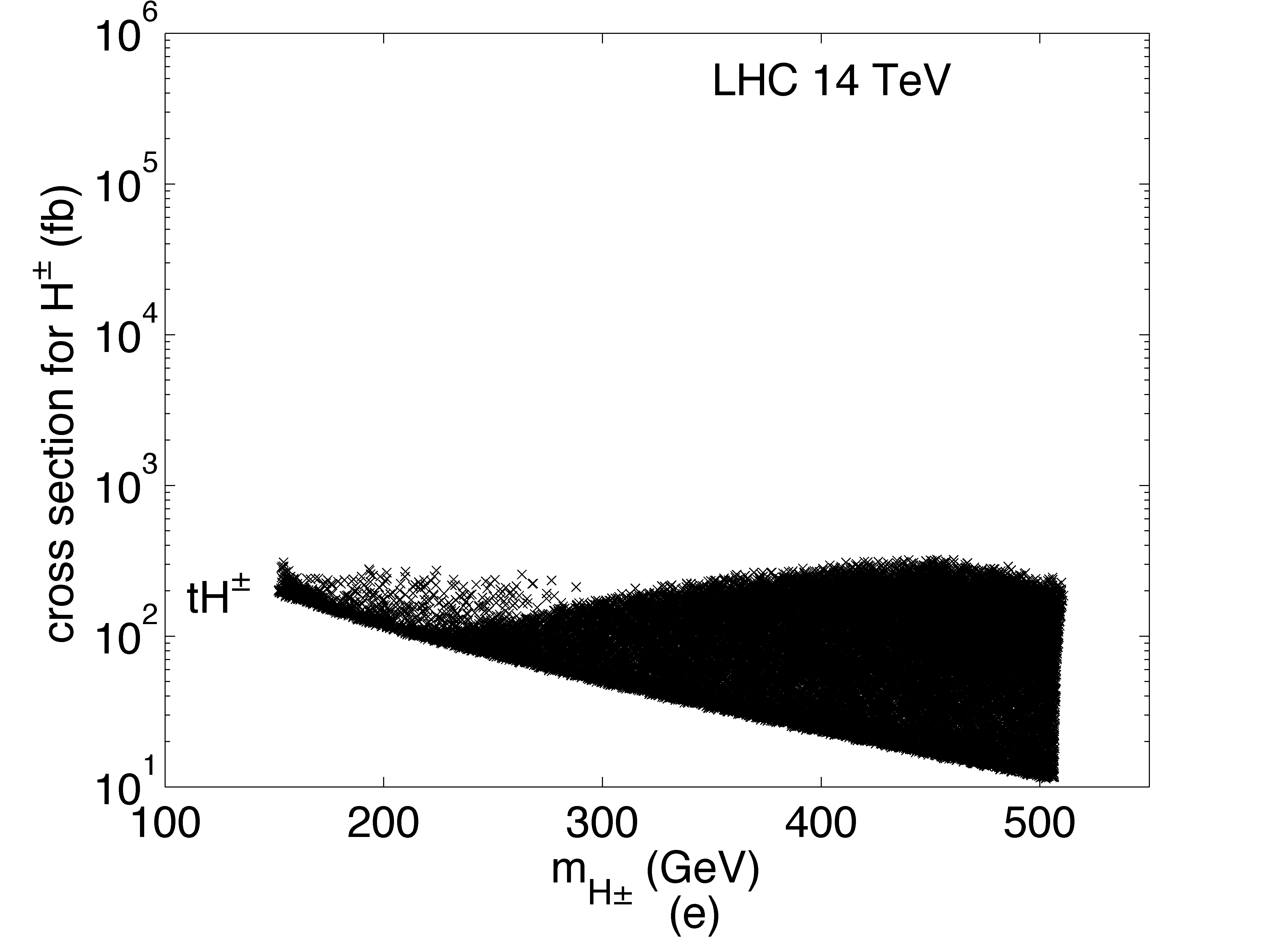}
\includegraphics[scale=1,width=7.6cm]{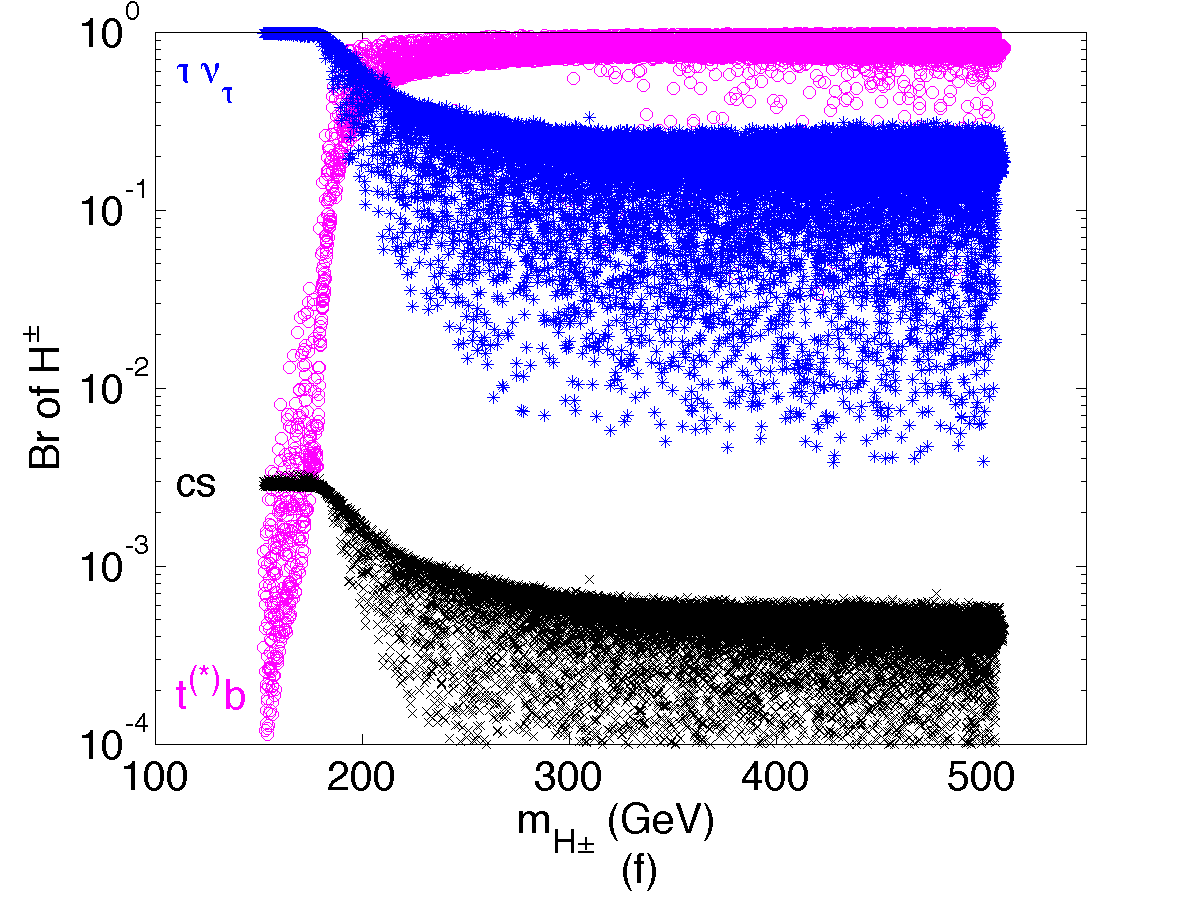}
\caption{Production cross sections at 14 TeV (left panels) and branching fractions (right panels) that satisfy all constraints for the non-SM-like Higgs bosons  in the decoupling region, (a) and (b) for $H^{0}$,  (c) and (d) for $A^{0}$,  (e) and (f) for $H^{\pm}$.   
%(g) for the combined production of $A^0$ and $h^0$, $H^0$ or $H^\pm$.
}
\label{fig:dec14tev}
\end{figure}
%
%\begin{figure}[tb]
%\includegraphics[scale=1,width=7.6cm]{sigma14_HH_loose_decouple.png}
%\includegraphics[scale=1,width=7.6cm]{decay_HH_loose_decouple.png}
%\includegraphics[scale=1,width=7.6cm]{sigma14_A0_loose_decouple.png}
%\includegraphics[scale=1,width=7.6cm]{decay_A0_loose_decouple.png}
%\includegraphics[scale=1,width=7.6cm]{sigma14_Hpm_loose_decouple_scale.png}
%\includegraphics[scale=1,width=7.6cm]{decay_Hpm_loose_decouple.png}
%\caption{Production cross sections at 14 TeV (left panels) and branching fractions (right panels) that satisfy all constraints for the non-SM-like Higgs bosons  in the decoupling region, (a) and (b) for $H^{0}$,  (c) and (d) for $A^{0}$,  (e) and (f) for $H^{\pm}$. 
%(g) for the combined production of $A^0$ and $h^0$, $H^0$ or $H^\pm$.
%}
%\label{fig:dec14tev}
%\end{figure}
%

%%%%%%%%%%%%%%%%%%
%\subsection{Leading signal channels and backgrounds: Decoupling region}
\subsection{Decoupling region: $\mH\sim \mHpm \sim \ma > 300\ \gev$}
\label{sec:channels_decouple}

Again motivated by the results seen as the light blue triangles in Figs.~\ref{fig:csR}(b) and \ref{fig:tanb}(b), a SM-like Higgs boson in the $\gaga$ mode with a sizable production rate could push $\ma$ toward the higher value in the decoupling regime.
In this region, the non-SM-like Higgs bosons are nearly degenerate and all heavier than 300 GeV. Their dominant couplings are those to the heavy fermions, that dictate production and decay channels. 
The six panels in Fig.~\ref{fig:dec14tev} show the total cross sections (left panels) at the 14 TeV LHC and decay branching fractions (right panels) for the leading channels. Again, the results for the production cross sections at a 8 TeV LHC will scale down by roughly a factor of $2.5-3.5$.
At tree level, the branching fractions are simply given by the mass ratios 
%at an appropriate scale 
and $\tan^{2}\beta$.
The band spreads are mainly due to the variation of $\tan\beta$ at tree level and to a lesser extent to other SUSY parameters at one loop. 

Similar to the non-decoupling case, the leading production channels are $b\bar b \to H^{0}, A^{0}$ at the order of $0.1-10$ pb,  and the next one for $gg\to H^{0}, A^{0}$ with a comparable or smaller rate. 
Although even smaller by another order of magnitude as seen in Fig.~\ref{fig:dec14tev}(e), 
the $t H^{\pm}$ channel is of unique kinematics and may be feasible to search for. 
%
%\begin{eqnarray}
%&& b\bar b \to H^{0},\ A^{0} \to \tau^{+}\tau^{-} + 0,\ 1,\ 2\ b's,\quad  gg \to H^{0},\ A^{0} \to \tau^{+}\tau^{-} \\
%&& g b \to t H^{\pm} \to Wb+ \tau \nu,\ Wb+ Wb\ b .
%\end{eqnarray}

Based on our results in the figures above, we summarize the leading signals in Table \ref{tab:decouple}, where we list the signal channels and their rates at the 8 TeV (14 TeV) LHC. 
There exist comprehensive studies for most of the signals listed above \cite{ATLASTDR,CMSTDR,Gennai:2007ys,Sven2}. 
There are also recent experimental searches for the neutral Higgs states at the LHC \cite{CMS-tautau,CMSA0}
and charged state at the Tevatron \cite{CDFD0}, which have been implemented in the previous figures. Efforts for the search are continuing in the LHC experiments.

\begin{table}
\begin{tabular}{|c|c|c|c|}
\hline 
\multirow{2}{*}{Production channels} & \multirow{2}{*}{$\tau$ decay BR ($\%$)} & \multicolumn{2}{|c|}{Signal events/1 fb$^{-1}$}\\
& &  at 8 TeV & at 14 TeV \\ \hline\hline
$b\bar b \to H^{0},\ A^{0}\to \tau^{+} \tau^{-}$  & pure leptonic: $12\%$& $0.04-96 $ & $0.2-480$\\
8 TeV: $2\times (20-2000)$ fb$\times (0.8-20)\%$ & semi leptonic:  $46\%$ & $0.15-370$ & $0.7-1840$ \\
 14 TeV: $2\times (10^{2}-10^{4})$ fb$\times (0.8-20)\%$ & pure
 hadronic: $42\%$  & $0.1-336$ & $0.7-1700$ \\ \hline\hline
$ gg \to H^{0},\ A^{0} \to \tau^{+} \tau^{-} $ & pure leptonic: $12\%$ & $0.05-24$ & $0.2-96$ \\
8 TeV: $2\times (25-500)$ fb$\times (0.8-20)\%$  &  semi leptonic:  $46\%$  & $0.2-92$ & $0.7-370$ \\
  14 TeV: $2\times (100-2000)$ fb$\times (0.8-20)\%$  & pure hadronic: $42\%$  & $0.2-84$ & $0.7-340$ \\ 
\hline
\hline
   $ g b \to t H^{\pm} \to W^{\pm}b\ \tau^{\mp} \nu$ &  & & \\
8 TeV: $2\times (5-60)$ fb$\times (0.5-30)\%$  & leptonic: $35\%$ & $0.02-13$ & $0.07-53$ \\
  14 TeV: $2\times (20-250)$ fb$\times (0.5-30)\%$  & hadronic
  $65\%$& $0.03-23.5$ & $0.1-98$ \\ 
 \hline
\end{tabular}
\caption{\label{tab:decouple}Signal channels and rates at the 8 TeV and 14 TeV LHC for the non-SM-like Higgs bosons in the decoupling region
with $\ma\approx 400$ GeV. The cross section ranges reflect the variation of $\tan\beta \approx 20-40$.}
\end{table}

%%%%%%%%%%%%%%%%%%%%%%%%%%%%%%%%%%%%%%%%%%%%% 

\section{Conclusions}
\label{sec:Conclude}

In light of the powerful results presented by ATLAS and CMS for the SM Higgs boson searches at the LHC,
along with the data from the LEP2 and Tevatron, we reexamined the MSSM Higgs sector for their masses, couplings and other related SUSY parameters. Instead of only presenting benchmark scenarios, we allowed variations of other SUSY parameters in a broad range.

If we accept the existence of a SM-like Higgs boson in the mass window of 123 GeV$-$127 GeV as indicated by the observed $\gamma\gamma$ events, we found that there are two distinctive mass regions left in the MSSM Higgs sector:  (a) the lighter CP-even Higgs boson being SM-like and the non-SM-like Higgs bosons all heavy and nearly degenerate above 300 GeV (an extended decoupling region); (b) the heavier CP-even Higgs boson being SM-like and the neutral non-SM-like Higgs bosons all nearly degenerate around 100 GeV (a small non-decoupling region). 
These features were shown in Figs.~\ref{fig:csR}(a) and~\ref{fig:csR}(b).

On the other hand, due to the strong positive correlation between the Higgs decays to $\ww$ and to $\gaga$ predicted in the MSSM, as seen in Fig.~\ref{fig:csR}(c) and Eq.~(\ref{eq:r}), the observed $\gaga$ signal and the apparent absence of the $\ww$ final state signal near the peak would be mutually exclusive to each other. Namely, the suppression to the $\ww$ channel would automatically reduce the $\gaga$ channel. In fact, the theoretical expectation for the $\gaga$ signal in the MSSM relative to that in the SM is even smaller than that for the $\ww$ channel [$\it e.g.$ Eq.~(\ref{eq:r})].
To accommodate both the $\ww$ deficit and the $\gaga$ enhancement, physics beyond the MSSM would be needed.
We also found another interesting inverse correlation between the Higgs decays to $\ww$ and to $\tautau$, as seen in Fig.~\ref{fig:csR}(d). The suppression to the $\ww$ channel would automatically force the $\tautau$ channel to be larger.

If the absence of the $\ww$ signal persists and the observation is strengthened for an extended mass range in the future run at the LHC, it  would imply that the SM-like Higgs boson has reduced couplings to $W^{\pm},\ Z$, rendering it less SM-like. Although less statistically significant, the lack of the $\tautau$ final state signal could also contribute to reach a consistent picture.  Consequently, the other non-SM-like Higgs bosons cannot be deeply into the decoupling regime, and thus cannot be too heavy, typically below 350 GeV, making them more accessible at the LHC.

%Both of the above cases could accommodate a SM-like Higgs boson with in the MSSM. However, 

Future searches for the SM-like Higgs boson at the LHC will provide critical tests for the MSSM predictions for those points, as presented in Sec.~\ref{sec:Future}. 
Guided by those observations, we studied the signals predicted for the non-SM-like Higgs bosons satisfying the current bounds. 
Along with the standard searching processes $q\bar q H^{0},\ W^{\pm} H^{0},\ ZH^{0}$ as shown in 
Fig.~\ref{fig:cs_h}(c), 
we emphasize the potential importance of the electroweak processes $pp\to H^{+}H^{-},\ H^{\pm} A^{0}$ in Fig.~\ref{fig:nondec14tev}(e), which are independent of the SUSY parameters except for their masses. In addition, there may be sizable contributions from $pp\to H^{\pm} h^{0},\ A^{0} h^{0}$ in the low-mass non-decoupling region, which may serve to discriminate the model parameters.
These cross sections can be as large as that of the $tH^\pm$ associated production, which sensitively depends on 
$\tan\beta$.

The stringent constraints also imply nontrivial correlation and prediction to some other SUSY parameters relevant to the Higgs sector, such as $\mu,\ A_{t},\ M_{3SQ},\ M_{3SU}$ etc. Further explorations may lead to predictions for other SUSY signals for gaugino and stops. Over all, the search for the SM Higgs boson will prove crucial in understanding the SUSY Higgs sector.

\acknowledgments
We would like to thank   S.~Heinemeyer, T.~Hahn, C. Wagner and H.~Baer for helpful discussions. The work of  N.C.~and T.H.~was supported in part by PITT PACC, and in part by the LHC-TI under U.S. National Science Foundation, grant NSF-PHY-0705682. The work of S.S.~was supported by the Department of Energy under 
Grant~DE-FG02-04ER-41298. 

%%%%%%%%%%%%%%%%%%%%%%%%%%%%%%%%%%%%%%%%%%%%%%

\end{document}